\documentclass[11pt, a4paper]{article}
\usepackage{jheppub}
\usepackage{amsmath}         
\usepackage{amssymb}          
\usepackage{bbold}
\usepackage{xcolor}
\usepackage{ulem}
\usepackage{afterpage}
\usepackage{youngtab}
\usepackage{multirow}
\usepackage[capitalise]{cleveref}
\usepackage{slashed}
\usepackage{booktabs}
\usepackage{subcaption}
\usepackage{comment}
\usepackage{rotating}
\usepackage{url}

\newcommand{\ra}[1]{\renewcommand{\arraystretch}{#1}}

\newcommand{\rescol}[2]{\textcolor{blue}{#1} \textcolor{red}{#2}}

\crefname{section}{Sec.}{Secs.}
\crefname{table}{Tab.}{Tabs.}
\crefname{figure}{Fig.}{Figs.}
\crefname{equation}{Eq.}{Eqs.}
\crefname{appendix}{Appendix}{Appendix}

\newcommand{\ord}[1]{\mathcal{O}\left({#1}\right)}

\newcommand{\hc}{\text{h.c.}}
\newcommand{\SO}{\text{SO}}
\newcommand{\SU}{\text{SU}}
\newcommand{\U}{\text{U}}
\newcommand{\Sp}{\text{Sp}}

\newcommand{\Tr}{\text{Tr}}

\newcommand{\br}{\mathrm{Br}}
\newcommand{\dlr}{\overleftrightarrow{\partial_\mu}}

\def\gsim{\raise0.3ex\hbox{$\;>$\kern-0.75em\raise-1.1ex\hbox{$\sim\;$}}}
\def\lsim{\raise0.3ex\hbox{$\;<$\kern-0.75em\raise-1.1ex\hbox{$\sim\;$}}}

\begin{document}

\title{Exploring extended Higgs sectors via pair production at the LHC}

\author[a,b]{G.\ Cacciapaglia,}
\author[c]{T.\ Flacke,}
\author[d]{M.\ Kunkel,}
\author[d]{W.\ Porod,}
\author[d]{L.\ Schwarze}

\affiliation[a]{Institut de Physique des Deux Infinis de Lyon (IP2I), CNRS/IN2P3, 4 rue Enrico Fermi, 69622 Villeurbanne Cedex, France}
\affiliation[b]{Universit\'e Claude-Bernard Lyon 1, Univ Lyon, Lyon, France}
\affiliation[c]{Center for AI and Natural Sciences, KIAS, Seoul 02455, Korea}
\affiliation[d]{Institut f\"ur Theoretische Physik und Astrophysik, Uni W\"urzburg, Emil-Hilb-Weg 22,
D-97074 W\"urzburg, Germany}

\emailAdd{g.cacciapaglia@ip2i.in2p3.fr}
\emailAdd{flacke@kias.re.kr}
\emailAdd{manuel.kunkel@physik.uni-wuerzburg.de}
\emailAdd{leonard.schwarze@stud-mail.uni-wuerzburg.de}
\emailAdd{porod@physik.uni-wuerzburg.de}

\abstract{Higgs sectors extended by electroweakly charged scalars can be explored by scalar pair production at the LHC. We consider a fermiophobic scenario, with decays into a pair of gauge bosons, and a fermiophilic one, with decays into top and bottom quarks. After establishing the current bounds on simplified models, we focus on an SU(5)/SO(5) composite Higgs model. This first exploration demonstrates the need for dedicated searches at current and future colliders.}
\preprint{KIAS-A22008}

\maketitle

\newpage

\section{Introduction}

The Standard Model (SM) of particle physics contains a single scalar field, a doublet of weak isospin $\SU(2)_L$ that is responsible for the breaking of the electroweak (EW) symmetry \cite{Englert:1964et,Higgs:1964pj}. Upon acquiring a vacuum expectation value (VEV), a massive physical scalar particle arises, the famous Higgs boson \cite{Higgs:1964pj} discovered in 2012 at the Large Hadron Collider (LHC) experiments \cite{ATLAS:2012yve,CMS:2012qbp}. However, most models of new physics feature extended Higgs sectors: for instance, minimal supersymmetric models~\cite{Martin:1997ns} and two Higgs doublet models~\cite{Branco:2011iw} feature a second doublet, type-II seesaw models~\cite{Schechter:1980gr} feature a zero hypercharge triplet, triplets appear also in the Georgi-Machacek model~\cite{Georgi:1985nv}, while larger representations appear in the custodial-preserving septet model~\cite{Hisano:2013sn}. In all these scenarios, the scalar fields acquire sizeable couplings to the SM gauge bosons and fermions via VEVs and/or via mixing with the SM Higgs boson. Hence, they are dominantly singly-produced at colliders, and most current searches are focused on these channels.

Single production, however, is always model dependent and it can be suppressed by tuning small single-scalar couplings. In contrast, pair production only depends on the gauge quantum numbers of the scalars, and cannot be tuned to be small. The couplings of two scalars from $\SU(2)_L \times \U(1)_Y$ multiplets to the EW gauge bosons arise from the covariant derivatives in the scalar kinetic terms and are, therefore, always present. They generate the dominant pair production via Drell-Yan, where two initial state quarks merge via an s-channel gauge boson. The kinetic term also yields a coupling of two scalars to two gauge bosons that, via vector boson fusion, contributes to the scalar pair production. However, this process is subdominant as compared to Drell-Yan \cite{Agugliaro:2018vsu}. If the EW symmetry were preserved, the  Drell-Yan pair production cross section for a given scalar multiplet would be a function of the scalar mass, only.  Via the EW breaking, the scalar gauge eigenstates can mix through the scalar potential, hence the couplings of two (physical) mass eigenstates to the EW gauge bosons (and thus the Drell-Yan pair production cross sections) acquire model dependence. Nevertheless, the mass mixing cannot reduce Drell-Yan pair production cross sections of all scalars at the same time and some channels are guaranteed to remain sizeable.  Other channels may be present via loops or scalar self couplings: famously, the Higgs boson itself is pair-produced  via the Higgs triple coupling as well as one-loop top box contributions. In this work, we focus on models where pair production is the dominant  mode for scalars charged under the EW gauge symmetry. This scenario appears naturally in composite Higgs models, where the Higgs boson is accompanied by additional light states, protected by parities internal to the strong sector~\cite{Ferretti:2016upr}.

In composite Higgs models, the Higgs boson emerges as a pseudo-Nambu-Goldstone boson (pNGB) \cite{Kaplan:1983fs} following the dynamical breaking of the EW symmetry triggered by misalignment in a condensing strong dynamics at the TeV scale \cite{Weinberg:1975gm,Dimopoulos:1979es}. It may well be accompanied by additional light meson-like scalars.
In fact, based only on the global symmetries, a minimal model $\SO(5)/\SO(4)$ with 4 pNGBs matching the Higgs doublet components can be constructed \cite{Agashe:2004rs}  based on holography \cite{Contino:2003ve}. However, it is not easy to obtain this symmetry pattern in an underlying gauge/fermion theory {\it \`a la} QCD. A fermion condensate $\langle \psi \psi \rangle$ can only generate the following patterns: $\SU(2N)/\Sp(2N)$, $\SU(N)/\SO(N)$ or $\SU(N)^2/\SU(N)$ depending on whether the representation of $\psi$ under the confining gauge symmetry is pseudo-real, real or complex \cite{Witten:1983tw,Kosower:1984aw}, respectively. Hence, from the point of view of the underlying gauge theory, the minimal model with custodial symmetry \cite{Georgi:1984af,Agashe:2006at} features $\SU(4)/\Sp(4)$ \cite{Ryttov:2008xe,Galloway:2010bp,Cacciapaglia:2014uja}, and has one additional pNGB besides the Higgs doublet. The next-to-minimal cases contain many more pNGBs: $15$ for $\SU(6)/\Sp(6)$ \cite{Low:2002ws,Cai:2019cow}, $14$ for $\SU(5)/\SO(5)$ \cite{Dugan:1984hq,Arkani-Hamed:2002ikv} and $15$ for $\SU(4)^2/\SU(4)$ \cite{Ma:2015gra,Vecchi:2015fma}. For other non-minimal patterns see Refs~\cite{Mrazek:2011iu,Bellazzini:2014yua}. Note that departure from minimality is not in contradiction with the null results of direct Beyond-the-Standard-Model (BSM) searches at colliders: the pNGBs are typically heavier than the Higgs and only have EW interactions, hence being very hard to discover at hadron colliders and too heavy for past $e^+ e^-$ colliders such as LEP. Other resonances, like baryon-like top partners needed for top partial compositeness \cite{Kaplan:1991dc}, can be much heavier.

Electroweak pNGBs have recently been studied in the context of exotic decays of top partners \cite{Bizot:2018tds,Xie:2019gya,Benbrik:2019zdp,Cacciapaglia:2019zmj,Banerjee:2022xmu}, as the latter have sizeable production cross sections at hadronic colliders like the LHC at CERN.
In this work, we instead focus on the pNGB direct production via their EW couplings.
The dominant channel is pair production via Drell-Yan:
The vector boson fusion (VBF) pair production via gauge couplings is found to be subleading to Drell-Yan \cite{Agugliaro:2018vsu}.  
Single production can also be phenomenologically relevant, however it is strongly model dependent. 
VBF single production is generated via topological anomalies, hence it is suppressed by a small anomaly coupling.
Drell-Yan single production could also be present if the pNGBs couple to quarks: However, for pNGBs in models with partial compositeness, couplings to light quark flavours are expected to be very small (roughly proportional to the quark mass). The dominant couplings, therefore, involve third generation quarks.
In this case, neutral pNGBs can be singly-produced via gluon fusion analogously to the Higgs boson.
Finally, both neutral and charged pNGBs can be singly-produced in association with $tt$ or $tb$, respectively, hence providing a relevant contribution if the couplings are large enough.
In this work, we will provide the first complete analysis of how current LHC searches probe the parameter space of the EW pNGBs via their pair production.  After providing bounds for simplified models, we will focus on a specific model to investigate the interplay between various channels.

Among the next-to-minimal models, the $\SU(5)/\SO(5)$ model has been studied since the early days of composite Higgs models \cite{Dugan:1984hq}. In the context of four-dimensional models with a microscopic description \cite{Ferretti:2013kya, Ferretti:2014qta, Belyaev:2016ftv}, it emerges as the minimal symmetry pattern from the condensate $\langle\psi\psi\rangle$ of two EW-charged fermions if $\psi$ live in a real irreducible representation (irrep) of the confining gauge group. 
Initial investigations of its LHC phenomenology were performed in Refs~\cite{Agugliaro:2018vsu,Banerjee:2022xmu}, and a detailed description of the model can be found in Ref.~\cite{Agugliaro:2018vsu}. For other models, we expect similar limits, with the caveat that they may contain a Dark Matter candidate \cite{Wu:2017iji,Cai:2020njb}, while this is not possible for $\SU(5)/\SO(5)$ due to the topological anomaly. We conclude the introduction by recalling that the singlet pNGB in the minimal case $\SU(4)/\Sp(4)$ is very hard to detect due to the small gauge couplings \cite{Galloway:2010bp,Arbey:2015exa}, unless it is lighter than the $Z$ boson \cite{Cacciapaglia:2021agf}. Finally, we recall that models with top partial compositeness also contain QCD-coloured pNGBs \cite{Cacciapaglia:2015eqa,Belyaev:2016ftv,Cacciapaglia:2020vyf,Cacciapaglia:2021uqh} and a ubiquitous, and potentially light, singlet associated with a global U$(1)$ symmetry \cite{Ferretti:2016upr,Belyaev:2016ftv,Cacciapaglia:2019bqz}.

The manuscript is organised as follows: In Sec.~\ref{sec:modelindependent} we present current bounds on various production and decay channels of a pair of scalars, which can apply to any model. In Sec.~\ref{sec:su5so5} we focus on the $\SU(5)/\SO(5)$ model and investigate both the fermiophobic case in Sec.~\ref{sec:boundsfermiophobic} and fermiophilic one in Sec.~\ref{sec:boundsfermiophilic}. Finally, we offer our conclusions in Sec.~\ref{sec:conclusions}.

\section{Simplified model bounds on Drell-Yan pair-produced scalars} \label{sec:modelindependent}

Many BSM models contain an extended scalar sector with $\SU(2)_L\times \U(1)_Y$ multiplets beyond the Higgs doublet. The bounds on (and signals of) these models are highly model dependent. Yukawa-type couplings of the additional scalars are subject to constraints from flavour physics, while the scalar potential influences the EW symmetry breaking and is, therefore, strongly constrained. The latter mainly occurs via VEVs of the new multiplets, while mixing with the Higgs through the scalar potential can also influence flavour physics. 
At the same time, Yukawa-type couplings and scalar VEVs and mixing patterns determine the single production cross sections of the BSM scalars at lepton and hadron colliders.
In the following we will only focus on pair production, via the dominant Drell-Yan channels.

\subsection{Simplified model Lagrangian}

For our phenomenological studies, we use parts of a simplified model which has already been introduced in \cite{Banerjee:2022xmu}. We extend the SM by colourless scalar states $S^0,S^{0\prime}, S^\pm,S^{\pm\pm}$ that are physical mass eigenstates labelled by their electric charge. We include the minimal set of states up to charge-2 that have all the possible couplings to the EW gauge bosons, hence including two neutral states with opposite parity.

We consider the simplified model Lagrangian
with kinetic and mass terms for the scalars as well as interaction terms
\begin{align}
\mathcal{L}_{\rm int}&=\mathcal{L}_{SS V}
+\mathcal{L}_{S V V}
+\mathcal{L}_{f f S}\,,
\label{eq:L_int}
\end{align}
where the first term contains the couplings of two scalars to an EW gauge boson, which determine the Drell-Yan pair production. The remaining terms contain the couplings of a scalar to two EW gauge bosons or to two SM fermions, which dictate the two-body decays into SM particles.

The first term arises from the $\SU(2)_L\times \U(1)_Y$ covariant derivative terms in full models and reads:
\begin{align}
\mathcal{L}_{SS V} &=\frac{ie}{s_W}W^{-\mu}\left(K^{S^0 S^+}_{W}S^0\overleftrightarrow{\partial_\mu}S^+ +K^{S^{0\prime} S^+}_{W}S^{0\prime}\overleftrightarrow{\partial_\mu}S^+ +K^{S^- S^{++}}_{W}S^-\overleftrightarrow{\partial_\mu}S^{++}\right)+{\mathrm{h.c.}} \nonumber\\
& +\frac{ie}{s_Wc_W}Z^\mu \left( K^{S^0 S^{0\prime}}_{Z} S^0\overleftrightarrow{\partial_\mu} S^{0\prime} + K^{ S^+ S^{-}}_{Z} S^+\overleftrightarrow{\partial_\mu} S^{-}+K^{ S^{++} S^{--}}_{Z} S^{++}\overleftrightarrow{\partial_\mu} S^{--} \right)\nonumber\\
&-ieA^\mu \left( S^+\overleftrightarrow{\partial_\mu} S^-+2 S^{++}\overleftrightarrow{\partial_\mu} S^{--}\right)\,,
\label{eq:L_SSV} 
\end{align}
where $\phi_1 \overleftrightarrow{\partial_\mu} \phi_2 \equiv \phi_1 (\partial_\mu \phi_2) - (\partial_\mu \phi_1) \phi_2$. The $K_V^{S S}$ parameters are determined by the $\SU(2)_L\times \U(1)_Y$ representations of the scalar multiplets as well as the mass mixing. The $K_V^{S S}$ coefficients for sample models, including the model discussed in \cref{sec:su5so5}, are given in \cref{app:coeffs}. The production cross section of each scalar pair is proportional to its respective $|K_V^{S S}|^2$.

The second term parameterises dimension-5 operators which yield the decay of the scalars into two EW gauge bosons, and reads
\begin{align}
\nonumber
\mathcal{L}_{ S V V} = \frac{e^2}{16 \pi^2 v} \bigg[ & S^0\left(\tilde{K}^{S^0}_{\gamma\gamma}F_{\mu\nu}\tilde{F}^{\mu\nu}+\frac{2}{s_Wc_W}\tilde{K}^{ S^0}_{\gamma Z}F_{\mu\nu}\tilde{Z}^{\mu\nu}+\frac{1}{s_W^2c_W^2}\tilde{K}^{ S^0}_{ZZ}Z_{\mu\nu}\tilde{Z}^{\mu\nu}\right.\nonumber\\
&\left. \hspace{6mm} +\frac{2}{s_W^2}\tilde{K}^{ S^0}_{WW}W^+_{\mu\nu}\tilde{W}^{-\mu\nu}\right)     \nonumber\\
 +& S^{0\prime}\left(K^{S^{0\prime}}_{\gamma\gamma}F_{\mu\nu}F^{\mu\nu}+\frac{2}{s_Wc_W} K^{ S^{0\prime}}_{\gamma Z}F_{\mu\nu}Z^{\mu\nu}+\frac{1}{s_W^2c_W^2}K^{ S^{0\prime}}_{ZZ}Z_{\mu\nu}Z^{\mu\nu}\right.\nonumber\\
&\left. \hspace{6mm} +\frac{2}{s_W^2}K^{ S^{0\prime}}_{WW}W^+_{\mu\nu}W^{-\mu\nu}\right)  \nonumber \\
+&\left( S^+ \left( \frac{2}{s_W}\tilde{K}^{ S^+}_{\gamma W}F_{\mu\nu}\tilde{W}^{-\mu\nu}+ \frac{2}{s_W^2c_W}\tilde{K}^{ S^+}_{ZW}Z_{\mu\nu}\tilde{W}^{-\mu\nu}\right)+ \mathrm{h.c.}\right)\nonumber\\
+& S^{++} \frac{1}{s_W^2} \tilde{K}^{ S^{++}}_{W^-W^-}W^-_{\mu\nu}\tilde{W}^{-\mu\nu}+\mathrm{ h.c.}\bigg]\,.
\label{eq:LpiVVtilde}
\end{align}
The couplings above are written assuming that all scalars are odd under parity, except for the even state $S^{0\prime}$ in order to allow the $Z$ couplings in Eq.~\eqref{eq:L_SSV}. This choice is motivated by matching to the composite models we consider in Sec.~\ref{sec:su5so5}, however the parity assignment can be flipped in a straightforward manner. Note that the parity assignment does not significantly affect the bounds we consider here, as the kinematics of the decay is untouched.  Hence we only study the case in Eq.~\eqref{eq:LpiVVtilde}.

The last term contains Yukawa-type couplings to the third generation quarks: 
\begin{align}
\mathcal{L}_{f f  S} & = S^0 \bigg[\bar{t}\left(\kappa^{ S^0}_{t}+i\tilde{\kappa}^{ S^0}_{t}\gamma_5\right)t+\bar{b}\left(\kappa^{ S^0}_{b}+i\tilde{\kappa}^{ S^0}_{b}\gamma_5\right)b\bigg] + \left(S^0\rightarrow S^{0\prime}\right) \nonumber \\
& +  S^+ \, \bar{t} \left(\kappa^{ S^+}_{tb,L}  P_L + \kappa^{ S^+}_{tb,R}  P_R \right) b + {\rm h.c.}\, ,
\label{eq:L_ffpi}
\end{align}
where, motivated by the SM structure, the couplings are allowed to violate parity. Couplings to other SM fermions could be included analogously: our choice here is motivated by the models of top partial compositeness from Sec~\ref{sec:su5so5}. 

\subsection{Di-scalar channels}\label{sec:channels}

\begin{figure}
    \centering
    \includegraphics{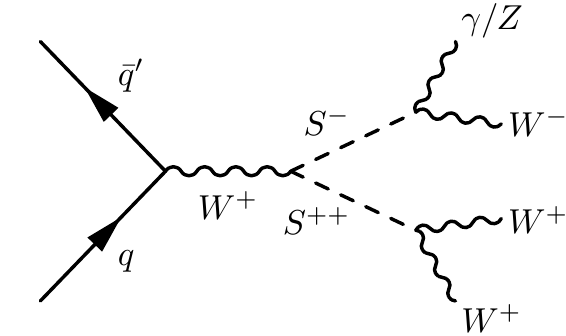} \qquad \includegraphics{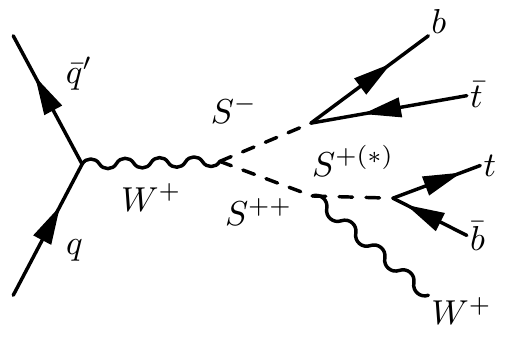}
    \caption{Examples of di-scalar channels from pair production via Drell-Yan processes with subsequent decays into SM particles.}
    \label{fig:pNGBpairgraphs}
\end{figure}

We investigate all scalar pairs produced at the LHC through the Drell-Yan processes:
\begin{equation}
\label{eq:DY}
pp \to  S^{\pm\pm} S^{\mp}\,,\, S^{\pm} S^{0(\prime)} \,,\,  
S^{++} S^{--} \,,\,S^{+} S^{-}
 \,,\,S^{0} S^{0\prime}\, .
\end{equation}
Together with the first tier decays of the scalar pairs into SM particles, these production processes yield many di-scalar channels, see \cref{fig:pNGBpairgraphs} for two examples. Charge-conjugated states belong to the same channel.
For the decays of the scalars, we consider two complementary scenarios: The fermiophobic case, where the dominant decays are into EW gauge bosons, and the fermiophilic case, where the scalars decay dominantly into a pair of third generation quarks. In both cases, we only consider narrow width resonances. The two choices are motivated by the different origins of the two sets of couplings in Eq.~\eqref{eq:LpiVVtilde} and Eq.~\eqref{eq:L_ffpi}: The former deriving from higher dimension operators or loops, the second from Yukawa-like couplings or (small) mixing to the Higgs boson.

\begin{table}[tb]
\centering
{\small \begin{tabular}{c|ccccc|}
fermiophobic & $\rescol{S^{++}}{S^{--}}$ & $\rescol{S^{\pm \pm}}{S^\mp}$ & $\rescol{S^+}{S^-}$ & $\rescol{S^\pm}{ S^{0(\prime)}}$ & $\rescol{S^0 }{S^{0\prime}}/\rescol{S^{0\prime}}{S^0}$ \\ \hline
$WWWW$ & $\rescol{W^+W^+}{W^-W^-}$ & - & - & - & $\rescol{W^+W^-}{W^+W^-}$ \\
$WWW\gamma$ & - & $\rescol{W^\pm W^\pm}{W^\mp \gamma}$ & - & $\rescol{W^\pm \gamma}{W^+ W^-}$ & - \\
$WWWZ$ & - & $\rescol{W^\pm W^\pm}{W^\mp Z}$ & - & $\rescol{W^\pm Z}{W^+ W^-}$ & - \\
$WW\gamma\gamma$ & - & - & $\rescol{W^+ \gamma}{W^- \gamma}$ & - & $\rescol{W^+W^-}{\gamma\gamma}$ \\
$WWZ\gamma$ & - & - & $\rescol{W^\pm \gamma}{ W^\mp Z}$ & - & $\rescol{W^+W^-}{\gamma Z}$  \\
$WWZZ$ & - & - & $\rescol{W^+ Z}{W^- Z}$ & - & $\rescol{W^+W^-}{ZZ}$ \\
$W\gamma\gamma\gamma$ & - & - & - & $\rescol{W^\pm \gamma}{\gamma \gamma}$ & - \\
$WZ\gamma\gamma$ & - & - & - & $\rescol{W^\pm}{} \{ Z \gamma\} \rescol{}{\gamma}$ & - \\
$WZZ\gamma$ & - & - & - & $\rescol{W^\pm}{} \{Z \gamma\} \rescol{}{Z}$ & - \\
$WZZZ$ & - & - & - & $\rescol{W^\pm Z}{ZZ}$ & - \\
$\gamma\gamma\gamma\gamma$ & - & - & - & - & $\rescol{\gamma\gamma}{\gamma\gamma}$\\
$Z\gamma\gamma\gamma$ & - & - & - & - & $\rescol{Z\gamma}{\gamma\gamma}$\\
$ZZ\gamma\gamma$ & - & - & - & - & $\rescol{Z}{}\{Z\gamma\}\rescol{}{\gamma}$\\
$ZZZ\gamma$ & - & - & - & - & $\rescol{ZZ}{Z\gamma}$ \\
$ZZZZ$ & - & - & - & - & $\rescol{ZZ}{ZZ}$ \\
 \end{tabular}}
 \caption{ \label{tab:feriophobic} Classification of the 24 di-scalar channels in terms of the 5 pair production cases (columns) and the 15 combinations of gauge bosons (rows) from decays. In the channels, the first two and second two bosons are resonantly produced. The notation $\{Z\gamma\} = \rescol{Z}{\gamma} + \rescol{\gamma}{Z}$ indicates the two permutations. Charge-conjugated states belong to the same di-scalar channel.}
\end{table}

In the fermiophobic case, we assume dominant decays of the scalars into EW gauge bosons via the couplings in Eq.~\eqref{eq:LpiVVtilde}, leading to\footnote{We do not consider the possible coupling of the neutral scalars to two gluons, as it can only be generated if they couple to states carrying QCD charges. We remark that Drell-Yan pair production of scalars with subsequent decays to a pair of dijets is targeted by experimental searches \cite{CMS:2022usq}, but public recasts of these are as of now not available.}
\begin{subequations}
	\begin{align}
		S^{++} &\to W^+ W^+\, , \\
		S^+ &\to W^+ \gamma, \, W^+Z \,,\\
		S^{0(\prime)} &\to W^+ W^-, \, \gamma \gamma, \, \gamma Z, \, ZZ.
	\end{align}
\end{subequations}
Combining the different Drell-Yan scalar pairs with the above decay channels leads to 24 di-scalar channels -- each containing four gauge bosons -- for which we present bounds in \cref{sec:simpresults}. One sample process is shown in the left diagram of \cref{fig:pNGBpairgraphs}, while a complete list of all channels is shown in Table~\ref{tab:feriophobic}.

\begin{table}[tb]
\centering
\begin{tabular}{c|ccccc|}
fermiophilic & $\rescol{S^{++}}{S^{--}}$ & $\rescol{S^{++}}{S^-}$ & $\rescol{S^+}{S^-}$ & $\rescol{S^+}{ S^{0(\prime)}}$ & $\rescol{S^0 }{S^{0\prime}}/\rescol{S^{0\prime}}{S^0}$ \\ \hline
$tttt$ & - & - & - & - & $\rescol{t\bar{t}}{t\bar{t}}$ \\
$tttb$ & - & - & - & $\rescol{t\bar{b}}{t\bar{t}}$ & - \\
$ttbb$ & - & - & $\rescol{t\bar{b}}{b\bar{t}}$ & - & $\rescol{t\bar{t}}{b\bar{b}}$ \\
$tbbb$ & - & - & - & $\rescol{t\bar{b}}{b\bar{b}}$ & - \\
$bbbb$ & - & - & - & - & $\rescol{b\bar{b}}{b\bar{b}}$ \\
$Wttbb$ & - & $\rescol{W^+t\bar{b}}{ b\bar{t}}$ & - & - & - \\
$WWttbb$ & $\rescol{W^+ t\bar{b}}{W^-b\bar{t}}$ & - & - & - & - \\
 \end{tabular}
 \caption{ \label{tab:feriophilic} Classification of the 8 di-scalar channels in terms of the 5 pair production cases (columns) and the 5 combinations of top and bottom from decays (rows). In cases with one or two doubly charged scalars, one always obtains $ttbb$ with one or two additional $W$'s, respectively. The charge-conjugated states are not shown.}
\end{table} 

In the fermiophilic scenario we assume dominant couplings of the scalars to third family quarks. Note that doubly charged scalars cannot decay to two quarks due to their charge, but if they are part of an $\SU(2)_L$ multiplet, the three-body decay $S^{++}\rightarrow W^+S^{+*} \rightarrow W^+t\bar{b}$ is allowed. The dominant decay channels we consider for the fermiophilic scenario are thus\footnote{Note that top and bottom loops generate effective couplings to gluons and EW gauge bosons, however they lead to subleading decay channels.}
\begin{subequations}
	\begin{align}
		S^{++} &\to W^+ t\bar b, \\
		S^+ &\to t\bar b, \\
		S^{0(\prime)} &\to t\bar t \,\,\,\mbox{ or }\,\, \, b\bar b.
	\end{align}
\end{subequations}
For pair-produced scalars, this yields 8 possible di-scalar channels in the fermiophilic scenario.
One sample process is shown in the right diagram of \cref{fig:pNGBpairgraphs}, while a complete list is showcased in Table~\ref{tab:feriophilic}.

\subsection{Simulation setup and determination of LHC bounds}

For the simulation of signal events, we use the publicly available \texttt{eVLQ} model first presented in Ref.~\cite{Banerjee:2022xmu}, which implements the simplified models in \cref{eq:L_int} as a \texttt{FeynRules} \cite{Alloul:2013bka} model at next-to-leading order in QCD.
The implementation contains one doubly charged, one singly charged and one neutral scalar, and we expanded it by another neutral scalar to allow for $S^0 S^{0\prime}$ production.

All events are generated at a centre-of-mass energy of $13$~TeV in proton-proton collisions.
For each di-scalar channel, we perform a parameter scan over the scalar mass $m_S$, starting at the decay mass threshold and up to $\left. m_S\right|_{\rm max} = 1$~TeV. For channels involving two different scalars, we assume them to be mass degenerate. 
For each scan point, we generate $10^5$ events of Drell-Yan scalar pairs with decay into the target channel, using \texttt{MadGraph5\_aMC@NLO} \cite{Alwall:2014hca} version 3.3.2 at NLO (including patches that were incorporated in version 3.4.0 after the completion of this work), in association with the parton densities in the \texttt{NNPDF 2.3} set \cite{Ball:2012cx,Buckley:2014ana}.
We then interface the events with \texttt{Pythia8} \cite{Sjostrand:2014zea} for SM particle decays, showering and hadronisation. 
The resulting showered signal events are analysed with  \texttt{MadAnalysis5}  \cite{Conte:2012fm,Conte:2014zja,Dumont:2014tja,Conte:2018vmg} version 1.9.60 and \texttt{CheckMATE} \cite{Drees:2013wra,Dercks:2016npn} version 2.0.34  (commit number \texttt{8952e7}).  
Both tools reconstruct the events using \texttt{Delphes 3} \cite{deFavereau:2013fsa} and the anti-$k_T$ algorithm \cite{Cacciari:2008gp} implemented in \texttt{FastJet} \cite{Cacciari:2011ma}.
The exclusion associated with the events is calculated with the CL$_s$ prescription \cite{Read:2002hq}.
We also run the events against the SM measurements implemented in \texttt{Rivet} \cite{Bierlich:2019rhm} version 3.1.5 and extract exclusions from the respective \texttt{YODA} files using \texttt{Contur} \cite{Butterworth:2019wnt, Buckley:2021neu} version 2.2.1.

To present simplified model bounds, we determine the signal cross section $\sigma_{95}$ which is excluded at 95\% CL.
The procedure differs between the tools:
\begin{itemize}
    \item \texttt{MadAnalysis5} explicitly calculates the upper limit on the signal cross section (both expected \texttt{sig95exp} and observed \texttt{sig95obs}) from the simulated signal events.
    We use the observed bound from the signal region to which \texttt{MadAnalysis5} ascribes the highest sensitivity (\texttt{best}).
    \item \texttt{CheckMATE} quotes upper limits on the signal, $S_{95}^\mathrm{exp}$ (expected) and $S_{95}^\mathrm{obs}$ (observed). From these, the input cross section $\sigma_\mathrm{in}$ and the signal $S$ that passed the cuts, we calculate the upper limit on the cross section as
    \begin{equation}
        \sigma_{95} = \frac{S_{95}^\mathrm{obs}}{S} \sigma_\mathrm{in}.
    \end{equation}
    We follow the default procedure recommended by the \texttt{CheckMATE} collaboration for determining the best signal region, i.e. we use the observed bound of the signal region with the strongest expected bound.
    \item \texttt{Contur} does not calculate bounds on the cross section, so we determine them manually by running \texttt{Rivet} and \texttt{Contur} multiple times on the same events and dynamically adjust the input cross section until we obtain $\mathrm{CL}_s = 0.05 \pm 0.01$. 
\end{itemize}

For each channel and each parameter point, we take the minimal value for $\sigma_{95}$ obtained from \texttt{MadAnalysis5}, \texttt{CheckMATE}, and \texttt{Contur} as the final bound, i.e. we do not attempt to combine them.

\subsection{Simplified model results and discussion}\label{sec:simpresults}

\begin{figure}[]
	\centering
	\begin{subfigure}{0.48\linewidth}
		\includegraphics[width=\linewidth]{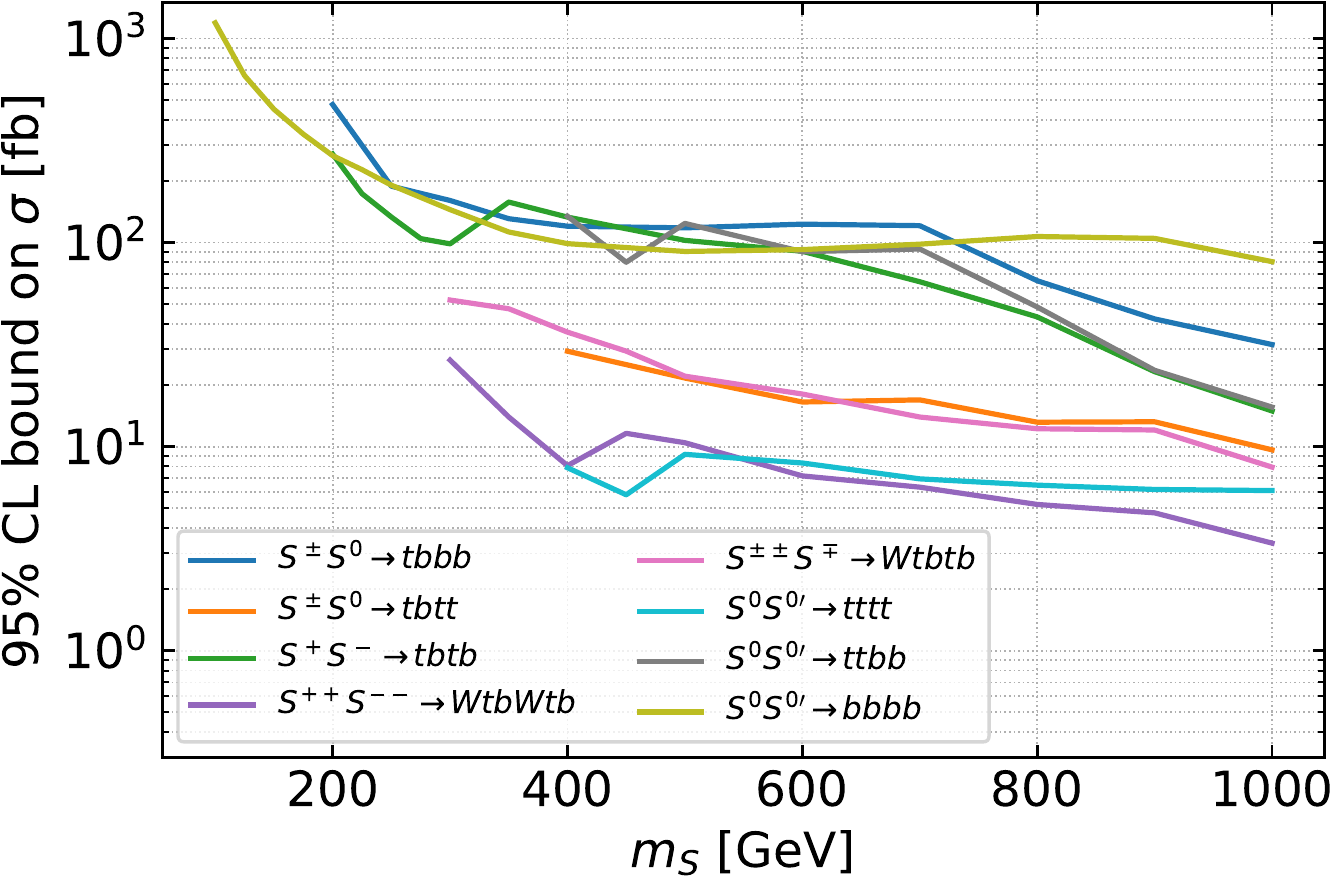}
		\caption{Scalar pair with decays to quarks}
		\label{fig:modelindependentquarks}
	\end{subfigure} \quad
	\begin{subfigure}{0.48\linewidth}
		\includegraphics[width=\linewidth]{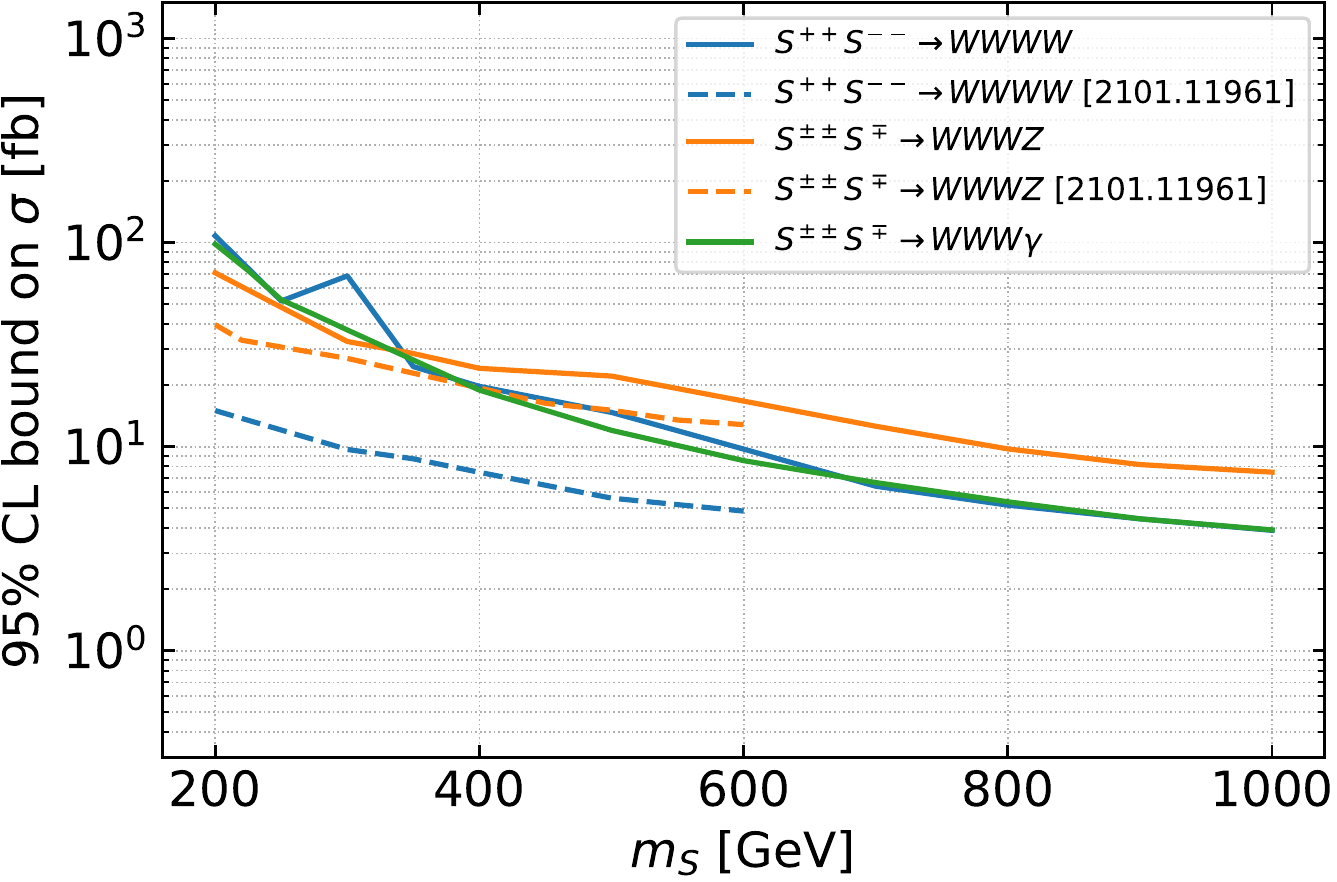}
		\caption{$S^{++} S^{--}$ and $S^{\pm\pm}S^\mp$ with di-boson decays}
		\label{fig:modelindependents12}
	\end{subfigure}  \vspace{2ex}

	\begin{subfigure}{0.48\linewidth}
		\includegraphics[width=\linewidth]{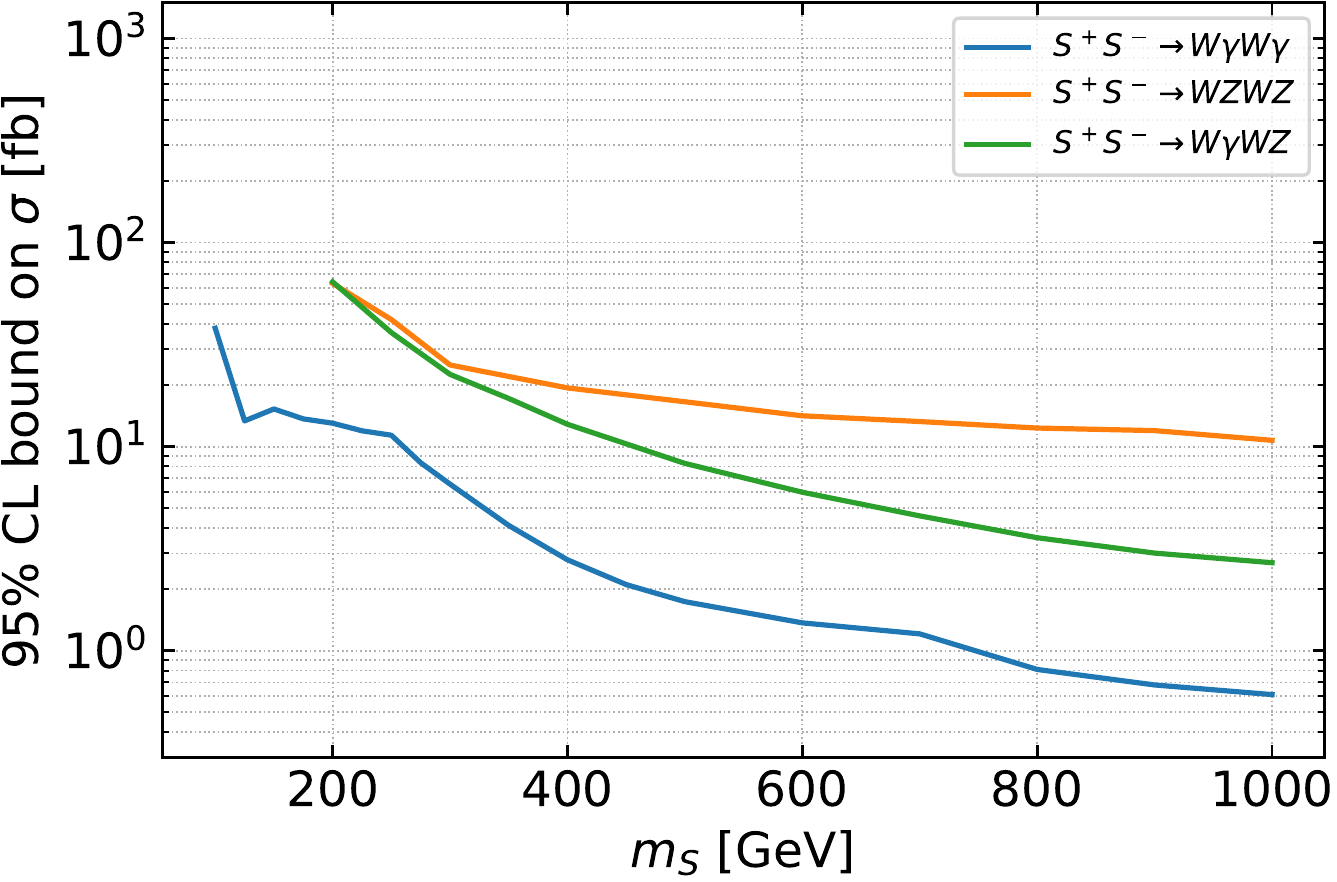}
		\caption{$S^{+} S^{-}$  with di-boson decays}
		\label{fig:modelindependents11s11}
	\end{subfigure} \quad
	\begin{subfigure}{0.48\linewidth}
		\includegraphics[width=\linewidth]{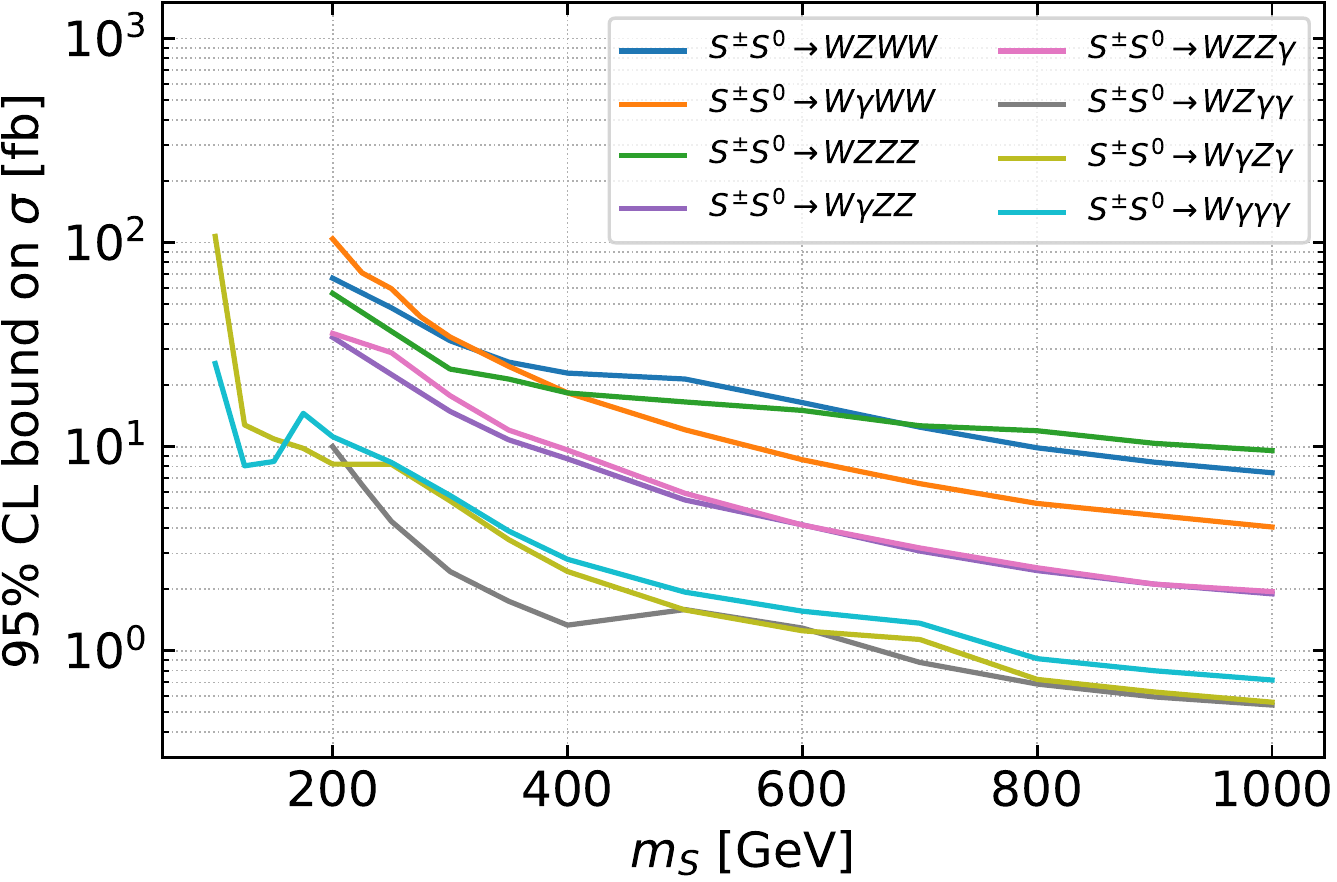}
		\caption{$S^{\pm} S^{0}$  with di-boson decays}
		\label{fig:modelindependents11s10}
	\end{subfigure} \vspace{2ex}
	
	\begin{subfigure}{0.48\linewidth}
		\includegraphics[width=\linewidth]{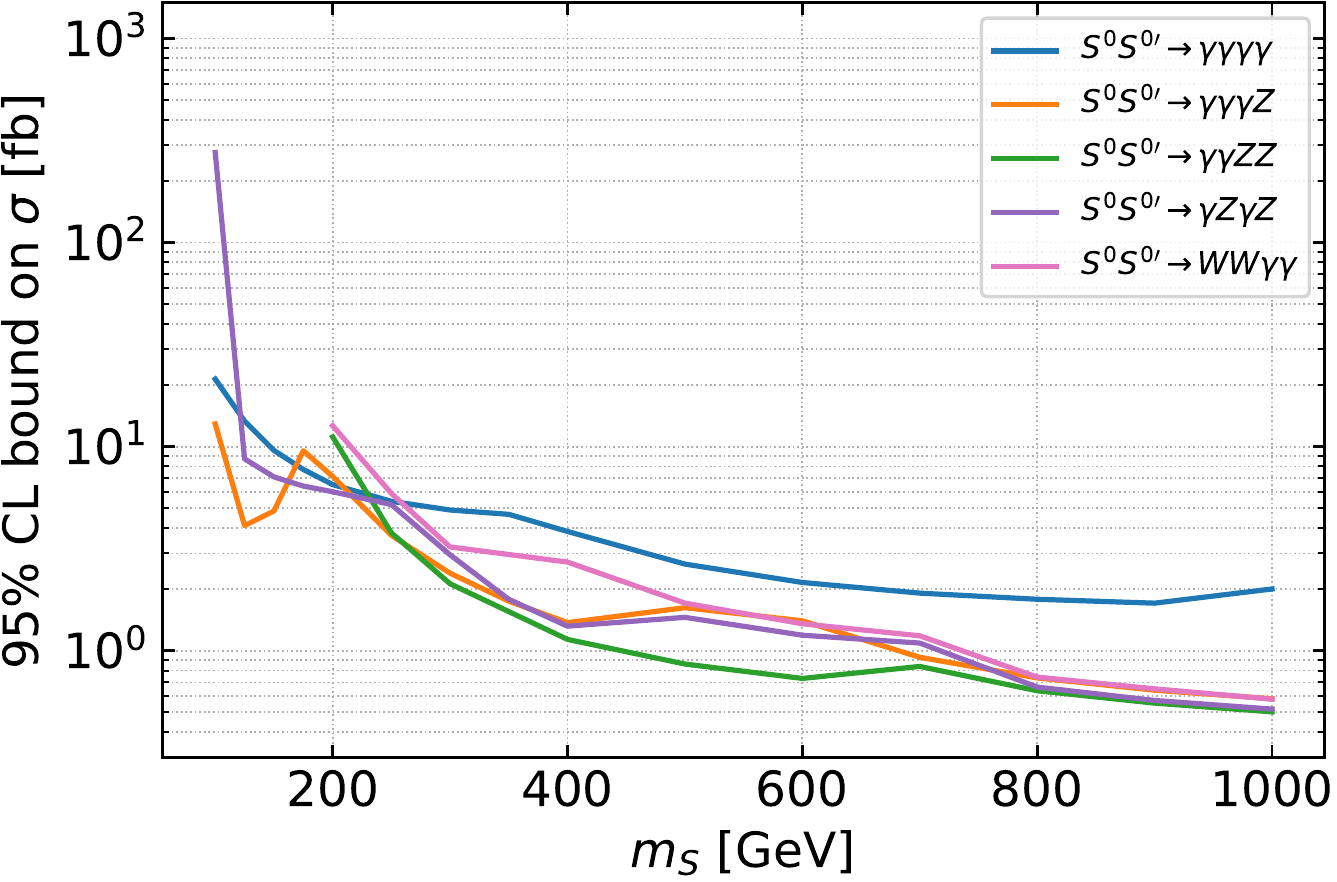}
		\caption{$S^{0} S^{0\prime}$  with di-boson decays with $\geq 2$ photons}
		\label{fig:modelindependents10s102aa}
	\end{subfigure} \quad
	\begin{subfigure}{0.48\linewidth}
		\includegraphics[width=\linewidth]{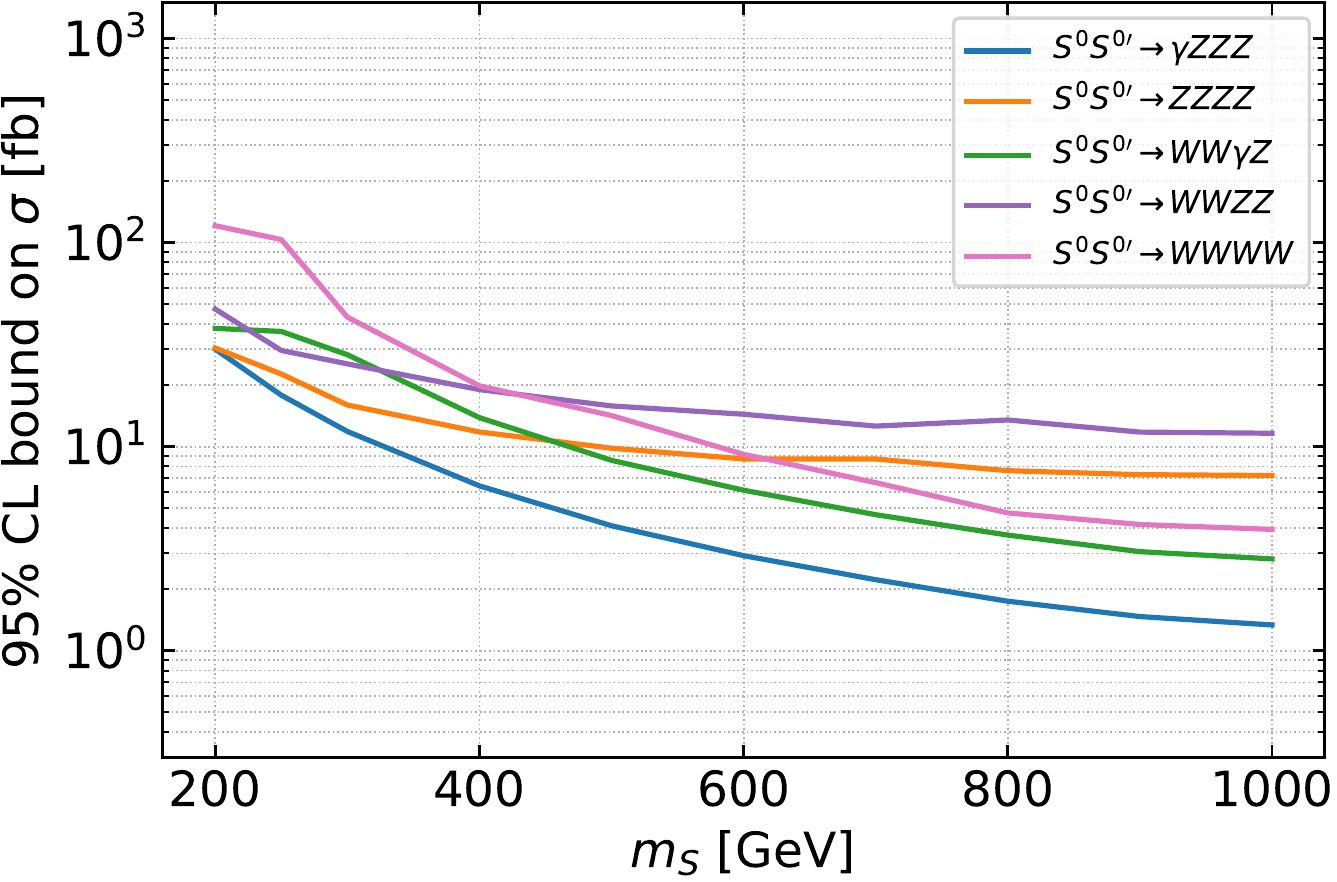}
		\caption{$S^{0} S^{0\prime}$  with di-boson decays with $\leq 1$ photons}
		\label{fig:modelindependents10s102a}
	\end{subfigure} 
	\caption{Upper limits on the cross section of the di-scalar channels from Drell-Yan pair production. The scalars decay to: (a) third generation quarks or (b)-(f) two vector bosons. Both scalars are assumed to have the same mass. The analyses contributing to the bounds are Refs.~\cite{ATLAS:2021jol, CMS:2019xud,  ATLAS:2018nud, ATLAS:2021mbt, ATLAS:2021fbt, ATLAS:2021twp, ATLAS:2018zdn, ATLAS:2019gdh, CMS:2019xjf, CMS:2017abv, ATLAS:2020qlk, ATLAS:2021kog, ATLAS:2019gey, ATLAS:2018nci, ATLAS:2016uwq, CMS:2017moi} (see \cref{tab:analyses} and \cref{tab:modelindependentanalyses} in \cref{app:technical} for details). The numerical values of the limits are available on \url{https://github.com/manuelkunkel/scalarbounds}.} 
	\label{fig:modelindependent}
\end{figure}

To the best of our knowledge, only very few of these di-scalar channels have been explicitly targeted by LHC searches. The doubly charged scalar $S^{\pm\pm}$ is the only one that cannot be singly-produced at a sizable rate, hence a direct search for its pair production or single production in association with a singly charged scalar is available \cite{ATLAS:2021jol}. The ATLAS search \cite{ATLAS:2021jol} can be directly applied to the channels $S^{++}S^{--}\rightarrow WWWW$ and  $S^{\pm \pm}S^{\mp}\rightarrow WWWZ$. Searches for di-Higgs production also contain final states of interest, like $WWWW$ and $bbbb$, however those searches strongly focus on scalar masses of $125$~GeV with production via SM processes \cite{CMS:2022cpr,CMS:2022nmn,CMS:2022kdx,ATLAS:2022ycx} or via a massive resonance \cite{ATLAS:2016paq,ATLAS:2018hqk}. Searches for pair production of two neutral scalars decaying to $WWWW$ and $bbbb$ final states are also available, but only via resonant production \cite{ATLAS:2018ili,CMS:2022qww}. These searches have limited applicability to the di-scalar channels discussed in this article because of the different kinematics. 

The di-scalar channels we obtained in \cref{sec:channels}, however, do populate the signal region of many BSM searches at the LHC as well as those used for SM cross section measurements.
To determine reliable bounds, it is required to recast the searches as the topology and/or kinematics can be very different. 
As a first step towards determining appropriate bounds we therefore simulate the processes and determine bounds from all LHC search recasts and measurements which are publicly available in \texttt{MadAnalysis5}, \texttt{CheckMATE}, and \texttt{Contur}.
The results are showcased in \cref{fig:modelindependent}, where we present the simplified model bounds on the cross section for each of the 32 di-scalar channels. For each channel, we simulate Drell-Yan produced pairs of bosons with subsequent decay into the target state (four EW gauge bosons, or four fermions with 0, 1, or 2 additional $W$ bosons) which are then further decayed, hadronised and analysed using the procedure described in the previous subsection. 
Further details on the dominant analyses are given in \cref{app:analyses}.

Figure \ref{fig:modelindependentquarks} shows the bounds on the 8 di-scalar channels in the fermiophilic scenario, consisting of third generation quarks plus one additional $W$ boson per doubly charged scalar due to the 3-body decay of $S^{\pm\pm}$.
In channels with multiple top quarks, dominant bounds arise from a search for $R$-parity violating supersymmetry \cite{ATLAS:2021fbt}, while various supersymmetric searches \cite{CMS:2019xjf,ATLAS:2018zdn,ATLAS:2021twp,ATLAS:2019gdh} and the generic search in Ref.~\cite{CMS:2017abv} are relevant for the multi-bottom channels.

Figures \ref{fig:modelindependents12} to \ref{fig:modelindependents10s102a} show the bounds for channels of the fermiophobic scenario that, for readability, are split into 5 figures and regrouped according to the charges of the di-scalar states. In case of  $S^0S^{0\prime}$, the channels are further sub-grouped according to the number of photons in the final state. 
\cref{fig:modelindependents12} is dedicated to di-scalar channels with at least one doubly-charged scalar, leading to at least 3 $W$ bosons plus a $W$, $Z$, or photon. The photon channel $WWW\gamma$ can be constrained using measurements of the $Z\gamma$ production cross section \cite{ATLAS:2019gey,ATLAS:2018nci}. The main searches for the $WWWW$ and $WWWZ$ channels look for multi-lepton final states \cite{CMS:2019xud, CMS:2017moi}. For these two channels, the results of the ATLAS search for doubly and singly charged Higgs bosons
decaying into vector bosons in multi-lepton final states \cite{ATLAS:2021jol} apply, and they are shown as blue and orange dashed lines. As is to be expected, the bounds from the ATLAS search dedicated to these final states are stronger than the bound we obtain from recasts of a large number of BSM searches targeting different signatures and scenarios. This also suggests that dedicated searches for the other di-scalar channels discussed in this article can lead to substantial improvement in covering their signatures.
In \cref{fig:modelindependents11s11} we show the di-scalar channels from $S^+S^-$ production. The bounds on $W\gamma W\gamma$ are by far the strongest, coming from a search for gauge-mediated supersymmetry in final states containing photons and jets \cite{ATLAS:2018nud}. 
The main bounds for the channels $WZWZ$ and $W\gamma WZ$ stem from a multi-lepton search \cite{CMS:2017abv} and the $Z\gamma$ cross section measurements \cite{ATLAS:2019gey,ATLAS:2018nci}, respectively.
\cref{fig:modelindependents11s10} is dedicated to the di-scalar channels from $S^\pm S^0$ production. As for the previous panel, the searches can be split by the number of photons, leading to the multi-lepton search \cite{CMS:2017abv} for channels containing 0 photons, measurements of the $Z\gamma$-cross section \cite{ATLAS:2019gey,ATLAS:2018nci} for single photon channels and Ref.~\cite{ATLAS:2018nud} for multi-photon channels.
In \cref{fig:modelindependents10s102aa} we present the $S^0 S^{0\prime}$ channels that contain at least 2 photons. 
The $\gamma\gamma\gamma\gamma$ channel is constrained by the generic search \cite{ATLAS:2018zdn} and the measurement of the $\gamma\gamma$-production cross section \cite{ATLAS:2021mbt}.
For the remaining channels, the dominant analysis is the (multi-)photon search \cite{ATLAS:2018nud}.
Finally, \cref{fig:modelindependents10s102a} contains the remaining $S^0 S^{0\prime}$ channels with at most one photon, which are less strongly constrained than the multi-photon channels.
The main searches contributing to the bounds are the multi-lepton search \cite{CMS:2017abv} and the $Z\gamma$-cross section measurements \cite{ATLAS:2018nci,ATLAS:2019gey} for the channels with 0 and 1 photon, respectively.

\subsection{Applicability and limitations of simplified model bounds}

The bounds presented in \cref{fig:modelindependent} are based on recasts of other searches and SM measurements, apart from the ATLAS direct searches for $S^{\pm\pm}$ production \cite{ATLAS:2021jol}.
As the most important and generic limitation, we wish to re-emphasise that our study is based only on searches and measurements by ATLAS, CMS, and LHCb, for which recasts in \texttt{MadAnalysis5}, \texttt{CheckMATE}, or \texttt{Contur} are available. This represents only a fraction of (in particular, the newest) LHC searches and measurements, implying that including recasts of additional searches will improve the bounds. Performing all these recasts is beyond the scope of this article. 

Another limitation of the simplified model approach stands in the fact that limits are extracted for each specific channel, in our case applying to 32 (24 fermiophobic and 8 fermiophilic) di-scalar channels. However, realistic models with an extended Higgs sector contain several scalar mass eigenstates, which can decay into more than one final state. 
How can the limits in \cref{fig:modelindependent} be used to extract reliable limits on a more complex extended Higgs sector? To answer this question, we consider below three template scenarios, which cover exhaustively all possibilities.

\begin{enumerate}
 \item \emph{Single scalar, Drell-Yan, several decay channels:}\\ 
   If only a single particle is produced with a single decay mode, the bounds on the mass of this particle can be immediately read off from
   \cref{fig:modelindependent}.
    If the Drell-Yan produced scalar has several decay channels, it is required to compute the cross section times branching ratio for each matching di-scalar channel, and compare them to the corresponding limit in \cref{fig:modelindependent}. The most conservative limit comes from the channel that has the strongest bound. 
    As different channels may contribute to the same signal region of the leading search, the actual bound on $m_S$ can be further improved by simulating the complete signal from the scalar pair production.

\item \emph{Several scalars, Drell-Yan, several decay channels:}\\ 
    If the model contains several scalars of similar masses,  even more di-scalar channels can be matched. Besides the most conservative bound described above, one can extract a more realistic bound by combining various channels. This is feasible if the scalars are relatively close in mass, so that the acceptances remain similar. Hence, the procedure would consist of summing the cross sections times branching ratios of all processes that contribute to the same di-scalar channel. An even more aggressive approach is to sum all the channels that contribute to the same search, as we will illustrate with an explicit example in the next section.
    
\item \emph{Non Drell-Yan and/or  new decay channels:}\\
    Finally, there are models where the dominant production is not Drell-Yan, in which case the limits in \cref{fig:modelindependent} cannot be directly applied. However, the impact of the different kinematics on the bound is typically limited because the searches are not dedicated to the specific final state and production mechanism. Hence, we expect the limits in \cref{fig:modelindependent} to provide a good estimate, while a full simulation is needed to extract a more reliable bound. The same consideration applies if additional decay channels are available, like for instance cascade or three-body decays.
    
\end{enumerate}

More generally, dedicated searches for the di-scalar channels could give stronger bounds than the ones we obtained in \cref{fig:modelindependent}, and detailed studies are needed to determine the most promising final states. We leave this investigation for future work.

\section{Bounds on the SU(5)/SO(5) pNGBs} \label{sec:su5so5}

The simplified model approach is very useful as the limits can be applied to a broad class of models, at least to a certain extent. In this section, we investigate a specific full model with an extended EW scalar sector, study the bounds on the full model and compare the results to estimates one can very quickly obtain by using the simplified model approach of \cref{sec:modelindependent}. 

We focus here on composite Higgs models based on gauge/fermionic underlying dynamics \cite{Ferretti:2013kya,Ferretti:2014qta,Ferretti:2016upr}. Minimal models feature one of the following cosets in the EW sector:
$\SU(4)/\Sp(4)$, $\SU(5)/\SO(5)$ or $\SU(4)\times\SU(4)/\SU(4)_D$. A first rough sketch of the LHC phenomenology of the pNGBs can be found in \cite{Ferretti:2016upr}. We  focus on the $\SU(5)/\SO(5)$ coset \cite{Agugliaro:2018vsu} in the following as it features a doubly charged scalar.

\subsection{The electroweak pNGBs and their LHC phenomenology}

The pNGBs from the $\SU(5)/\SO(5)$ coset have been investigated in detail in Ref.~\cite{Agugliaro:2018vsu} (see also Refs.~\cite{Dugan:1984hq,Ferretti:2014qta}). Here, we summarise the key elements and discuss in some detail the underlying LHC phenomenology. A complete summary of the pNGB couplings to vector bosons, which are relevant for this study, can be found in Ref.~\cite{Banerjee:2022xmu}.
The pNGBs of the EW sector form a $\mathbf{14}$ of $\SO(5)$, which decomposes with respect to the custodial $\SU(2)_L \times \SU(2)_R$ as 
\begin{equation}
	\mathbf {14} \to (\mathbf 3, \mathbf 3) + (\mathbf 2, \mathbf 2) + (\mathbf 1, \mathbf 1) \,.
\end{equation}
We identify the $(\mathbf 2, \mathbf 2)$ with the Higgs doublet (bi-doublet of the custodial symmetry).
Following the notation of Ref.~\cite{Agugliaro:2018vsu}, the bi-triplet can be decomposed under the custodial $\SU(2)_D \subset \SU(2)_L \times \SU(2)_R$ as
\begin{equation}
	(\mathbf 3, \mathbf 3) \to \mathbf 1+\mathbf 3+\mathbf 5 \equiv \eta_1 + \eta_3 + \eta_5\,,
\end{equation}
where
\begin{equation} \label{eq:SU2Dbasis}
	\eta_1 = \eta_1^0, \quad \eta_3 = (\eta_3^+, \eta_3^0, \eta_3^-), \quad \, \eta_5 = (\eta_5^{++}, \eta_5^+, \eta_5^0, \eta_5^-, \eta_5^{--}).
\end{equation}
This basis is suggested by the fact that the vacuum of the strong sector preserves the custodial $\SU(2)_D$.\footnote{Note that $(\mathbf 2, \mathbf 2) \rightarrow \mathbf 3 + \mathbf 1$, where $\mathbf 3$ are the longitudinal $W$ and $Z$ components and $\mathbf 1$ is the Higgs boson.} Nevertheless, a mixing among the states is induced by the terms in the scalar potential that violate it. To simplify the analysis, in the following we neglect the mixing and assume that the three multiplets have common masses $m_1$, $m_3$ and $m_5$, respectively.
Mass differences are due to the EW symmetry breaking, hence one naively expects a relative mass split of the order $v/m_i$ ($i=1,3,5$) where $v$ is the VEV of the Higgs boson. 
The precise values depend on the details of the scalar potential: here, we consider the mass differences as free parameters, and allow them to vary up to $200$~GeV.
Besides the masses, there is an additional parameter that is important for the phenomenology:
$\sin\theta =v/f_\psi$, with $f_\psi$ being the decay constant of this pNGB sector. Electroweak precision data give a lower bound of  about 1~TeV on $f_\psi$ \cite{Agugliaro:2018vsu}. Last but not least, we assume that the vacuum is only misaligned along the Higgs direction in order to avoid large breaking of the custodial symmetry. 
We remark that, while the EW quantum numbers of the scalars are similar to those of the Georgi-Machacek model~\cite{Georgi:1985nv}, all states in \cref{eq:SU2Dbasis} are parity-odd, except for $\eta_3$ which is parity even. Hence, in the composite model only the custodial triplet can develop a VEV without breaking CP.

\begin{figure}[t]
	\centering
	\begin{subfigure}{0.48\linewidth}
		\includegraphics[width=\linewidth]{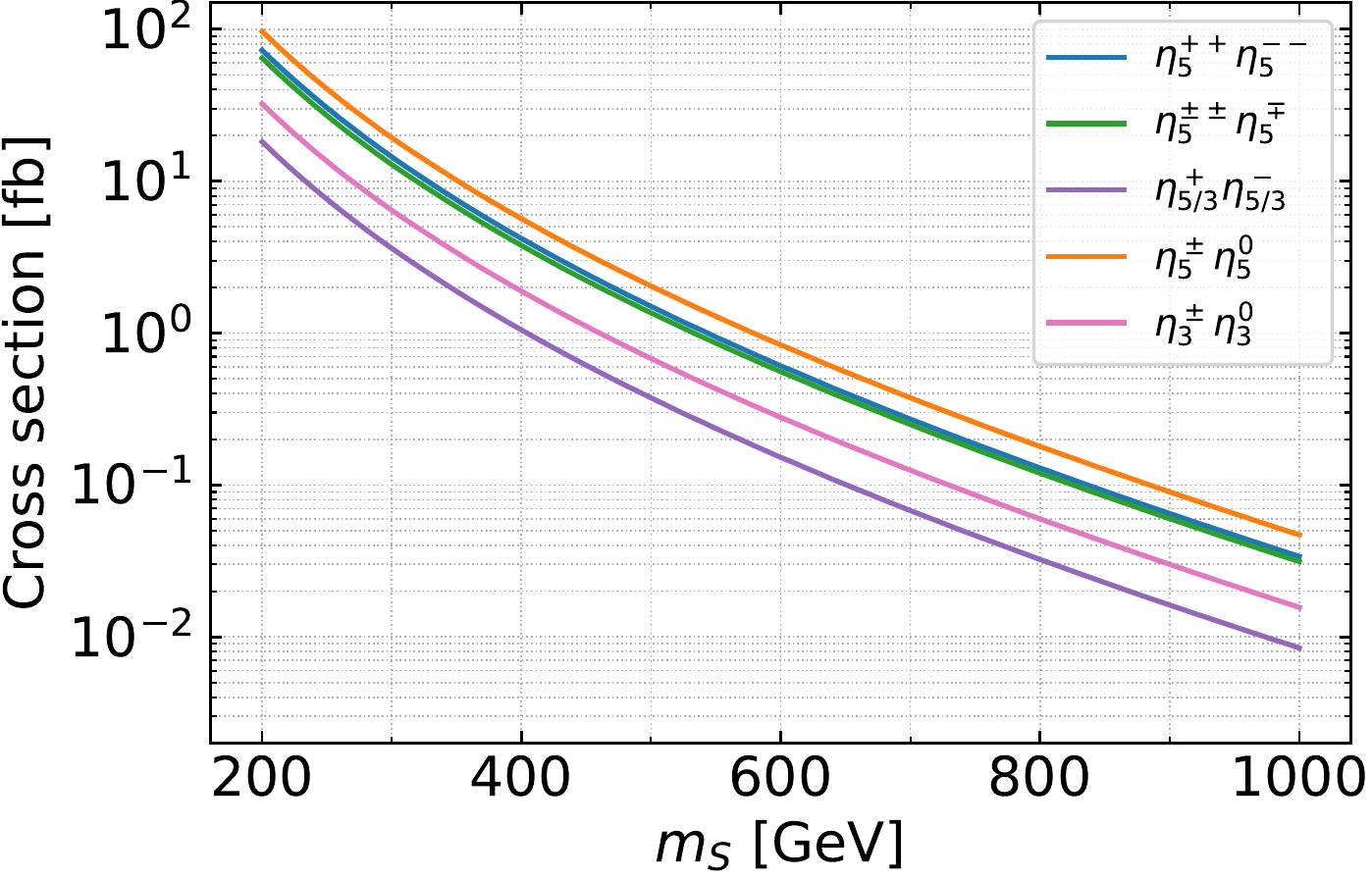}
	\end{subfigure} \quad
	\begin{subfigure}{0.48\linewidth}
		\includegraphics[width=\linewidth]{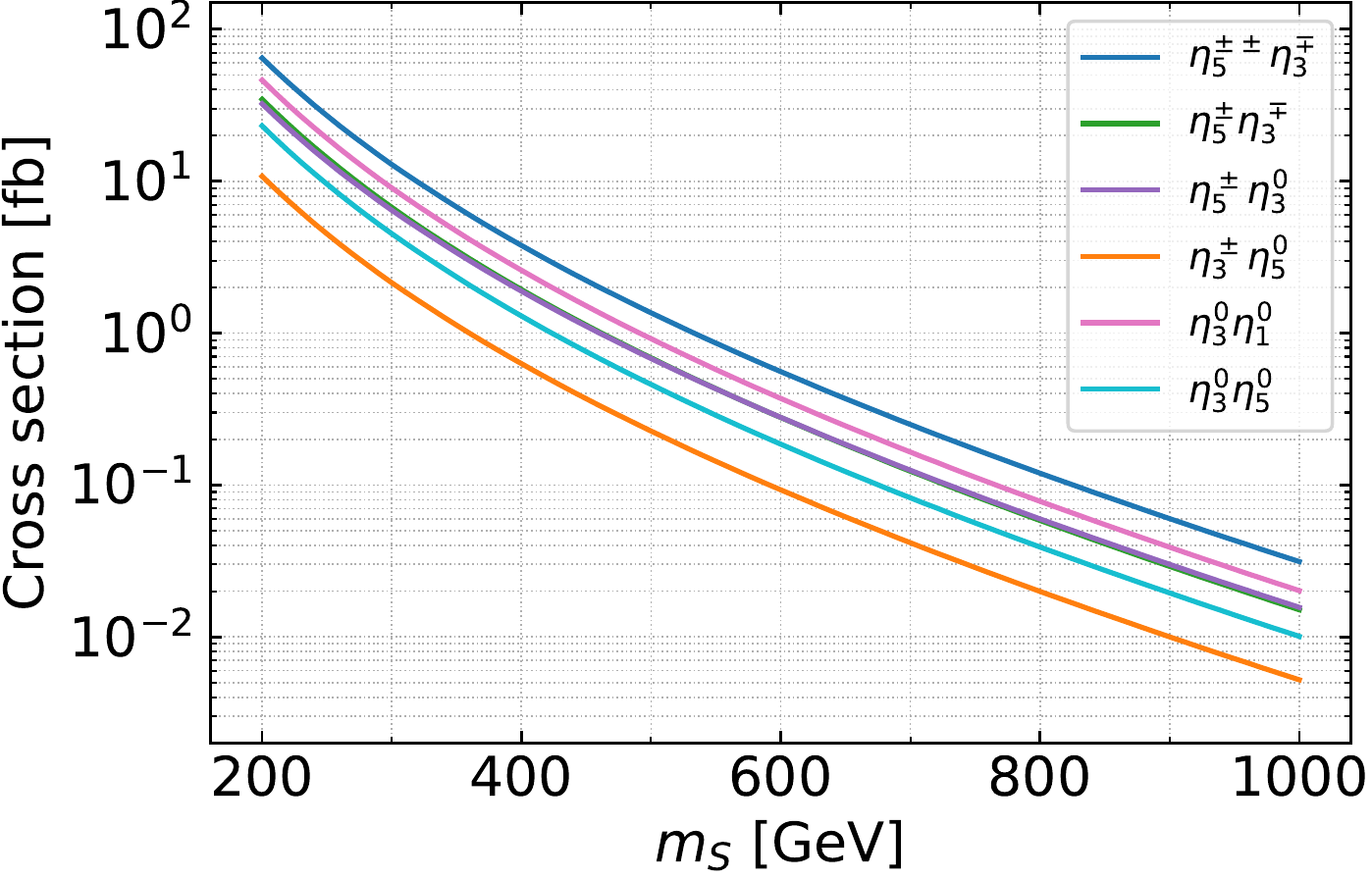}
	\end{subfigure} 
	\caption{Cross sections for the Drell-Yan production of $\SU(5)/\SO(5)$ pNGBs at the LHC with $\sqrt{s} = 13\,$TeV, assuming the same mass for all states of the custodial singlet, triplet, and quintuplet. Note that the $\eta_1^0 \eta_5^0$ combination is not allowed as they are both parity-odd.}
	\label{fig:xseta}
\end{figure}

\begin{figure}
    \centering
    \includegraphics[scale=1.0]{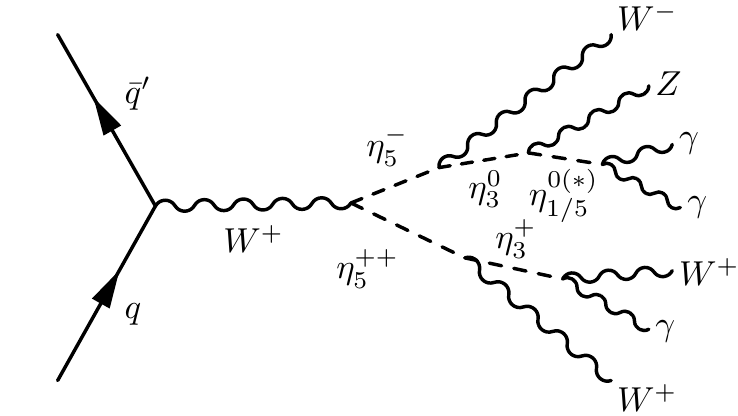} 
    \caption{Example of Drell-Yan production of two pNGBs with cascade and anomaly decays. If the triplet is the lightest multiplet, the $\eta_3^0$ undergoes three-body decays via  off-shell pNGBs.}
    \label{fig:feynmaneta5}
\end{figure}

In composite Higgs models with an extended pNGB sector, there are three types of couplings that determine the phenomenology of the scalars:
\begin{itemize}
    \item[(i)] Gauge interactions due to the EW quantum numbers of the pNGBs. In absence of VEVs, they lead to couplings of two scalars with one (and two) gauge boson(s), along the lines of \cref{eq:L_SSV}. For the $\SU(5)/\SO(5)$ coset, a complete list of these couplings is reported in Ref.~\cite{Banerjee:2022xmu}.
    \item[(ii)] Couplings of one pNGB to two EW gauge bosons generated by the topological anomaly of the coset, in the form of \cref{eq:LpiVVtilde}. They correspond to dimension-5 operators and are suppressed by one loop. For $\SU(5)/\SO(5)$, the coefficients are listed in Refs.~\cite{Dugan:1984hq,Ferretti:2014qta}. Note that the parity-even state $\eta^0_3$ lacks these couplings, gluons do not appear as the underlying fermions are only charged under the EW symmetry, and the model dependence is contained in a pre-factor that depends on the gauge group of the underlying confining dynamics. 
    \item[(iii)] Couplings of one pNGB to SM fermions, in the form of \cref{eq:L_ffpi}, where only top and bottom appear following top partial compositeness. These couplings depend on the properties of the top partners, and they are classified in Ref.~\cite{Agugliaro:2018vsu}.
\end{itemize}
The couplings (i) are responsible for Drell-Yan pair production, which dominate as (ii) and (iii) lead to very small cross sections. The cross sections of all pNGB pairs as a function of a common mass are shown in Fig.~\ref{fig:xseta}, which include a K-factor of 1.15 arising from QCD corrections \cite{Fuks:2013vua}.
Finally, all types of couplings determine the decay patterns of the scalar pair. We illustrate an example in Fig.~\ref{fig:feynmaneta5}. Besides the cascade decays, which are relevant for large enough mass splits between multiplets, the final states match the di-scalar channels discussed in Sec.~\ref{sec:modelindependent}. In particular, when couplings to fermions are present, they tend to dominate over the decays to gauge bosons.

\begin{figure}[th]
	\centering
	\begin{subfigure}{0.48\linewidth}
		\includegraphics[width=\linewidth]{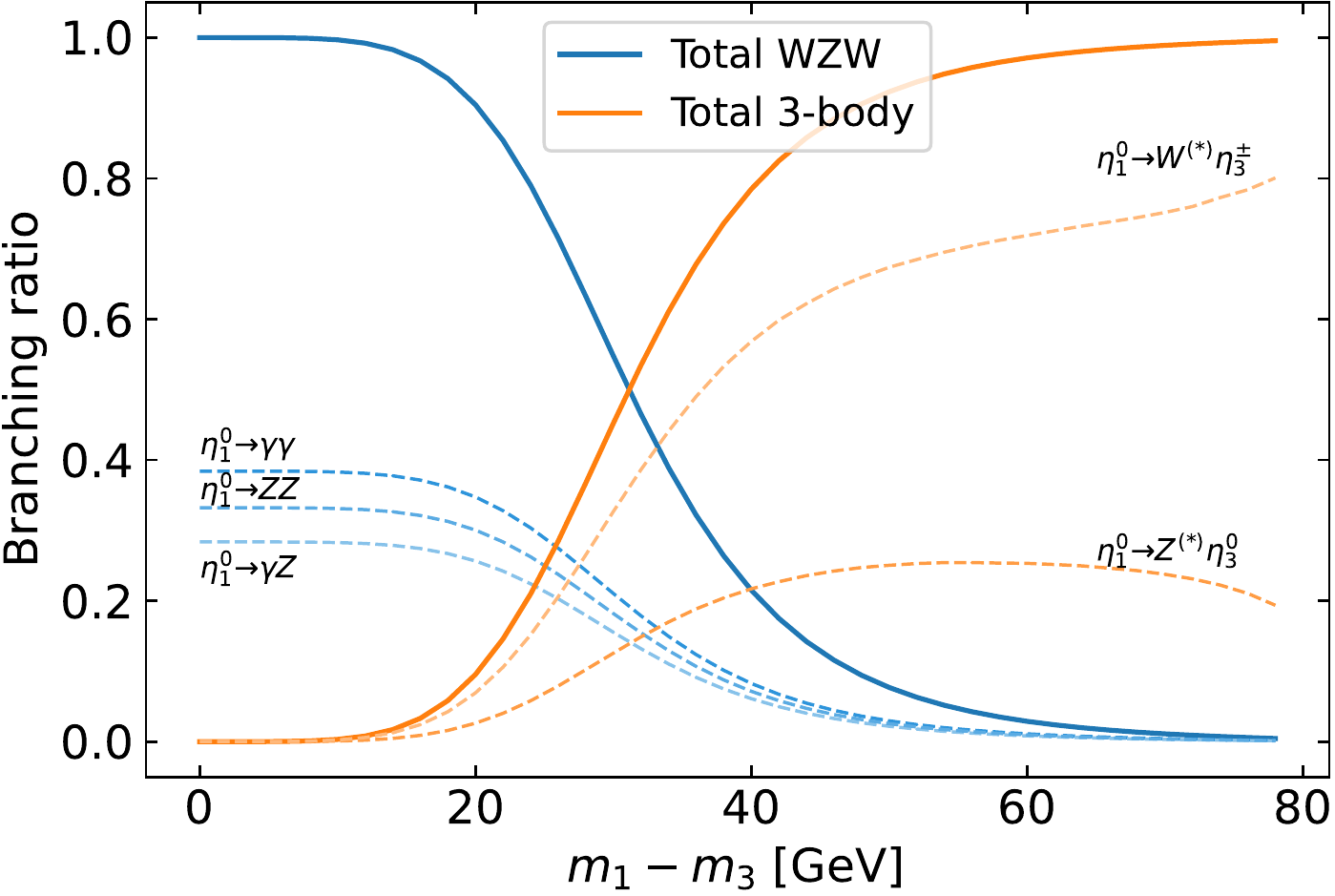}
		\caption{Decays of $\eta_{1}^0$ for $m_1=600~\mathrm{GeV}>m_3$}
		\label{fig:eta10brs}
	\end{subfigure} \quad
	\begin{subfigure}{0.48\linewidth}
		\includegraphics[width=\linewidth]{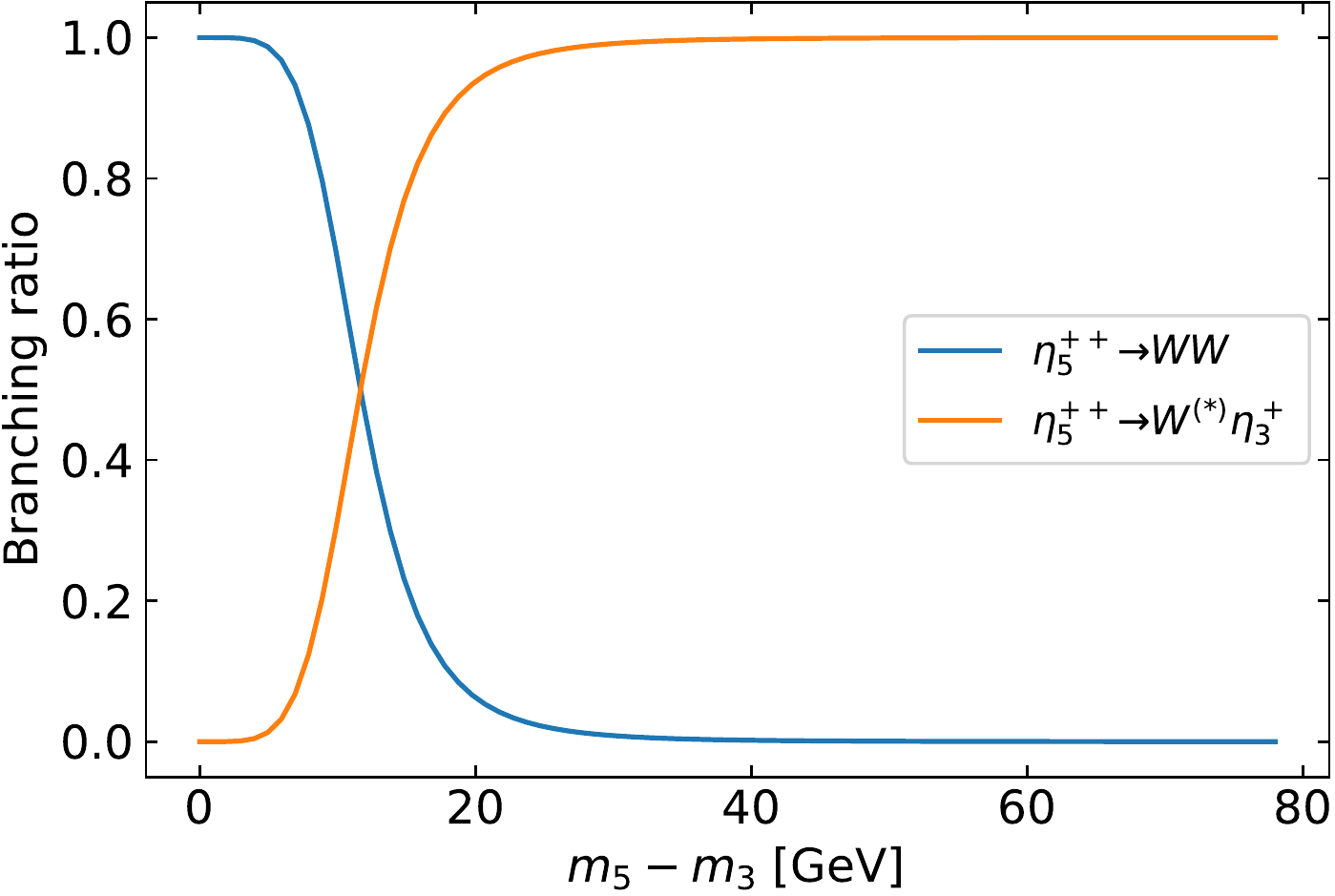}
		\caption{Decays of $\eta_{5}^{++}$ for $m_5=600~\mathrm{GeV}>m_3$}
		\label{fig:eta5ppbrs}
	\end{subfigure}  \vspace{1ex}
	
	\begin{subfigure}{0.48\linewidth}
		\includegraphics[width=\linewidth]{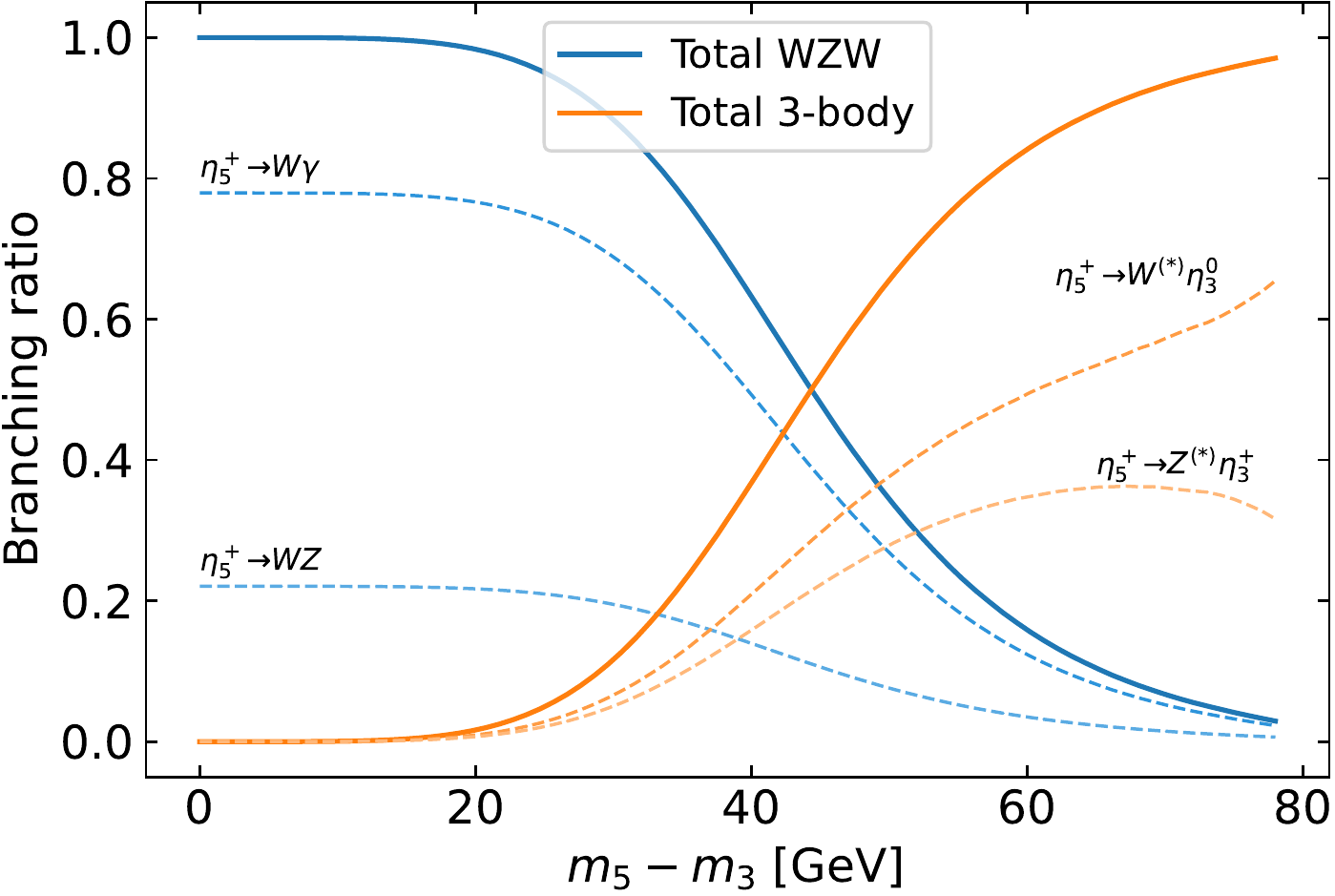}
		\caption{Decays of $\eta_{5}^+$ for $m_5=600~\mathrm{GeV}>m_3$}
		\label{fig:eta5pbrs}
	\end{subfigure} \quad
	\begin{subfigure}{0.48\linewidth}
		\includegraphics[width=\linewidth]{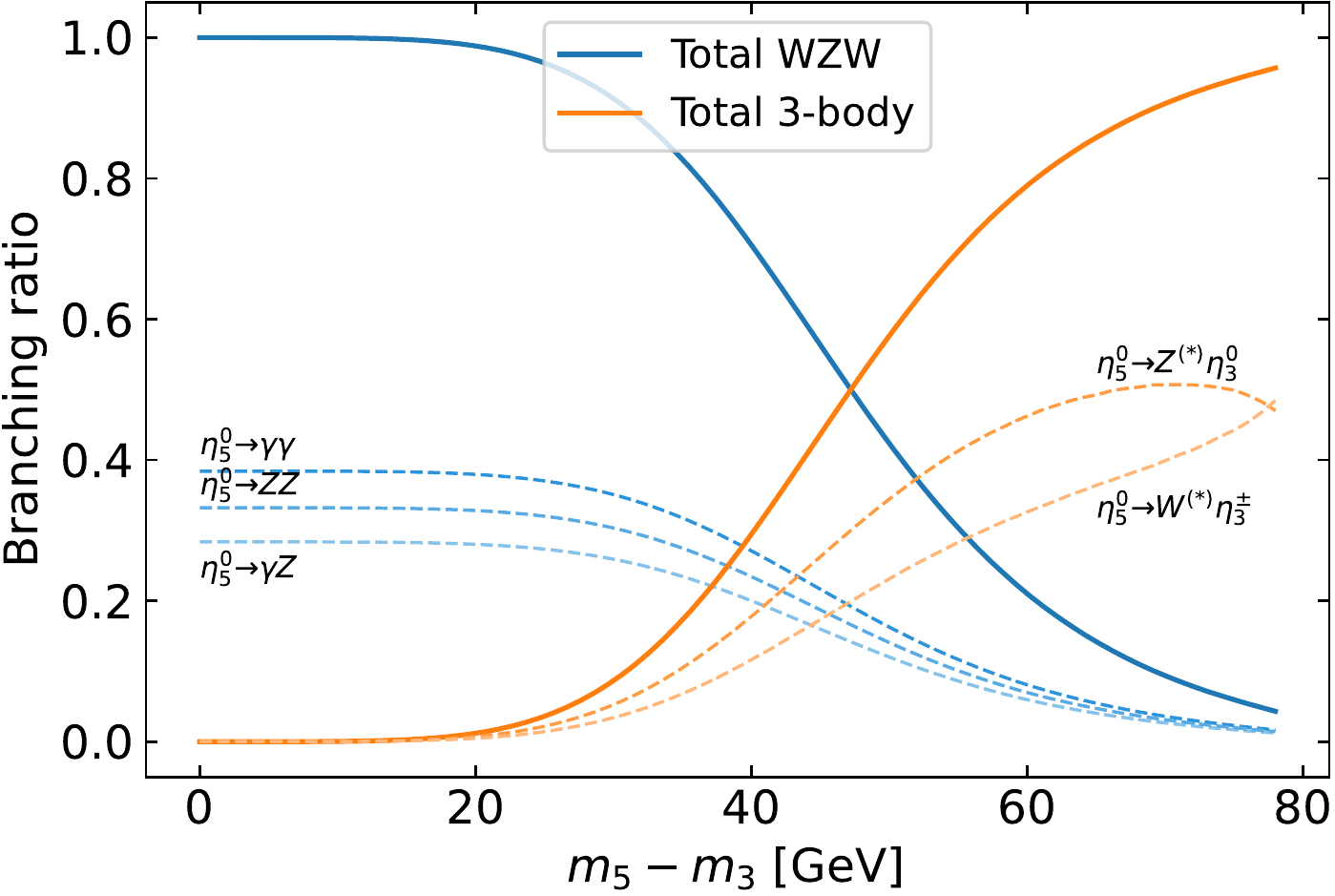}
		\caption{Decays of $\eta_{5}^{0}$ for $m_5 = 600~\mathrm{GeV}>m_3$}
		\label{fig:eta50brs}
	\end{subfigure} 
	\caption{Overview of the pNGB decays in the fermiophobic case. The mass of the decaying particles is set to 600~GeV. The heavier state decays either via the anomaly into di-boson final states or via an (off-shell) gauge boson into a lighter pNGB.}
	\label{fig:breta}
\end{figure}

\begin{figure}[th]
	\centering
	\begin{subfigure}{0.48\linewidth}
		\includegraphics[width=\linewidth]{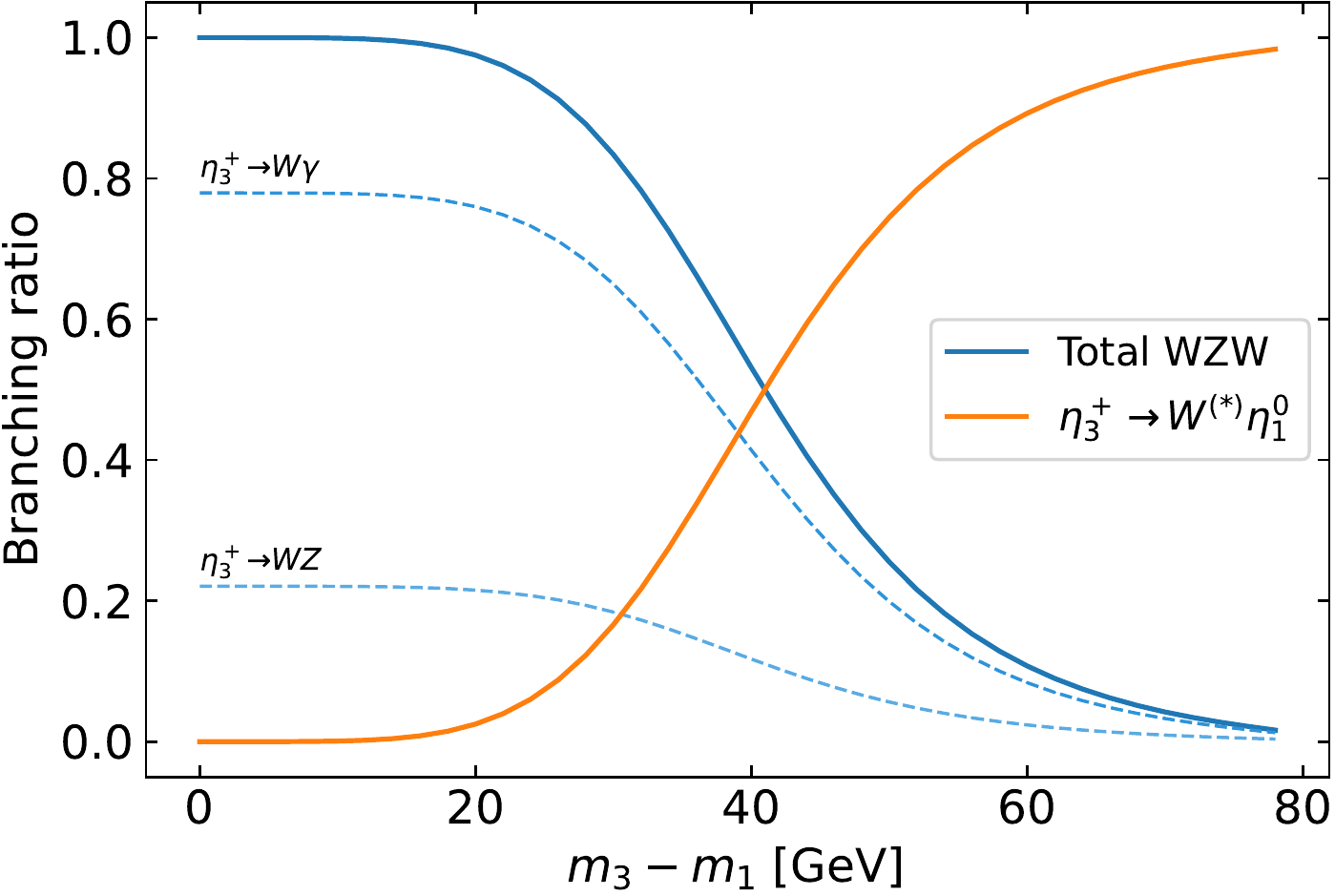}
		\caption{Decays of $\eta_{3}^+$ for $m_5 \gg m_3 = 600~\mathrm{GeV} >m_1$}
		\label{fig:eta3pbrs-m3m1}
	\end{subfigure} \quad
	\begin{subfigure}{0.48\linewidth}
		\includegraphics[width=\linewidth]{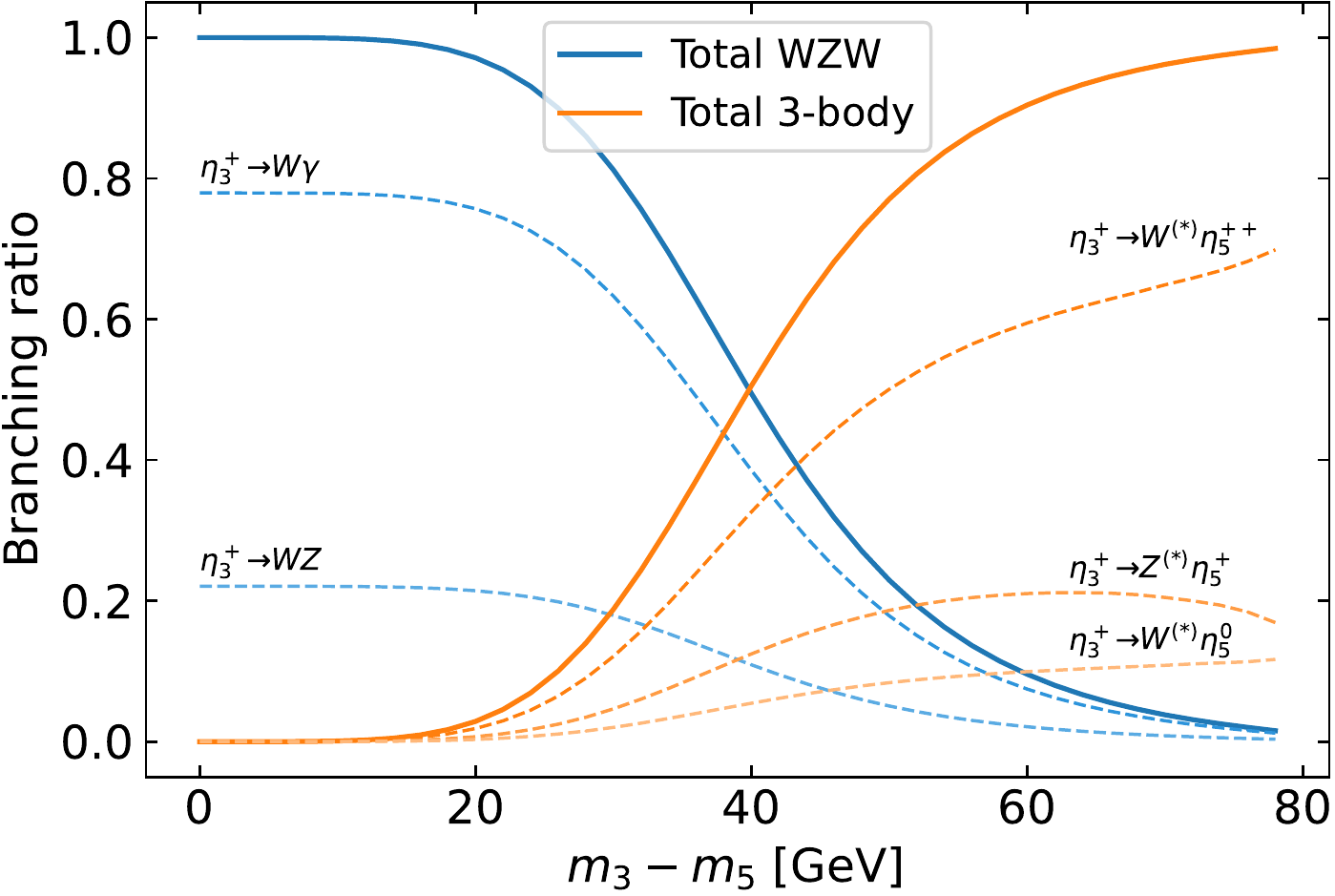}
		\caption{Decays of $\eta_{3}^+$ for $m_1 \gg m_3 = 600~\mathrm{GeV} >m_5$}
		\label{fig:eta3pbrs-m3m5}
	\end{subfigure}  \vspace{1ex}
	
	\begin{subfigure}{0.48\linewidth}
		\includegraphics[width=\linewidth]{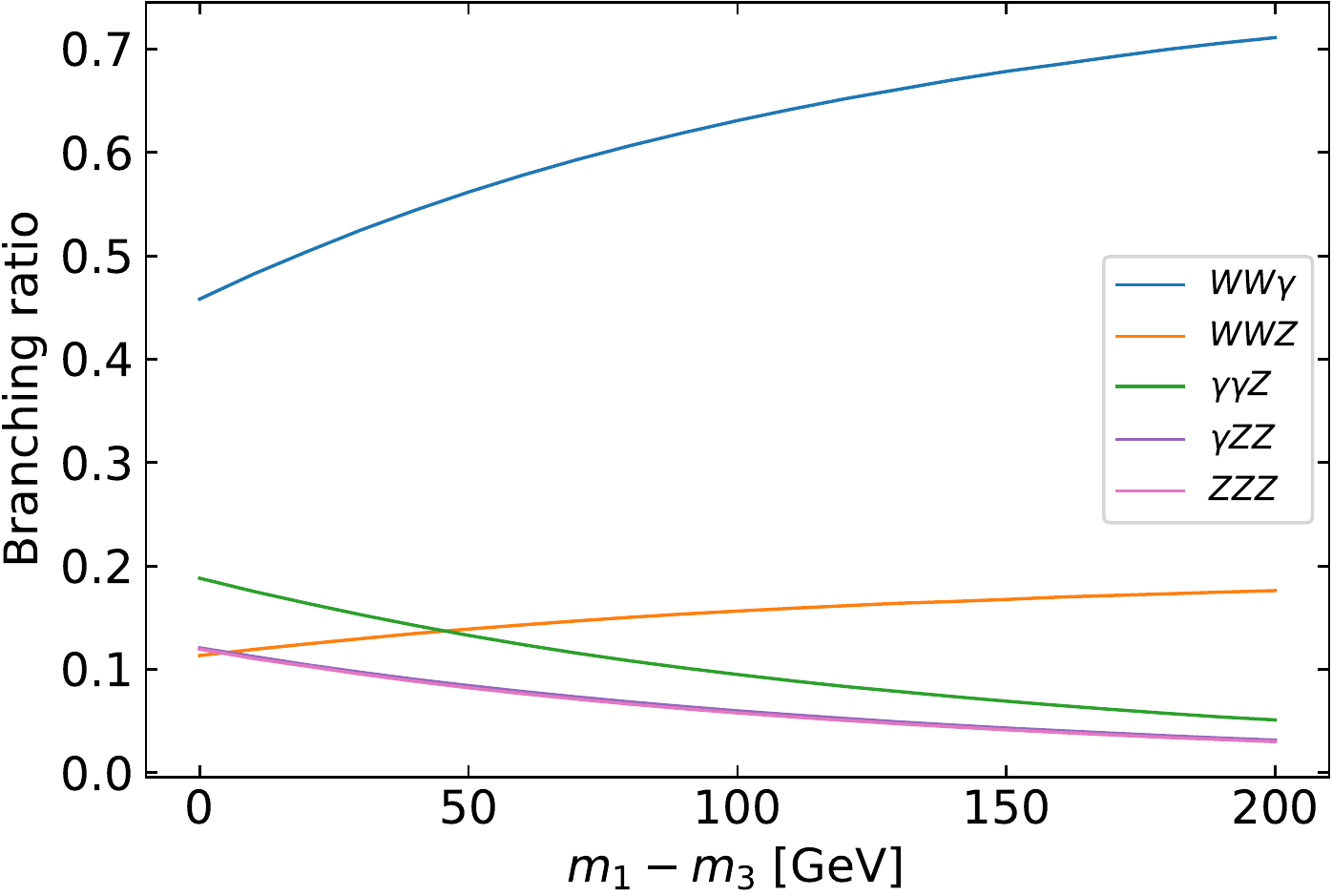}
		\caption{Decays of $\eta_{3}^0$ for $m_5\gg m_1 > m_3=600$~GeV}
		\label{fig:eta30brs-m1m3}
	\end{subfigure} \quad
	\begin{subfigure}{0.48\linewidth}
		\includegraphics[width=\linewidth]{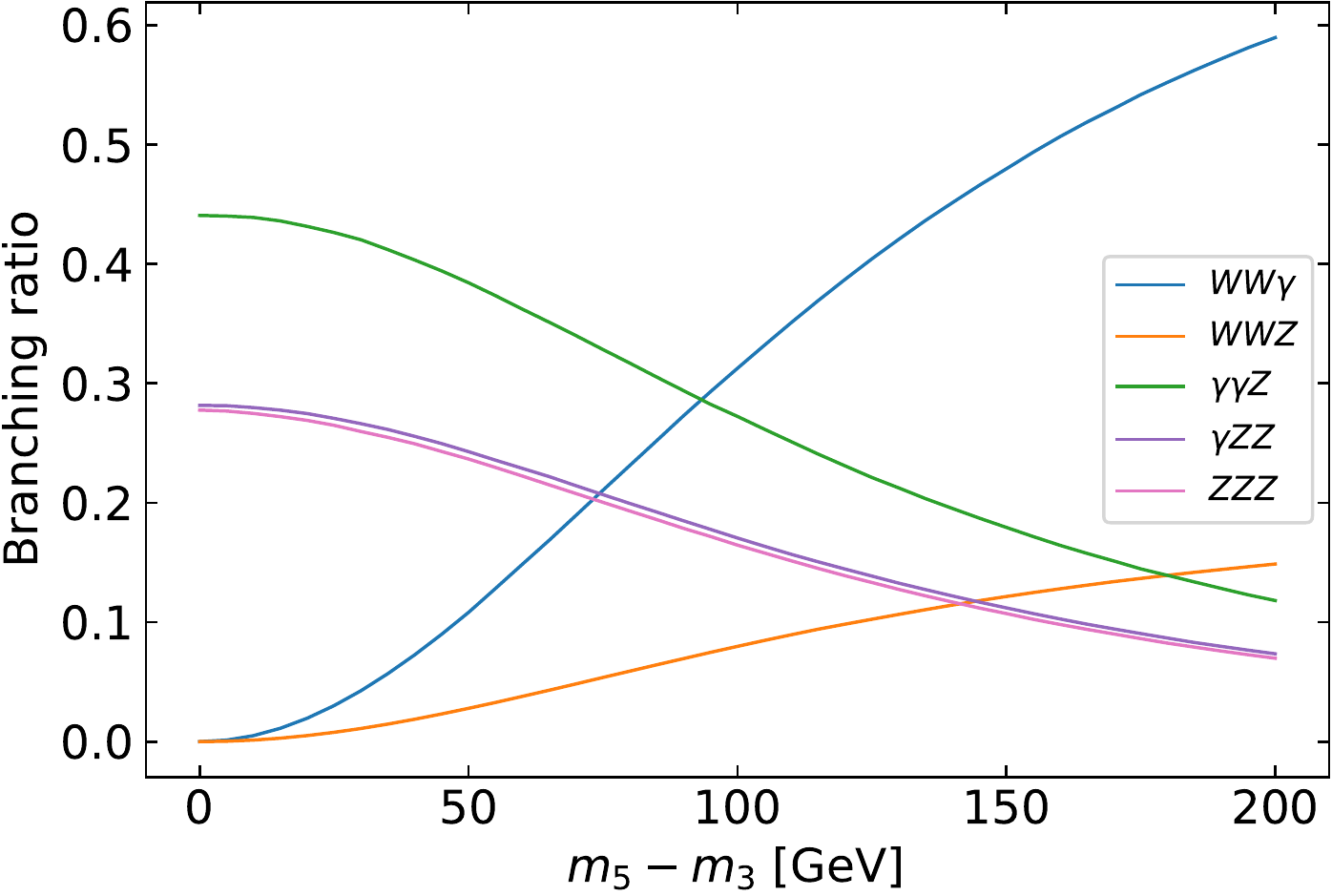}
		\caption{Decays of $\eta_{3}^0$ for $m_1\gg m_5 > m_3=600$~GeV}
		\label{fig:eta30brs-m5m3}
	\end{subfigure} 
	\caption{Overview of the pNGB decays in the fermiophobic case (continued from \cref{fig:breta}). The neutral triplet component decays into three gauge bosons, as it does not couple to the anomaly.}
	\label{fig:breta2}
\end{figure}

The LHC signatures of pNGB pair production depend strongly on whether the pNGBs are fermiophilic or fermiophobic.
We start the discussion with the fermiophobic case, in which case interactions to the EW gauge bosons are relevant.
The corresponding branching ratios are shown in \cref{fig:breta,fig:breta2}.
For the lightest multiplet and near-degenerate masses, the anomaly couplings determine decays into a pair of EW gauge bosons, with the exception of $\eta_3^0$. At the leading order in $v/f_\psi$, only decays involving neutral gauge bosons appear.\footnote{This is due to the fact that the only gauge-invariant operator appears for the neutral triplets, $\phi^a W^a_{\mu\nu} \tilde{B}^{\mu\nu}$, where $B$ contains the hypercharge gauge boson. Couplings with only $W^\pm$ need two insertion of the Higgs VEV, hence they are suppressed by $v^2/f_\psi^2$.}
Hence, the singly charged states decay as
\begin{equation}\label{eq:eta35pdec}
    \eta_{3,5}^+ \to W^+ \gamma,\, W^+ Z\,,
\end{equation}
with dominant photon channel as $\br(\eta_{3,5}^+\to W^+\gamma)\approx \cos^2 \theta_W \approx 78\%$ \cite{Agugliaro:2018vsu} for both multiplets, as shown in \cref{fig:eta5pbrs,fig:eta3pbrs-m3m1,fig:eta3pbrs-m3m5} for small mass split.
The neutral singlet and quintuplet can decay as
\begin{equation}\label{eq:eta150dec}
    \eta_{1,5}^0 \to \gamma\gamma,\, \gamma Z,\, ZZ\,,
\end{equation}
with comparable branching ratios, see for example the $\eta^0_1$ decays in \cref{fig:eta10brs} and the $\eta^0_5$ decays in \cref{fig:eta50brs} for small mass split.
Couplings to charged $W^\pm$ are suppressed by the EW scale, hence they lead to $\eta_{1,5}^0 \to W^+ W^-$ branching ratios suppressed by $(v/f_\psi)^4 \lsim 10^{-3}$, which we neglect. Instead, while still suppressed, this provides the only available decay channel for the doubly charged pNGB in the quintuplet:
\begin{equation}\label{eq:eta5ppdec}
    \eta_5^{++} \to W^+ W^+.
\end{equation}

Finally, the $\eta_3^0$ is CP-even and thus has no couplings to the anomaly.
It therefore undergoes three-body decays via off-shell pNGBs:
\begin{subequations}\label{eq:eta30dec}
\begin{alignat}{2}
    &\eta_3^0 \to W^+ W^- \gamma, \, W^+ W^- Z \quad &\text{ via } \eta_{3,5}^{\pm (*)}\,, \mbox{ and } \label{eq:eta30dec+}\\
    &\eta_3^0 \to Z \gamma \gamma, \, Z Z \gamma, \, ZZZ \quad &\text{ via } \eta_{1,5}^{0 (*)}\,. \label{eq:eta30dec0}
\end{alignat}
\end{subequations}
These processes contribute to the upper tier in \cref{fig:feynmaneta5}.
There is an interesting cancellation taking place in the three-body decays:
In the limit $\theta \to 0$, the contributions to \cref{eq:eta30dec+} cancel exactly if $m_3=m_5$.
The same holds for \cref{eq:eta30dec0} if $m_1=m_3=m_5$.
Thus, if the pNGBs are mass-degenerate, the $\eta_3^0$ becomes rather long-lived and leaves the detector before it decays.
In practice, however, we expect at least a small split, so $\eta_3^0$ decays promptly to three vector bosons.
The main effect on the phenomenology is that the decays through the charged channel \cref{eq:eta30dec+} are suppressed if $m_1\gg m_5 \gtrsim m_3$, which we explore further in \cref{sec:boundsfermiophobic} and which is illustrated in \cref{fig:eta30brs-m5m3}.

The discussion so far applies to the lightest multiplet and also covers the case where the multiplets are very close in mass. However, there can be a sizeable mass split, in which case cascade decays from one multiplet into a lighter one and a (potentially off-shell) vector boson become important. 
Assuming for example $m_5>m_3>m_1$, we have
\begin{subequations}\label{eq:etachaindec}
\begin{gather}
    \eta_5^{++} \to W^{+(*)} \eta_3^+, \qquad
    \eta_5^+ \to Z^{(*)} \eta_3^+,\, W^{+(*)} \eta_3^0, \qquad
    \eta_5^0 \to W^{\pm(*)} \eta_3^\mp, \, Z^{(*)} \eta_3^0, \\
    \eta_3^+ \to W^{+(*)} \eta_1^0, \qquad \eta_3^0 \to Z^{(*)} \eta_1^0.
\end{gather}
\end{subequations}
We find that both classes of decays are of similar importance once the mass split is between $30$ and $50$~GeV, see \cref{fig:breta,fig:breta2}, while cascade decays dominate for larger mass splits.
The two exceptions to this rule of thumb are $\eta_5^{++}$ as shown in \cref{fig:eta5ppbrs}, whose anomaly coupling is suppressed by $v^2/f_\psi^2$, and $\eta_3^0$, for which the anomaly-induced three-body decays are irrelevant as soon as any cascade decay is accessible. We note, for completeness, that the quintuplet does not couple to the singlet
in the model considered.

We turn now to the fermiophilic case. We assume here that only couplings to quarks are present. 
One expects that the couplings in \cref{eq:L_ffpi} scale like the quark masses, e.g.
\begin{align}\label{eq:couplingscaling}
\kappa^{\eta^0_i}_{t} = c_t^i\,
\frac{m_t}{f} \quad , \quad
\kappa^{\eta^0_i}_{b} = c_b^i\,
\frac{m_b}{f} \quad \text{and} \quad
\kappa^{\eta^+_j}_{tb} = c_{tb}^j\,
\frac{m_t}{f}\,,
\end{align}
where the $c$ coefficients are of order one.
In this case the decays to third generation quarks dominate over the loop-level anomaly-induced decays into two vector bosons or the three-body decays discussed above. Hence, we  consider for this scenario the decays
\begin{align}
    \eta_{3,5}^+ \to t \bar b, \qquad \eta_{1,3,5}^0 \to t \bar t,\, b\bar b\,.
\end{align}
From \cref{eq:couplingscaling}, the $t\bar t$ channel dominates over $b\bar b$ above threshold.
In the case of $\eta^{++}_5$, it turns out that the three-body decay
\begin{equation}
    \eta_5^{++} \to W^+ t \bar b
\end{equation}
via an off-shell $\eta_{3,5}^+$ is dominant over the decay to $W^+ W^+$. In case of $m_5 > m_3$ also
the decay $\eta^{++}_5 \to W^{+(*)} \eta^+_3$ becomes important. We have checked that for mass differences below $25$~GeV the decay into quarks clearly dominates and for a mass difference of $50$~GeV the modes $W^+ t \bar b$ and $ W^{+(*)} \eta^+_3$ are of equal importance. For larger mass differences the latter mode is the most important one. Here we have assumed that the coefficients $c$ are equal to one.

\subsection{LHC bounds in the fermiophobic case}\label{sec:boundsfermiophobic}

\begin{figure}
    \centering
    \begin{subfigure}{0.48\linewidth}
        \includegraphics[width=\linewidth]{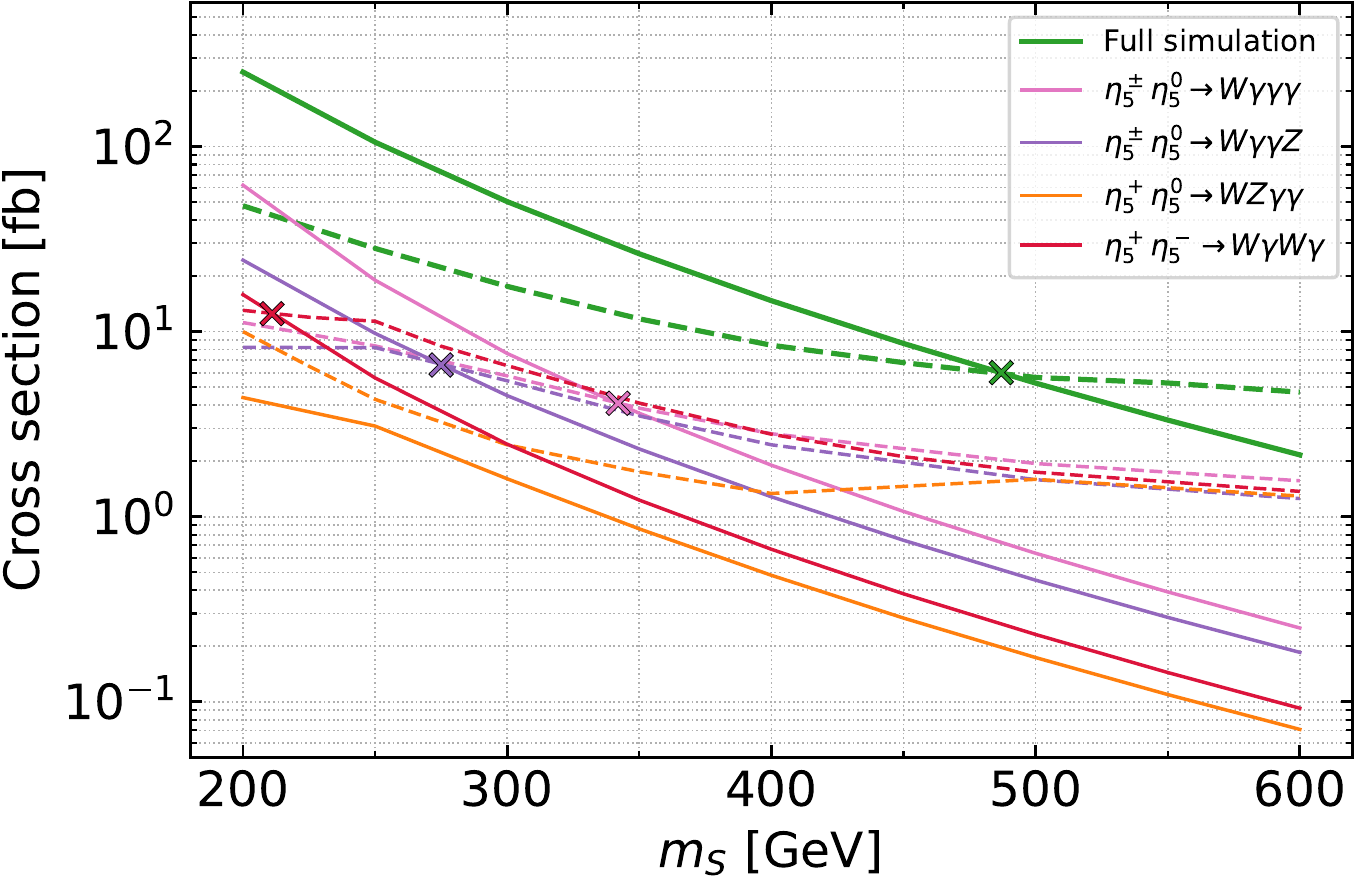}
        \caption{Bounds from individual channels}
        \label{fig:illustrationindividual}
    \end{subfigure} \quad
    \begin{subfigure}{0.48\linewidth}
        \includegraphics[width=\linewidth]{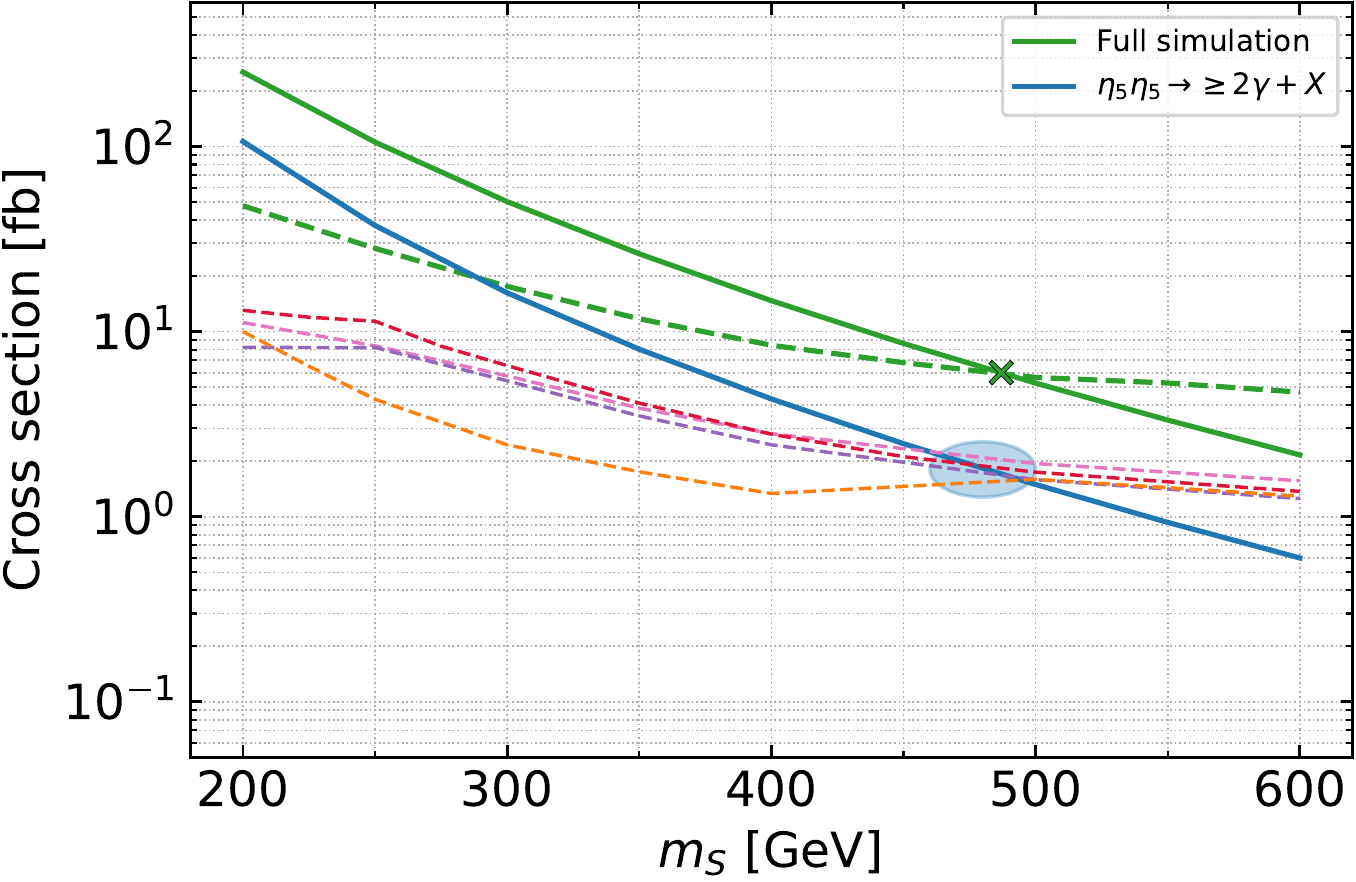}
        \caption{Bounds from sum of multiphoton channels}
        \label{fig:illustrationcombined}
    \end{subfigure} \vspace{2ex}
    
    \caption{Application of the model-independent bounds to a specific model, the custodial quintuplet $\eta_5$ from the $\SU(5)/\SO(5)$ coset.
    In (a) we determine the bounds from the dominant individual channels by comparing the cross section time branching ratio from the model (solid) with the upper limits from \cref{fig:modelindependent} (dashed).
    In green we show the results of a full simulation. The blue line in (b) is the sum of the individual multi-photon cross sections shown in (a).
    Further details are given in the text.}
    \label{fig:illustration}
\end{figure}

As a first step, we consider only the quintuplet $\eta_5$ and apply the simplified model bounds from \cref{sec:modelindependent}, where we found that final states with multiple photons and at least one $W/Z$ yield the strongest constraints.
In \cref{fig:illustrationindividual} we compare the cross section times branching ratio of all multi-photon final states (solid lines) with the corresponding bounds from \cref{fig:modelindependent} (dashed lines).
From the individual channels we find that masses below $340$~GeV are excluded, with the strongest bound coming from $\eta_5^\pm \eta_5^0 \to W\gamma\gamma\gamma$. In addition, we perform a full simulation in which all states contained in the quintuplet are pair-produced and decayed according to the specific model under study.  
The solid green line denotes the sum over all pair production cross sections of the quintuplet. The dashed green line shows the corresponding bound, i.e.\ the sum of scalar pair production cross sections that would be needed in order to exclude the convolution of all decay channels from quintuplet states. As can be seen, the bound on the mass $m_S$ is $485$~GeV and thus significantly stronger than the bounds obtained from individual channels. 
The apparent discrepancy between simplified models and the full simulation stems from the fact that all  multi-photon channels populate the same signal region of the search~\cite{ATLAS:2018nud} that yields the dominant bound. Also, all multi-photon channels have a similar upper limit, indicating that the signal acceptances are comparable. Adding up the various signal cross sections with two or more photons results in the blue line shown in \cref{fig:illustrationcombined}.
Comparing this summed cross section with the bounds from different multi-photon channels (see the shaded area in \cref{fig:illustrationcombined}) yields an estimated bound on $m_S$ of $460-500$~GeV, in agreement with the result of the full simulation. This example shows the usefulness (and limitations) of the simplified model bounds and how they can be combined in the context of a particular model.

In a second step, we take all multiplets into account and consider scenarios with fixed mass differences. We study the following benchmark scenarios, characterised by varying a single mass scale $m_S$:
\begin{subequations}\label{eq:scenarios}
\begin{alignat}{4}
    \text{S-eq:}\quad &m_3 = m_S -2~\mathrm{GeV},\quad &&m_5 = m_S,\quad &&m_1 = m_S+2~\mathrm{GeV}\,; \\
    \text{S-135:}\quad &m_1 = m_S - 50~\mathrm{GeV},\quad &&m_3 = m_S,\quad &&m_5 = m_S + 50~\mathrm{GeV}\,; \\
    \text{S-531:}\quad &m_5 = m_S - 50~\mathrm{GeV},\quad &&m_3 = m_S,\quad &&m_1 = m_S + 50~\mathrm{GeV}\,; \\
    \text{S-351:}\quad &m_3 = m_S-50~\mathrm{GeV} ,\quad &&m_5 = m_S , \quad &&m_1 = m_S + 50~\mathrm{GeV}\,.
\end{alignat}
\end{subequations}
The choice of $50$~GeV is motivated by the fact that the mass splits are expected to be a fraction of the Higgs VEV.
The phenomenology differs in each case:
In S-eq, all particles decay via the anomaly and $\eta_3^0$ exhibits three-body decays.
We introduce a small mass split of 2~GeV to avoid the cancellation for some $\eta_3^0$ decays discussed below \cref{eq:eta30dec}.
In S-135 and S-531, the heavier states decay into the next lighter states or di-bosons, while the lightest states only have anomaly decays.
Finally, in S-351 both $\eta_1$ and $\eta_5$ decay into the triplet, and $\eta_3^0$ decays into three vector bosons.

\begin{figure}
    \centering
    \begin{subfigure}{0.48\linewidth}
        \includegraphics[width=\linewidth]{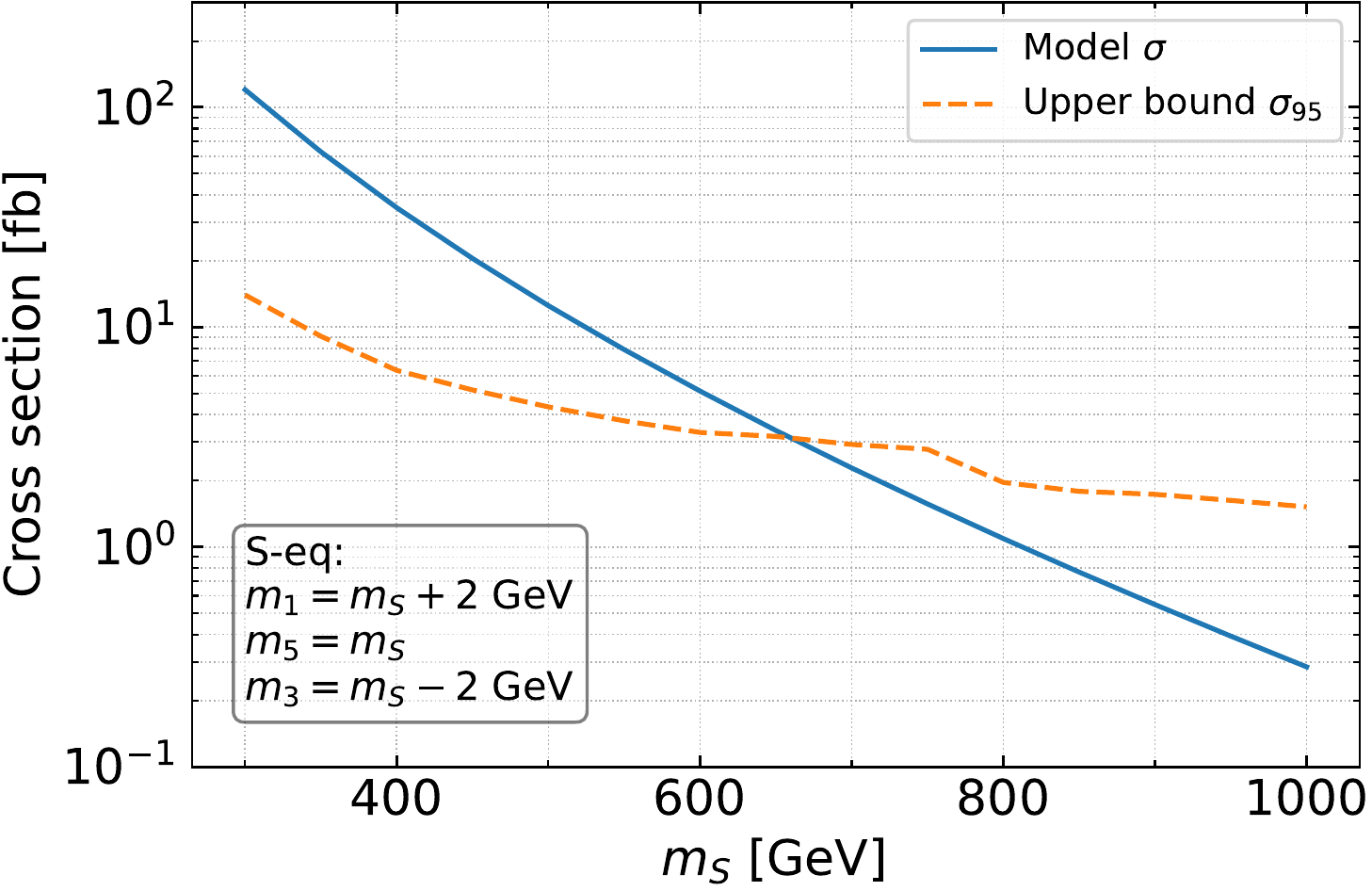}
        \caption{Scenario S-eq}
    \end{subfigure} \quad
    \begin{subfigure}{0.48\linewidth}
        \includegraphics[width=\linewidth]{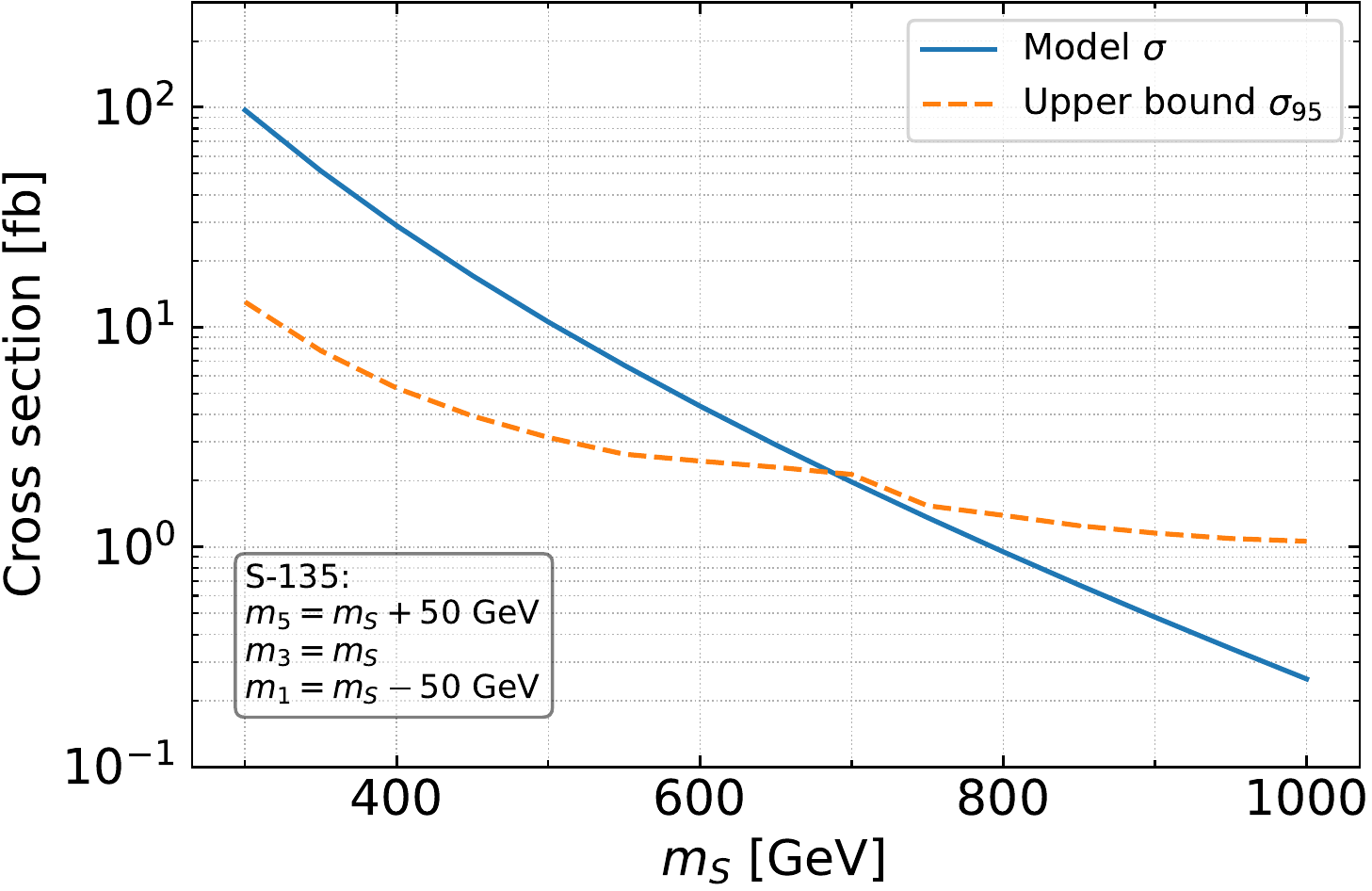}
        \caption{Scenario S-135}
    \end{subfigure} \vspace{2ex}
    
    \begin{subfigure}{0.48\linewidth}
        \includegraphics[width=\linewidth]{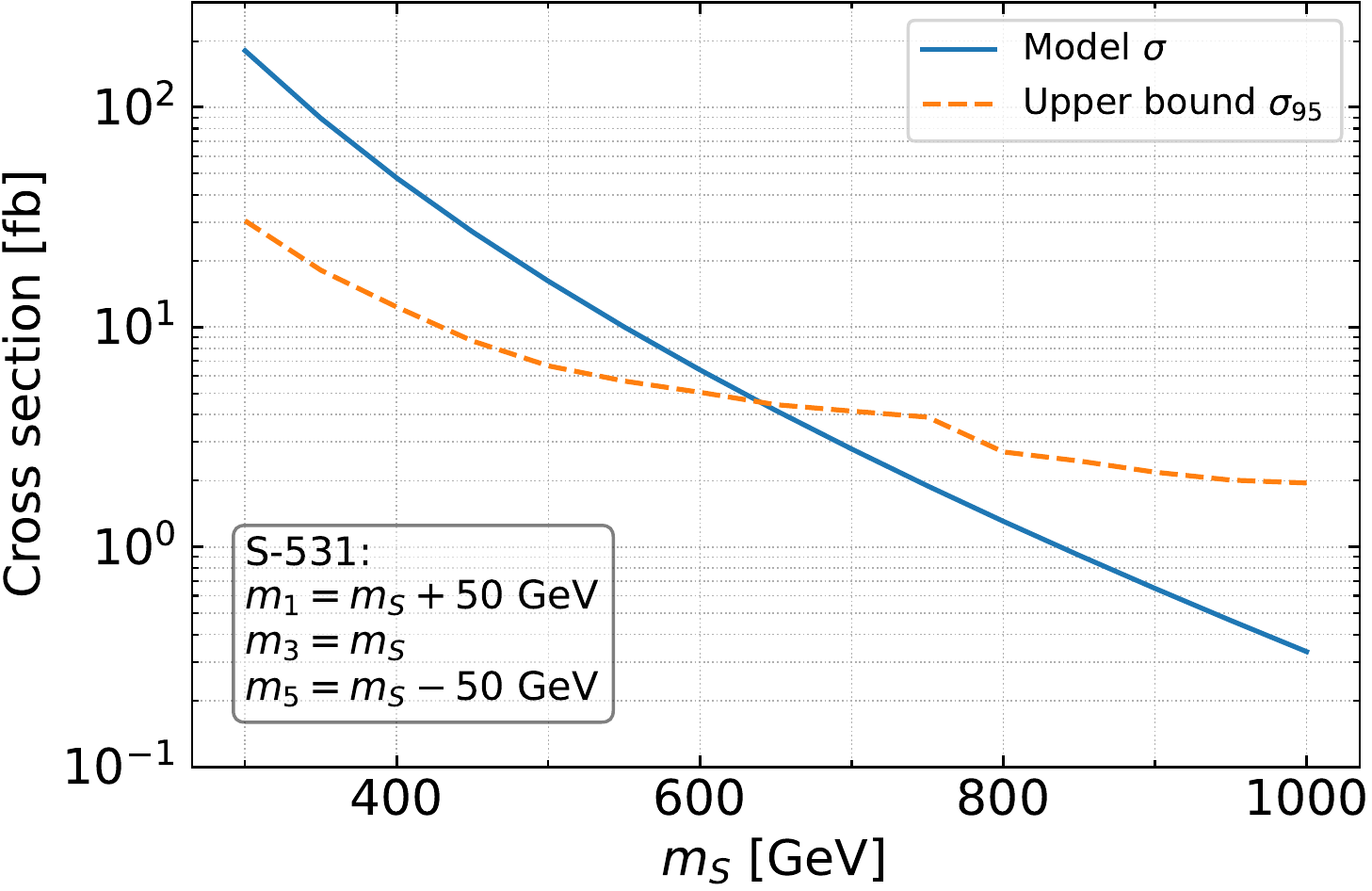}
        \caption{Scenario S-531}
    \end{subfigure} \quad
    \begin{subfigure}{0.48\linewidth}
        \includegraphics[width=\linewidth]{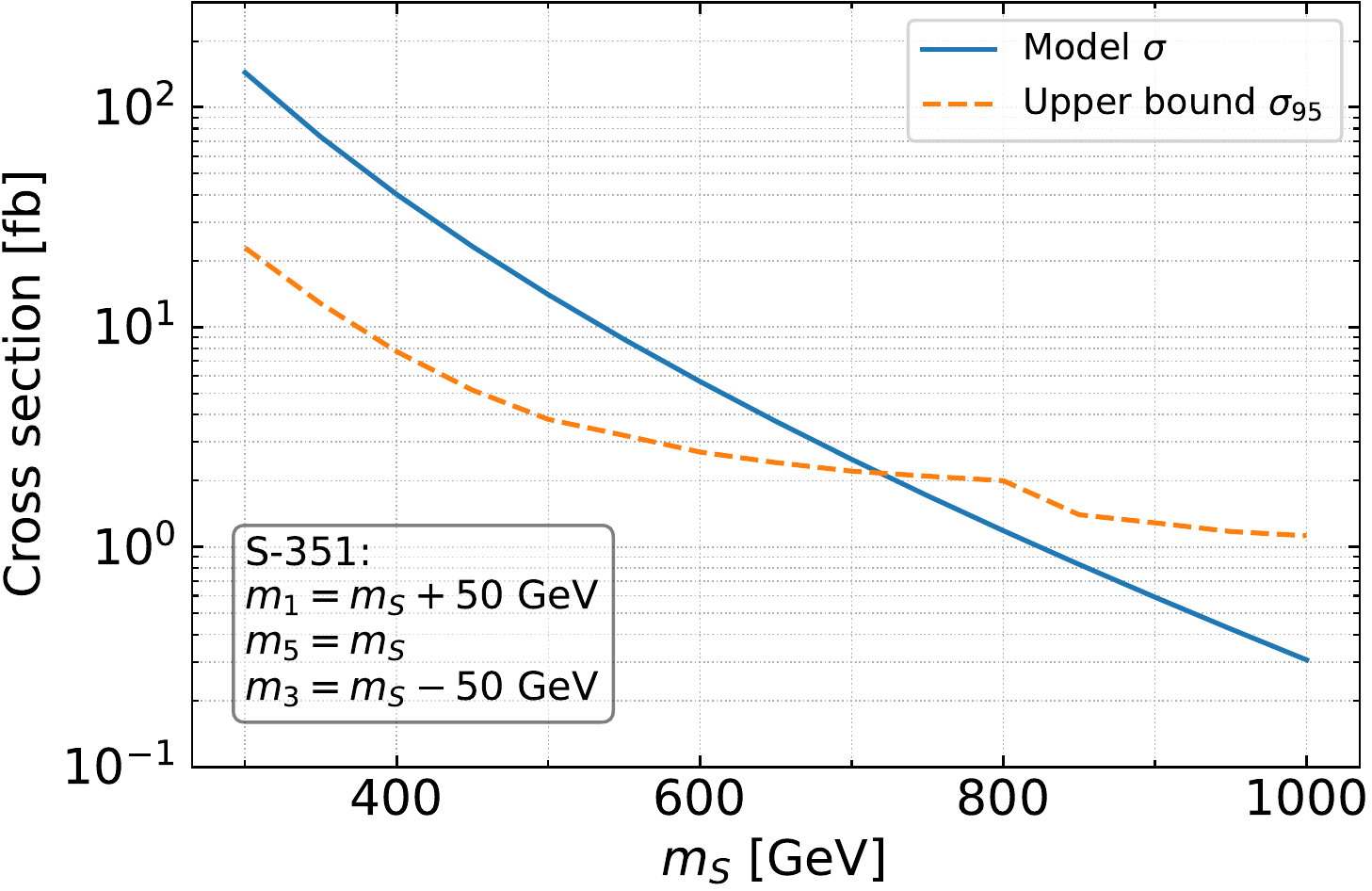}   
        \caption{Scenario S-351}
    \end{subfigure}
    \caption{Bounds on the pNGB masses for the Drell-Yan production of the full bi-triplet for multiple benchmark mass spectra defined in \cref{eq:scenarios}. In (a), all masses are approximately equal. In the remaining panels, there is a 50~GeV mass split between the multiplets.}
    \label{fig:1dimbounds-wzw}
\end{figure}

We present the bounds on the mass parameter $m_S$ for the four benchmark scenarios in \cref{fig:1dimbounds-wzw}.
In orange, we show the sums over all scalar pair production cross sections $\sigma_{95}$ that would be needed to exclude the model at 95\% CL at each parameter point.
As discussed above, the strongest bounds come from multi-photon channels, with Ref.~\cite{ATLAS:2018nud} being the dominant analysis, cf.\ \cref{tab:modelindependentanalyses} in \cref{app:analyses}.
The kink in $\sigma_{95}$ is due to a change in dominant signal region within the same analysis.
The actual sum over all pair production cross sections is drawn in blue.
The bounds range from $640$~GeV for S-135 to $720$~GeV for S-153.
The case S-eq can be understood by adding the additional channels due to the triplet and using the same procedure as in case of the pure quintuplet. The fact that the $\eta^0_3$ decays only via three-body modes is of lesser importance for final states containing photons.
The different bounds for the other scenarios considered are due to the relative size of the cross section for the triplet and quintuplet. In the case where the quintuplet is heavier than the triplet, the decay $\eta^{++}_5 \to W^{+*} \eta^+_3$ leads to additional photons stemming from the $\eta^+_3$ decays that increase the bound compared to the scenarios in which $\eta^{++}_5$ decays only into $W^+W^+$.

\begin{figure}
    \centering
    \includegraphics[width=\linewidth]{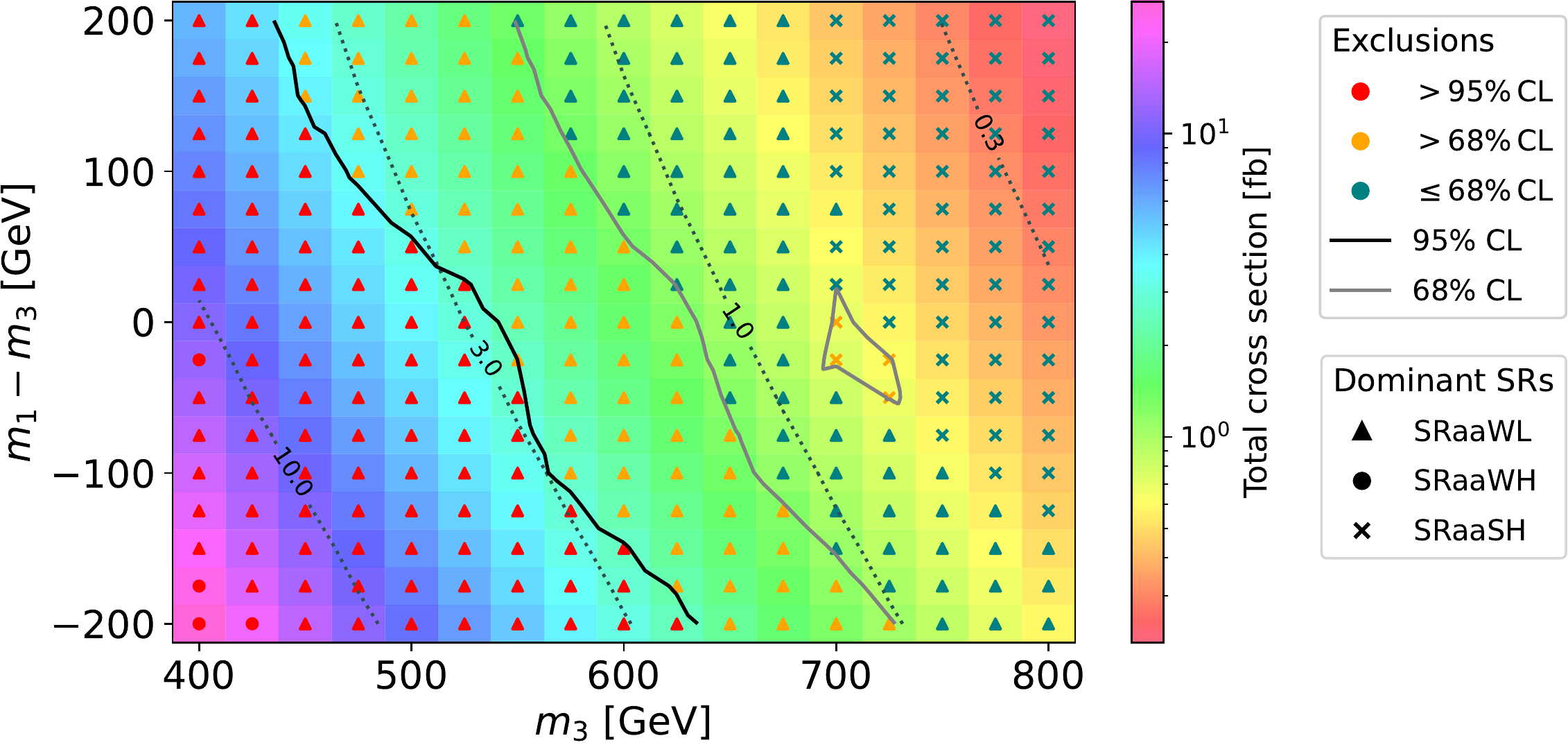}
    \caption{Bounds on the pNGB masses for the Drell-Yan production of the custodial triplet $\eta_3$ and singlet $\eta_1$ with the quintuplet $\eta_5$ decoupled (scenario S-31). Depending on the mass hierarchy, the pNGBs decay either into di-bosons or into one vector boson and a lighter pNGB. The heatmap and the dotted contours show the total cross section. The bounds are obtained from Ref.~\cite{ATLAS:2018nud}, with the dominant signal region indicated by the marker symbol. The 95\% and 68\% CL exclusion contours are drawn in solid black and gray, respectively.}
    \label{fig:bounds-m3m1}
\end{figure}

Finally, we consider a third case where one of the multiplets is effectively decoupled, and define two benchmarks: 
\begin{equation}
    \text{S-31}: \quad m_5\gg m_{3,1}\,; \qquad \qquad \text{S-35}: \quad m_1 \gg m_{3,5}\,.
\end{equation}
The case $m_3\gg m_{1,5}$ is already covered by our first example of this section since the singlet and quintuplet do not couple and only the quintuplet members are produced via Drell-Yan processes.
For both scenarios, we scan over the two light masses with a mass split of up to $200$~GeV and simulate the Drell-Yan production of two pNGBs.
In \cref{fig:bounds-m3m1}, we show the results for S-31 in the $m_3$-$\Delta m_{13}$ plane, where $\Delta m_{13} = m_1-m_3$.
In addition to the exclusion contours at 95\% CL (solid black) and 68\% CL (solid gray), we also show the sum over pair production cross sections as a heatmap with dotted contours.
This highlights interesting features in the form of regions where the bounds deviate from the cross section contours.
Following the 95\% CL bound, we identify three such regions: 
In the lower half, the triplets decay to the singlet and the final state is determined by the anomaly decays of $\eta_1^0$, see \cref{fig:eta10brs}.
From $\Delta m_{13} = - 200$~GeV to $-100$~GeV, the bounds grow weaker as the $W$ and $Z$ bosons from $\eta_3$-cascade decays get softer, followed by an increase towards $\Delta m_{13}=0$ as the $\eta_3^+\to W^+ \gamma$ decay sets in.
Finally, when the singlet is heavier than the triplet, the bounds are weaker again due to the decreasing $\br(\eta_3^0 \to \gamma\gamma Z)$.

\begin{figure}
    \centering
    \includegraphics[width=\linewidth]{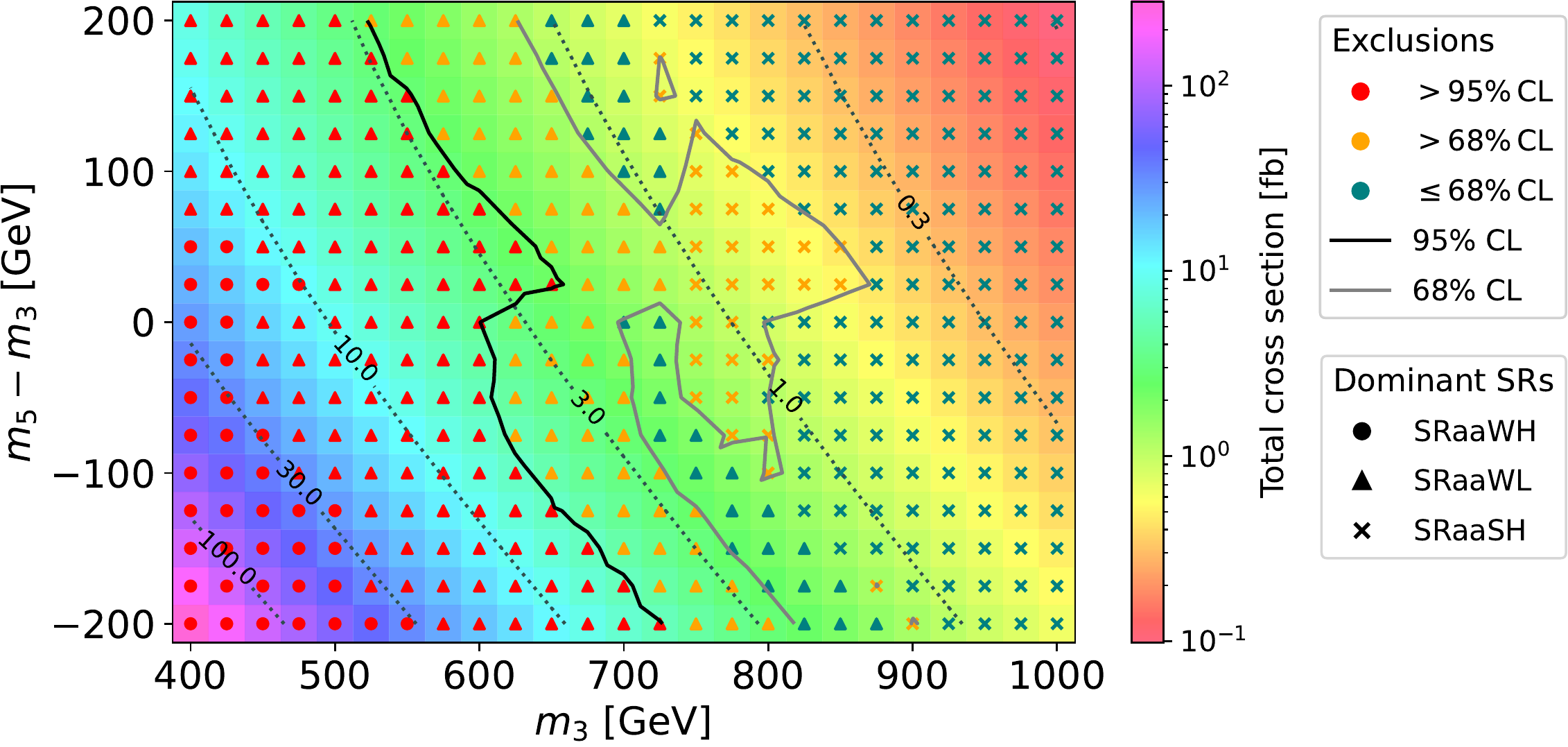}
    \caption{Bounds on the pNGB masses for the Drell-Yan production of the custodial triplet $\eta_3$ and quintuplet $\eta_5$ with the singlet $\eta_1$ decoupled (scenario S-35). Depending on the mass hierarchy, the pNGBs decay either into di-bosons or into one vector boson and a lighter pNGB. The heatmap and the dotted contours show the sum over all scalar pair production cross sections. The bounds are obtained from Ref.~\cite{ATLAS:2018nud}, with the dominant signal region indicated by the marker symbol. The 95\% and 68\% CL exclusion contours are drawn in solid black and grey, respectively.}
    \label{fig:bounds-m3m5}
\end{figure}

In \cref{fig:bounds-m3m5}, we show the bounds on $m_3$ and $m_5$ with the singlet decoupled, S-35.
To understand the features, we again follow the 95\% CL exclusion contour.
For negative $\Delta m_{53} = m_5 - m_3$, the quintuplet states decay via the anomaly.
The $\eta_3^{+}$ dominantly decays into the $\eta_5^{++}$, which cannot contribute photons to the final state.
Thus, the bounds increase relative to the cross section from $\Delta m_{53}=-50$~GeV as the $\eta_3^+$ anomaly decays become relevant.
For a positive mass split, the bounds rapidly increase.
The reason for this is that the $\eta_5^{++}$, which is produced with a large cross section, has a very small anomaly coupling so that already at $25$~GeV mass split the branching ratio of the cascade decay to $\eta_3^+$ is almost 100\%, resulting in a large photon production.
With increasing mass split, the bounds become weaker again. This is mainly
due to the dependence of the $\eta^0_3$ decays on the mass difference, see \cref{fig:eta30brs-m1m3}. In the nearly mass degenerate case the decays into $W$ bosons are strongly suppressed, leading to an enhancement of photons from the $\eta^0_3$ decays.

\subsection{LHC bounds in the fermiophilic case} \label{sec:boundsfermiophilic}

\begin{figure}[t]
	\centering
	\begin{subfigure}{0.48\linewidth}
		\includegraphics[width=\linewidth]{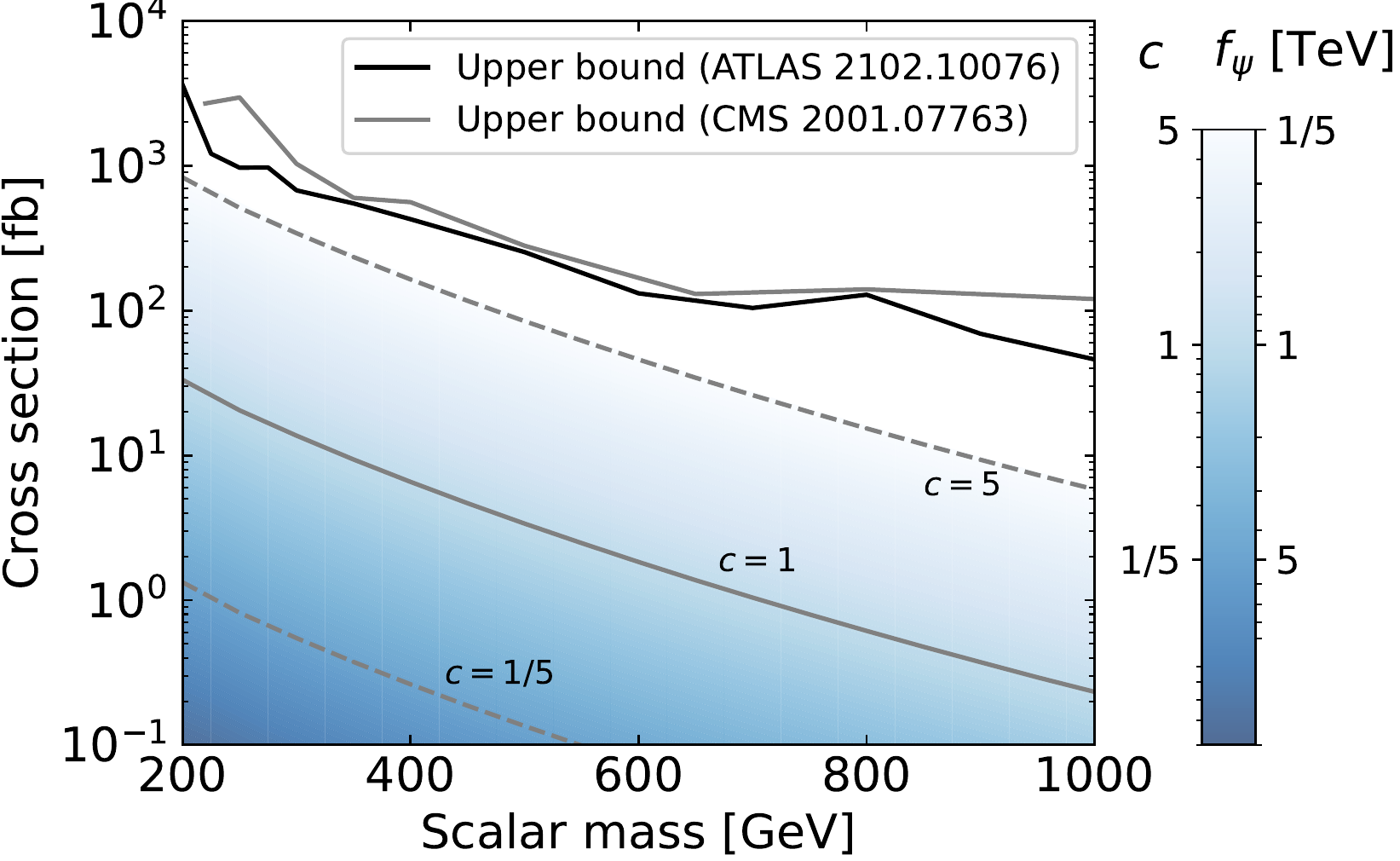}
		\caption{$S^+ tb \to tbtb$}
		\label{fig:s11tb}
	\end{subfigure} \quad
	\begin{subfigure}{0.48\linewidth}
		\includegraphics[width=\linewidth]{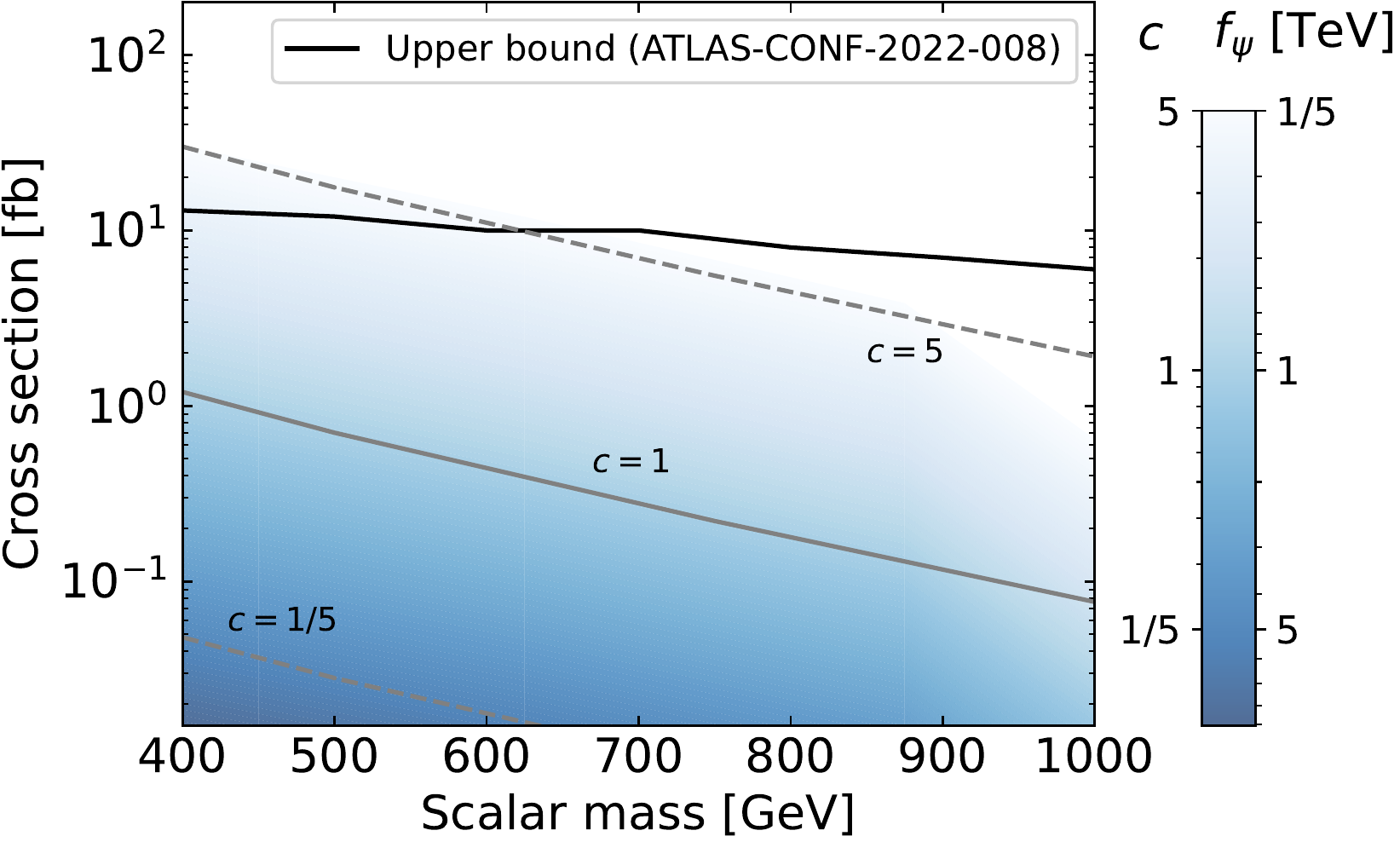}
		\caption{$S^0 tt \to tttt$}
		\label{fig:s10tt}
	\end{subfigure} \vspace{2ex}
	
	\begin{subfigure}{0.48\linewidth}
		\includegraphics[width=\linewidth]{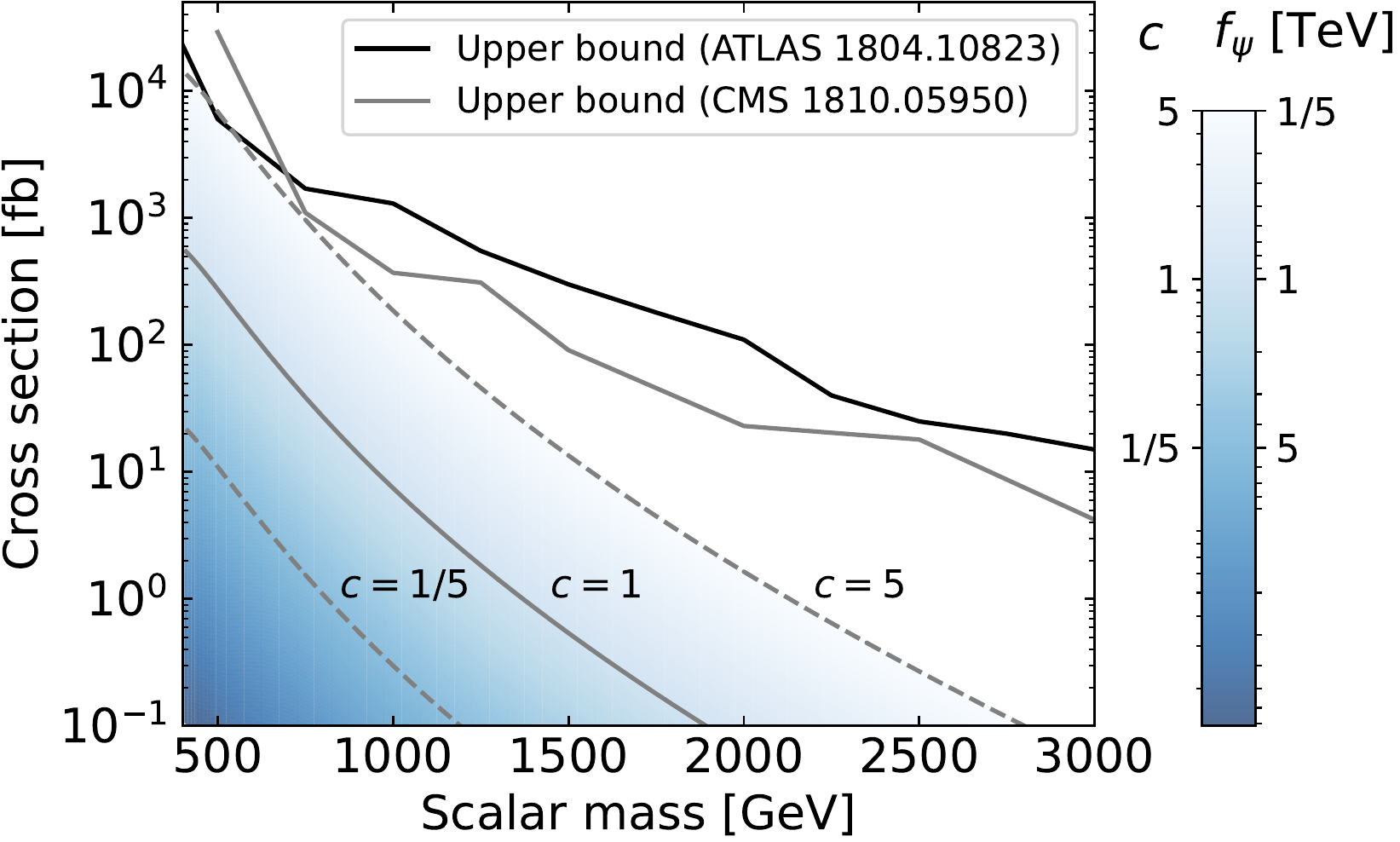}
		\caption{$S^0 \to tt$}
		\label{fig:s10ggF}
	\end{subfigure} \quad
        \begin{subfigure}{0.48\linewidth}
		\includegraphics[width=\linewidth]{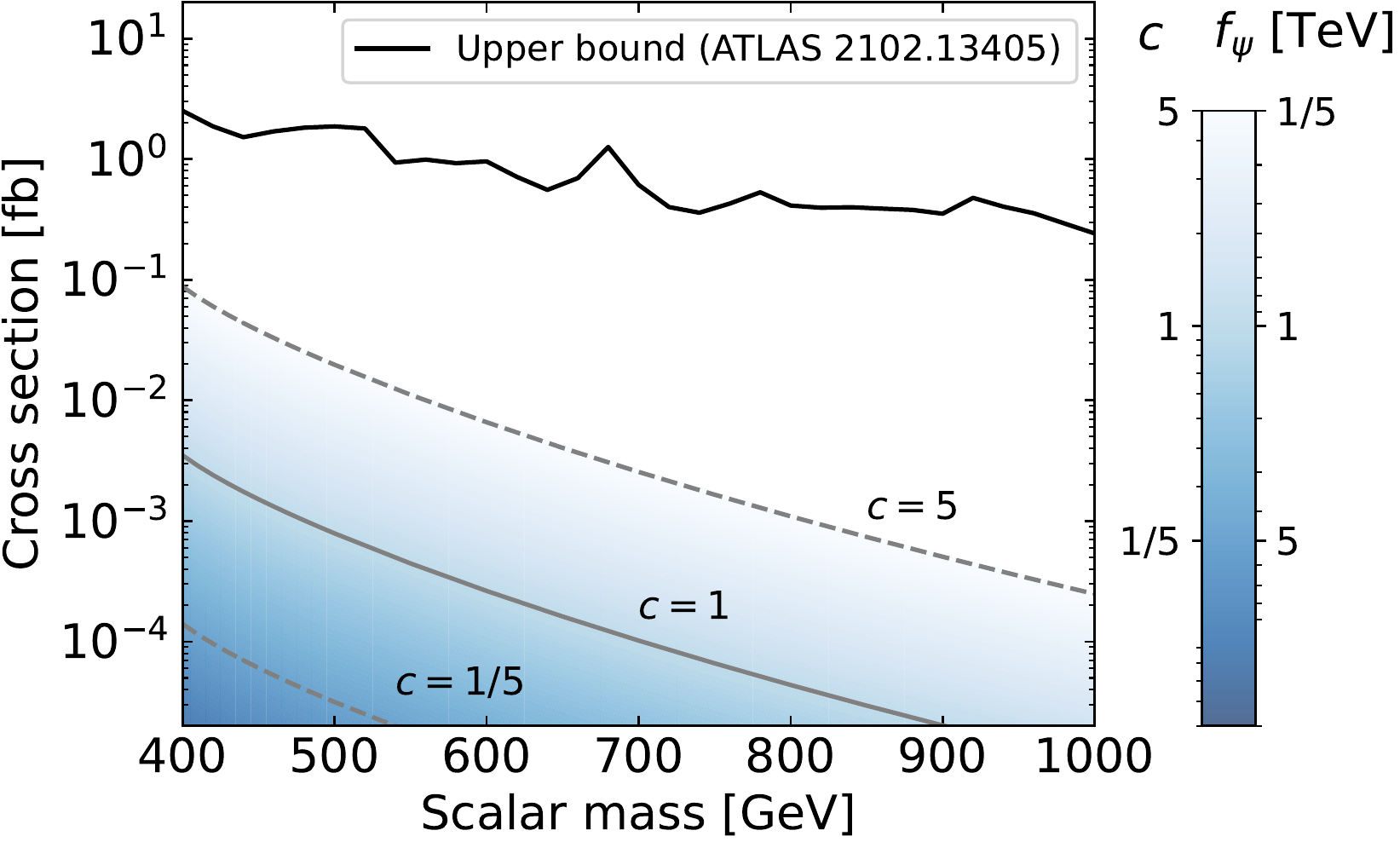}
		\caption{$S^0 \to \gamma \gamma$}
		\label{fig:s10ggFaa}
	\end{subfigure}
	\caption{Bounds on the single production of pNGBs in association with two third- generation quarks. The blue shaded area indicates typical cross sections assuming that only one scalar is present.  The coupling to the quarks is given by $c\, \frac{m_t}{f_\psi}$ 
	and the side band maps the blue shade to the corresponding value of $c$ (for $f_\psi = 1$~TeV) and $f_\psi$ (for $c=1$). From a model-building point of view, it is always possible to choose a larger scale,  $f_\psi > 1$~TeV.
	Hence, as $c \sim \mathcal{O}(1)$, the darker shaded regions correspond to theoretically more favourable parameter points than the light and white regions.}
	\label{fig:associated}
\end{figure}

We turn now to the scenarios in which the pNGBs couple to quarks, where decays via the anomaly are strongly suppressed and can be neglected, as already discussed in \cref{sec:modelindependent}.
In these scenarios, one has single scalar production
via the processes
\begin{align}
    pp\to S^0 t\bar{t} \quad
    \text{ and } \quad 
    pp\to S^\pm t b \,.
\end{align}
Moreover, the couplings of the neutral scalars to bottom and top quarks induce couplings to gluons and photons at the one-loop level. This leads to processes like
\begin{align}
    pp \to S^0  \to t\bar{t} \quad \text{ and } \quad
      pp \to S^0  \to \gamma \gamma \,.
\end{align}
We show in \cref{fig:associated} bounds on various processes
for $f_\psi=1$~TeV  and three different values of the factors $c$ defined in Eq. \eqref{eq:couplingscaling}: 1/5, 1 and 5. Note, that different values of $c$ and $f_\psi$ can be obtained by a simple rescaling of the $c=1$ line by a factor $(c/f_\psi)^2$, with $f_\psi$ in TeV. We compare available searches for
$pp \to S^\pm t b \to \bar t b t \bar b$ \cite{CMS:2019rlz,ATLAS:2021upq}, $pp \to H/A t \bar{t} \to t \bar t t \bar t$ \cite{ATLAS:2022ohr},
$pp \to Z' \to t\bar{t}$  \cite{ATLAS:2018rvc,CMS:2018rkg}
and $pp \to S^0 \to \gamma \gamma$ \cite{ATLAS:2021uiz}, from which we extract a limit on the respective signal cross section. For \cref{fig:s11tb} we use the renormalisation and factorisation scales $\mu_R=\mu_F=(m_t+m_b+m_{S^+})/3$ as this gives a K-factor very close to 1 \cite{Dittmaier:2009np}. For the other plots we have taken the cross sections from the Higgs Xsection working group \cite{Higgs:Xsec,LHCHiggsCrossSectionWorkingGroup:2016ypw} and have rescaled the Yukawa couplings accordingly.
We see that currently we do not get any bounds except for $c=5$ and $f_\psi=1$~TeV
in the 4 $t$ channel, \cref{fig:s10tt}, which gives
a bound of about 640~GeV on $m_{S^0}$. This corresponds to a rather small fraction of the available parameter space and if one reduces $(c=5)/(f_\psi=1$~TeV) by a factor $\simeq 1/\sqrt{3}$ one does not get any bound.

\begin{figure}
    \centering
    \includegraphics[width=0.6\linewidth]{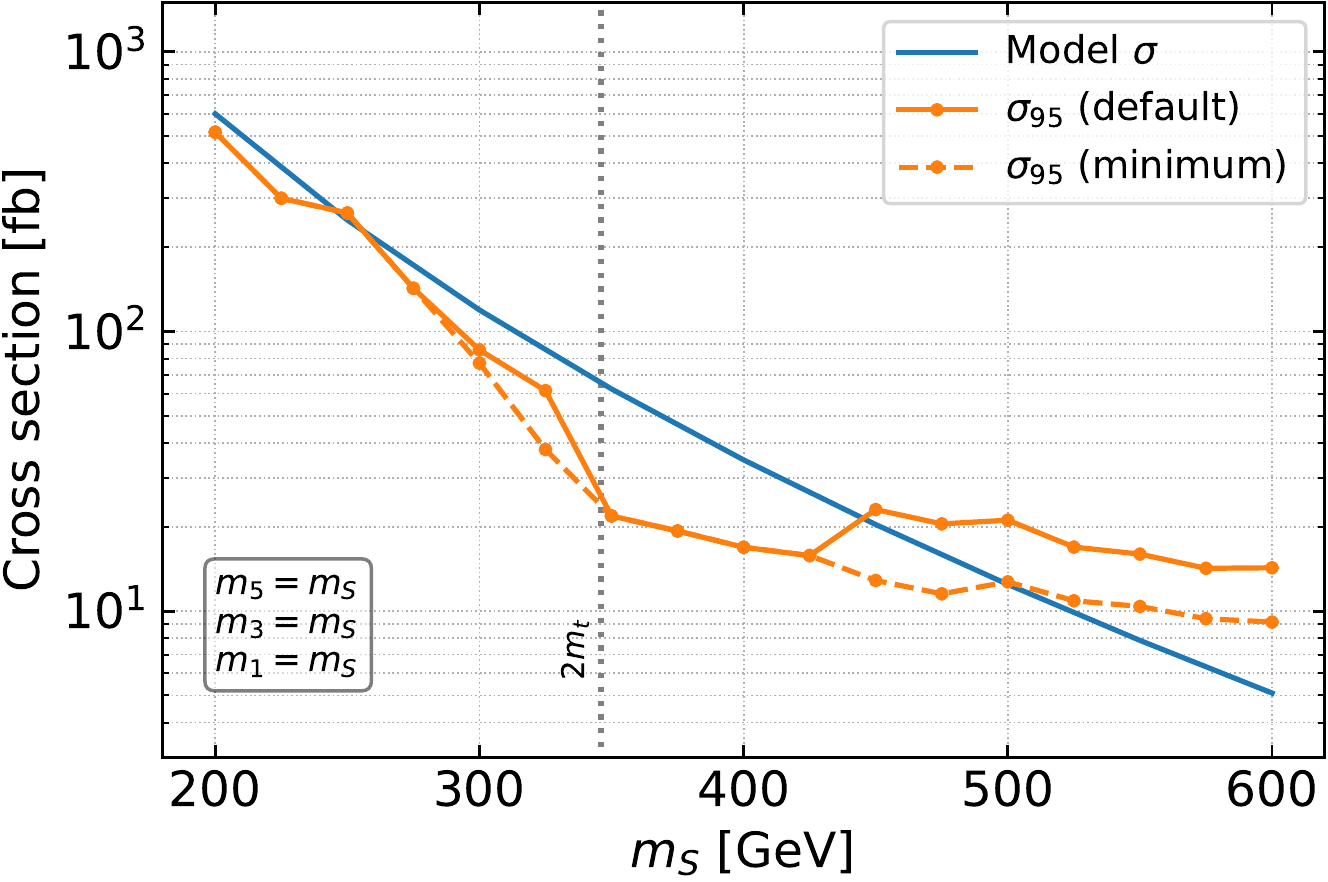}    
    \caption{Bounds on the pNGB masses for the Drell-Yan production of the full bitriplet with decays to third-generation quarks.}
    \label{fig:1dimbounds-quarks}
\end{figure}

We now turn to Drell-Yan pair production, for which we give our results in \cref{fig:1dimbounds-quarks}. Here we have assumed that all pNGBs have the same mass and all factors $c=1$ (neither branching ratios nor production cross sections depend on $f_\psi$). The blue line gives the total cross section summing over all pNGBs irrespective of their decay modes. The orange lines give the exclusion when considering all possible channels. They are dominated by Ref.~\cite{ATLAS:2021fbt} implemented in \texttt{CheckMATE}. Note that \texttt{CheckMATE} uses
the signal region with the strongest expected bound and reports the corresponding observed
bound as the final result. Using this standard procedure, one obtains the bound given by the solid orange line. However, this can lead to difficulties if observed and expected bounds differ significantly leading to the kinks at $m_S=350$~GeV and $450$~GeV. Modifying the procedure such that always the strongest observed bound is taken, one obtains a smoother curve for the limit, shown by the dashed orange line. This yields a somewhat stronger bound of about $500$~GeV. We detail the differences of these procedures in \cref{app:bestsignalregion}.

\section{Conclusions and outlook} 
\label{sec:conclusions}

In this work, we investigate the bounds on the Drell-Yan pair production of scalar bosons that carry electroweak charges at the LHC. We first consider all possible channels in a simplified model approach, leading to 32 distinct channels: 24 containing four vector bosons, and 8 with top and bottom quarks. The two scenarios arise from fermiophobic and fermiophilic models, respectively. The only channels that have dedicated searches contain doubly charged scalars decaying into a pair of same-charge $W$ bosons. For other channels, we use all the available recast searches for new physics and measurements of SM cross sections. These limits, showcased in \cref{fig:modelindependent}, can be applied to any model with an extended Higgs sector dominated by pair production.

As a concrete example, we focus on a composite Higgs model based on the coset $\SU(5)/\SO(5)$, which features a custodial bi-triplet. We show that the limits on individual channels lead to relatively weak bounds on the scalar masses. Instead, stronger bounds can be obtained by combining various pair production channels. Considering several benchmark scenarios, we establish limits on the scalar mass scale around $500-700$~GeV in the fermiophobic case. For decays into top and bottom quarks, the bounds are around $500$~GeV.

The main limitation of the simplified model approach is the restriction to searches and measurements that have been recast.
By determining limits from \texttt{MadAnalysis5}, \texttt{CheckMATE} and \texttt{Rivet/Contur}, we cover a considerable amount of analyses.
Still, there are many searches, not yet implemented, that have the potential to significantly improve these bounds.
Another limitation is in the combination of different searches, which is not possible without detailed knowledge of the experimental correlations between the various signal regions. 
Designing simple combination procedures, like the one proposed in Ref.~\cite{Araz:2022vtr}, could mitigate this issue.
Furthermore, the combination removes the ambiguity in choosing the most sensitive signal region.

Within its limitations, our analysis proves that current non-dedicated searches and standard model measurements impose significant bounds on extended Higgs sectors, which contain many scalar bosons with electroweak charges. Nevertheless, the variety of production channels and available final states leaves open the possibility to improve the coverage of this large class of models by means of dedicated searches. In the fermiophobic case, final states with photons are remarkable, as they already lead to the best bounds via generic multi-photon searches. The reach could be improved by searches targeting photon resonances or kinematic features related to the mass of the decaying scalar bosons. These channels are particularly relevant for composite models, where final states with photons are naturally abundant. In the fermiophilic case, final states with multi tops could be targeted by resonance searches.
A dedicated experimental search programme by ATLAS, CMS, and LHCb could immediately improve the experimental coverage of extended Higgs sectors. 

\begin{figure}[t]
	\centering
	\includegraphics[width=0.6\linewidth]{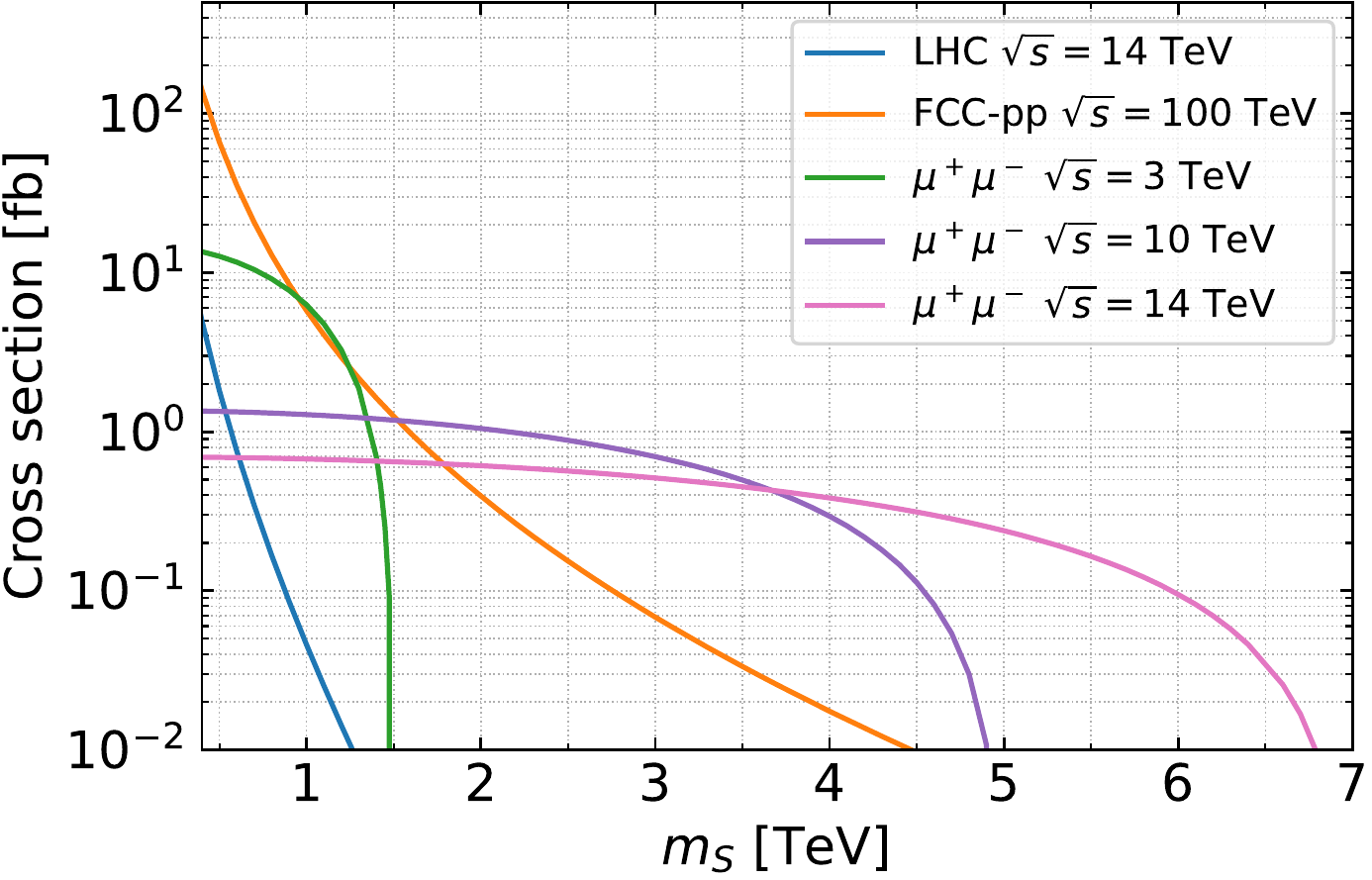} 
	\caption{Cross sections for the $\eta^{++}_5\eta^{--}_5$ pair production  at the $14$~TeV HL-LHC and some of the proposed future collider options.}
	\label{fig:xsfuturecolliders}
\end{figure}

The reach can clearly be extended further by prospective future colliders. We show in  \cref{fig:xsfuturecolliders} the cross section of a typical process, the pair production of the doubly charged $\eta^{++}_5$, for various colliders:  Besides the High-Luminosity LHC at $14$~TeV, we consider a $100$~TeV proton-proton collider and a muon-collider with a centre-of-mass energy of $3$, $10$, and $14$~TeV. For a $100$~TeV $pp$-collider with a typical integrated luminosity of $30$~ab$^{-1}$
\cite{Hinchliffe:2015qma,Arkani-Hamed:2015vfh},
by naive re-scaling we estimate its
reach to cover pNGB masses up to $4$~TeV. For the various muon-collider options, the reach should be close to $m_S\sim \sqrt{s}/2$ assuming that the integrated luminosity scales as $(\sqrt{s}/(10~ \mathrm{TeV}))^2\ 10^4$~ab$^{-1}$ \cite{Delahaye:2019omf,Han:2022edd}. Dedicated studies will be necessary to obtain more realistic values for the reach of the different collider options.

\section*{Acknowledgement}
We thank A.~Banerjee, J.~Butterworth, G.~Ferretti, K.~Rolbiecki and R.~Str\"ohmer for useful discussions.
This work has been supported by the ``DAAD, Frankreich'' and ``Partenariat Hubert Curien (PHC)'' PROCOPE 2021-2023, project number 57561441 as well as the international cooperation program ``GenKo'' managed by the National Research Foundation of Korea (No. 2022K2A9A2A15000153, FY2022) and DAAD, P33 - projekt-id 57608518.
G.C. and T.F. acknowledge support from the Campus-France STAR project n. 43566UG, ``Higgs and Dark Matter connections''.
T.F. is supported by a KIAS Individual Grant (AP083701) via the Center for AI and Natural Sciences at Korea Institute for Advanced Study.

\appendix

\section{Technical Notes}\label{app:technical}

\subsection{Choosing the best signal region}\label{app:bestsignalregion}

\begin{figure}
    \centering
    \includegraphics[width=0.7\linewidth]{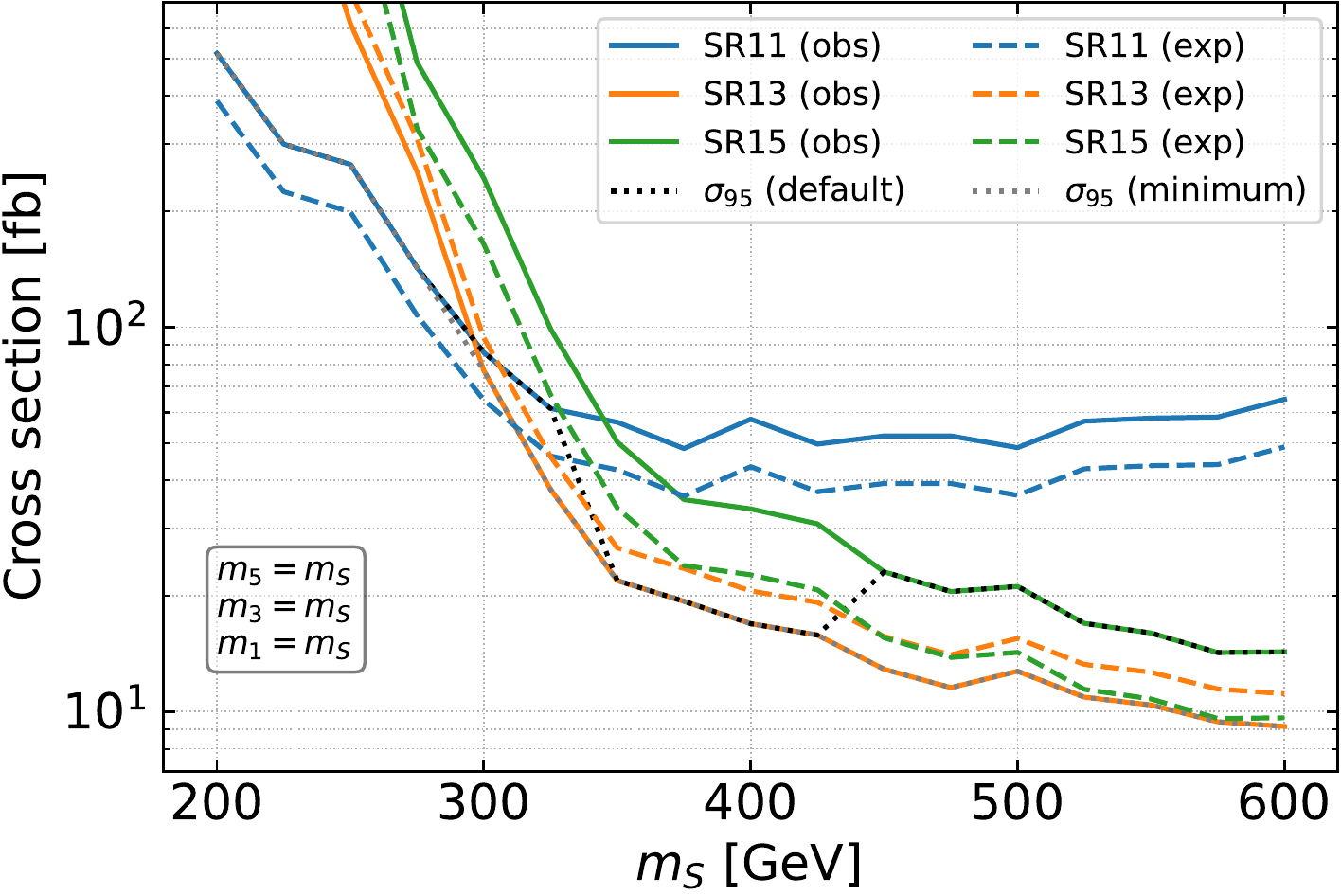}
    \caption{Bounds on the pNGB masses for the Drell-Yan production of the full bitriplet with decays to third-generation quarks.}
    \label{fig:1dimbounds-quarks-expobs}
\end{figure}

When choosing the most sensitive signal region for a given analysis, \texttt{CheckMATE} uses the signal region with the strongest expected bound but reports the corresponding observed bound as the final result.
This can lead to some unintuitive results when there is a large difference between the expected and observed bound, such as the sudden increase in $\sigma_{95}$ in \cref{fig:1dimbounds-quarks} at $m_S=450$~GeV.
To illustrate what causes this behaviour, we show the bounds from all relevant signal regions in \cref{fig:1dimbounds-quarks-expobs}:
The dominant analysis for the decays to quarks is Ref.~\cite{ATLAS:2021fbt}, and there are three important signal regions, SR11 (blue), SR13 (orange) and SR15 (green).
For each signal region, we show the observed bounds as a solid line and the expected bounds as a dashed line.
The black dotted line indicates the ``strongest'' bound $\sigma_{95}$ using the default method described above, while the grey dotted line is the $\sigma_{95}$ that is obtained by choosing the minimum of the observed bounds for each parameter point.

When there is one signal region that clearly dominates, such as SR11 for small masses, the default and minimum procedures coincide.
However, for $m_S\geq 450$~GeV, the expected bounds from SR13 and SR15 are very similar with SR15 being marginally more sensitive.
The default procedure then dictates using the observed bounds from SR15 for $\sigma_{95}$, although they are significantly weaker than the ones from SR13.
Given that the difference in the expected significance is small, we find it justified to use SR13 instead.

\subsection{List of dominant analyses}\label{app:analyses}

In \cref{fig:modelindependent} in the main text, we present upper limits on the Drell-Yan production cross section of electroweak scalars for a variety of decay channels.
Due to the different topologies of the resulting final states, the analyses that yield the strongest constraints differ among the various channels.
In this appendix we break down which analyses contribute to which decay channel.
\cref{tab:analyses} gives a brief description of the relevant analyses, including the recasting tool they are implemented in and their respective tool-internal name.
In \cref{tab:modelindependentanalyses}, we then list for each channel the analyses that give the dominant bound for at least one mass point. 
The full information is available on \url{https://github.com/manuelkunkel/scalarbounds}.

\begin{table}[]
\centering
\small
\begin{tabular}{lll}
    \toprule
    Analysis & Description & Recast \\ \midrule
    \begin{tabular}[c]{@{}l@{}}ATLAS JHEP \cite{ATLAS:2021jol}\\ $139~\mathrm{fb}^{-1}$\end{tabular} & \begin{tabular}[c]{@{}l@{}}$S^{++}S^{--}\to 4W$, $S^{++} S^- \to WW WZ$; \\ 2, 3 or 4 leptons, MET and jets\end{tabular} & -- \\[3ex]
    \begin{tabular}[c]{@{}l@{}}CMS PAS EXO-19-002 \cite{CMS:2019xud}\\ $137~\mathrm{fb}^{-1}$\end{tabular} & \begin{tabular}[c]{@{}l@{}}Type-III seesaw and light scalars;\\ at least 3 charged leptons\end{tabular} & \begin{tabular}[c]{@{}l@{}}\texttt{MadAnalysis5}\\ \texttt{cms\_exo\_19\_002}\end{tabular} \\[3ex]
    \begin{tabular}[c]{@{}l@{}}ATLAS PRD 97 \cite{ATLAS:2018nud}\\ $36.1~\mathrm{fb}^{-1}$\end{tabular} & \begin{tabular}[c]{@{}l@{}}Gauge mediated SUSY breaking;\\ (multi)photon and jets\end{tabular} & \begin{tabular}[c]{@{}l@{}}\texttt{CheckMATE}\\ \texttt{atlas\_1802\_03158}\end{tabular} \\[3ex]
    \begin{tabular}[c]{@{}l@{}}ATLAS JHEP \cite{ATLAS:2021mbt}\\ $139~\mathrm{fb}^{-1}$\end{tabular} & \begin{tabular}[c]{@{}l@{}}Measurement of prompt photon-pair \\ production\end{tabular} & \begin{tabular}[c]{@{}l@{}}\texttt{Rivet/Contur}\\ \texttt{ATLAS\_2021\_I1887997}\end{tabular} \\[3ex]
    \begin{tabular}[c]{@{}l@{}}ATLAS EPJ C 81 \cite{ATLAS:2021fbt}\\ $139~\mathrm{fb}^{-1}$\end{tabular} & \begin{tabular}[c]{@{}l@{}}RPV SUSY; many jets,\\ $\geq 1$ leptons and 0 or $\geq 3$ $b$-jets\end{tabular} & \begin{tabular}[c]{@{}l@{}}\texttt{CheckMATE}\\ \texttt{atlas\_2106\_09609}\end{tabular} \\[3ex]
    \begin{tabular}[c]{@{}l@{}}ATLAS EPJ C 81 \cite{ATLAS:2021twp}\\ $139~\mathrm{fb}^{-1}$\end{tabular} & \begin{tabular}[c]{@{}l@{}}Squarks and gluinos;\\ 1 lepton, jets and MET\end{tabular} & \begin{tabular}[c]{@{}l@{}}\texttt{CheckMATE}\\ \texttt{atlas\_2101\_01629}\end{tabular} \\[3ex]
    \begin{tabular}[c]{@{}l@{}}ATLAS EPJ C 79 \cite{ATLAS:2018zdn}\\ $3.2~\mathrm{fb}^{-1}$\end{tabular} & General search for new phenomena & \begin{tabular}[c]{@{}l@{}}\texttt{CheckMATE}\\ \texttt{atlas\_1807\_07447}\end{tabular} \\[3ex]
    \begin{tabular}[c]{@{}l@{}}ATLAS JHEP \cite{ATLAS:2019gdh}\\ $139~\mathrm{fb}^{-1}$\end{tabular} & \begin{tabular}[c]{@{}l@{}}Bottom-squark pair production; \\ no leptons, $\geq 3$ $b$-jets and MET\end{tabular} & \begin{tabular}[c]{@{}l@{}}\texttt{CheckMATE}\\ \texttt{atlas\_1908\_03122}\end{tabular} \\[3ex]
    \begin{tabular}[c]{@{}l@{}}CMS PAS SUS-19-006 \cite{CMS:2019xjf}\\ $137~\mathrm{fb}^{-1}$\end{tabular} & \begin{tabular}[c]{@{}l@{}}Gluinos and squarks;\\ no leptons, multiple jets and MET\end{tabular} & \begin{tabular}[c]{@{}l@{}}\texttt{MadAnalysis5}\\ \texttt{cms\_sus\_19\_006}\end{tabular} \\[3ex]
    \begin{tabular}[c]{@{}l@{}}CMS-SUS-16-033 \cite{CMS:2017abv}\\ $35.9~\mathrm{fb}^{-1}$\end{tabular} & \begin{tabular}[c]{@{}l@{}}Gluinos and stops;\\ no leptons, multiple jets and MET\end{tabular} & \begin{tabular}[c]{@{}l@{}}\texttt{MadAnalysis5}\\ \texttt{cms\_sus\_16\_033}\end{tabular} \\[3ex]
    \begin{tabular}[c]{@{}l@{}}ATLAS JHEP \cite{ATLAS:2020qlk}\\ $139~\mathrm{fb}^{-1}$\end{tabular} & \begin{tabular}[c]{@{}l@{}}Chargino-neutralino production;\\ MET and $h\to \gamma\gamma$\end{tabular} & \begin{tabular}[c]{@{}l@{}}\texttt{CheckMATE}\\ \texttt{atlas\_2004\_10894}\end{tabular} \\[3ex]
    \begin{tabular}[c]{@{}l@{}}ATLAS JHEP \cite{ATLAS:2021kog}\\ $139~\mathrm{fb}^{-1}$\end{tabular} & \begin{tabular}[c]{@{}l@{}}Measurements of four-lepton \\ differential cross sections\end{tabular} & \begin{tabular}[c]{@{}l@{}}\texttt{Rivet/Contur}\\ \texttt{ATLAS\_2021\_I1849535}\end{tabular} \\[3ex]
    \begin{tabular}[c]{@{}l@{}}ATLAS JHEP \cite{ATLAS:2019gey}\\ $139~\mathrm{fb}^{-1}$\end{tabular} & \begin{tabular}[c]{@{}l@{}}Measurement of the $Z(\to\ell^+ \ell^-)\gamma$ \\production cross section\end{tabular} & \begin{tabular}[c]{@{}l@{}}\texttt{Rivet/Contur}\\ \texttt{ATLAS\_2019\_I1764342}\end{tabular} \\[3ex]
    \begin{tabular}[c]{@{}l@{}}ATLAS JHEP \cite{ATLAS:2018nci}\\ $36.1~\mathrm{fb}^{-1}$\end{tabular} & \begin{tabular}[c]{@{}l@{}}Measurement of the $Z(\to\nu \bar\nu)\gamma$\\production cross section\end{tabular} & \begin{tabular}[c]{@{}l@{}}\texttt{Rivet/Contur}\\ \texttt{ATLAS\_2018\_I1698006}\end{tabular} \\[3ex]
    \begin{tabular}[c]{@{}l@{}}ATLAS-CONF-2016-096 \cite{ATLAS:2016uwq}\\ $13.3~\mathrm{fb}^{-1}$\end{tabular} & \begin{tabular}[c]{@{}l@{}}Electroweakino production; \\ 2 to 3 leptons, MET and no jets\end{tabular} & \begin{tabular}[c]{@{}l@{}}\texttt{CheckMATE}\\ \texttt{atlas\_conf\_2016\_096}\end{tabular} \\[3ex]
    \begin{tabular}[c]{@{}l@{}}CMS PAS SUS-16-039 \cite{CMS:2017moi}\\ $35.9~\mathrm{fb}^{-1}$\end{tabular} & \begin{tabular}[c]{@{}l@{}}Electroweakino production; \\ $\geq 2$ leptons and MET\end{tabular} & \begin{tabular}[c]{@{}l@{}}\texttt{CheckMATE}\\ \texttt{cms\_sus\_16\_039}\end{tabular} \\\bottomrule
\end{tabular}
\caption{Summary of the analyses that contribute to the simplified model bounds in \cref{fig:modelindependent}.}
\label{tab:analyses}
\end{table}

\begin{table}[]
	\centering
	\ra{1.15}
	\begin{tabular}{ccccc}\hline
		Production & Channel & \texttt{MadAnalysis5} & \texttt{CheckMATE} & \texttt{Rivet/Contur} \\ \hline
		$S^{++} S^{--}$ & $WWWW$ & \cite{CMS:2019xud} & \cite{CMS:2017moi} &  \\ \hline
		\multirow{2}{*}{$S^{\pm\pm} S^{\mp}$} & $WWWZ$ &  \cite{CMS:2019xud} & \cite{CMS:2017moi} &  \\
		& $WWW\gamma$ &  &  & \cite{ATLAS:2019gey, ATLAS:2018nci} \\ \hline
		\multirow{3}{*}{$S^+ S^-$} & $WZWZ$ &  & \cite{CMS:2017moi} &  \\
		& $WZW\gamma$ &  &  & \cite{ATLAS:2019gey, ATLAS:2018nci} \\
		& $W\gamma W\gamma$ & &\cite{ATLAS:2018nud, ATLAS:2020qlk}  &  \\ \hline
		\multirow{8}{*}{$S^\pm S^0$} & $WZWW$ & \cite{CMS:2019xud} & \cite{CMS:2017moi} &  \\
		& $W\gamma WW$ &  & & \cite{ATLAS:2019gey, ATLAS:2018nci} \\
		& $WZZZ$ &  & \cite{CMS:2017moi} &  \\
		& $(W\gamma)(ZZ)$ & & & \cite{ATLAS:2019gey, ATLAS:2021kog, ATLAS:2018nci} \\
		& $(WZ)(Z\gamma)$ & & & \cite{ATLAS:2019gey, ATLAS:2018nci} \\
		& $(WZ)(\gamma\gamma)$ &  & \cite{ATLAS:2018nud} &  \\
		& $(W\gamma)(Z\gamma)$ &  & \cite{ATLAS:2020qlk, ATLAS:2018nud} &  \\
		& $W\gamma\gamma\gamma$ &  & \cite{ATLAS:2020qlk, ATLAS:2018zdn, ATLAS:2018nud} & \\ \hline
		\multirow{10}{*}{$S^0 S^{\prime 0}$} & $WWWW$ & \cite{CMS:2019xud} & \cite{CMS:2017moi} &  \\
		& $WWZZ$ &  & \cite{CMS:2017moi} &  \\
		& $WW\gamma Z$ &  &  & \cite{ATLAS:2019gey, ATLAS:2018nci} \\
		& $WW\gamma\gamma$ &  & \cite{ATLAS:2018nud} &  \\
		& $ZZZZ$ &  & \cite{CMS:2017moi} & \cite{ATLAS:2021kog} \\
		& $\gamma ZZZ$ &  &  & \cite{ATLAS:2018nci, ATLAS:2019gey, ATLAS:2021kog} \\
		& $(\gamma Z)(\gamma Z)$ &  & \cite{ATLAS:2018nud,ATLAS:2020qlk} & \cite{ATLAS:2021kog, ATLAS:2019gey} \\
		& $(\gamma\gamma) (ZZ)$ &  & \cite{ATLAS:2018nud} &  \\
		& $\gamma\gamma\gamma Z$ &  & \cite{ATLAS:2020qlk, ATLAS:2018nud} &  \\
		& $\gamma\gamma\gamma\gamma$ &  & \cite{ATLAS:2018zdn} & \cite{ATLAS:2021mbt} \\ \hline \hline
		$S^{++} S^{--}$ & $WtbWtb$ &  & \cite{ATLAS:2021fbt}  &  \\ \hline
		$S^{\pm\pm} S^\mp$ & $Wtbtb$ &  & \cite{ATLAS:2021fbt} &  \\ \hline
		$S^+ S^-$ & $tbtb$ &  & \cite{ATLAS:2021fbt,ATLAS:2021twp} &  \\ \hline 
		\multirow{2}{*}{$S^\pm S^0$} & $tbtt$ &  & \cite{ATLAS:2021fbt} &  \\
		& $tbbb$ & \cite{CMS:2019xjf} & \cite{ATLAS:2018zdn, ATLAS:2021fbt, ATLAS:2021twp, ATLAS:2019gdh} &  \\ \hline
		\multirow{3}{*}{$S^0 S^{\prime 0}$} & $tttt$ &  & \cite{ATLAS:2021fbt} &  \\
		& $ttbb$ & \cite{CMS:2017abv} &  \cite{ATLAS:2021fbt, ATLAS:2021twp} & \\
		& $bbbb$ & \cite{CMS:2017abv} & \cite{ATLAS:2018zdn} & \\ \hline
	\end{tabular}
	\caption{Experimental analyses contributing to the simplified model bounds in \cref{fig:modelindependent}. More details are available on \url{https://github.com/manuelkunkel/scalarbounds}.}
	\label{tab:modelindependentanalyses}
\end{table}

\newpage

\section{$K^{SS}_V$ coefficients from the scalar kinetic term}\label{app:coeffs}

As outlined in the main text, the Drell-Yan pair production process of two gauge eigenstate scalars of an $\SU(2)_L$ multiplet arises from a coupling in the kinetic term of the scalar, and as such, it depends only on the $\SU(2)_L\times \U(1)_Y$ quantum numbers of the scalar. 

As a first example, we review the calculation for a complex scalar $\SU(2)_L$ triplet with hypercharge $Y$ which we denote by $\phi_{3,Y}$.
We write $\phi_{3,Y}$ as
\begin{align*}
\phi_{3,Y} = \frac{1}{2}\begin{pmatrix}
                \phi_{3,Y}^3 & \phi_{3,Y}^1 - i \phi_{3,Y}^2\\
                \phi_{3,Y}^1 + i \phi_{3,Y}^2 & -\phi_{3,Y}^3\\
             \end{pmatrix} = \frac{1}{2}\begin{pmatrix}
                \phi_{3,Y}^0 & \sqrt{2}\phi_{3,Y}^+\\
                \sqrt{2}\phi_{3,Y}^- & -\phi_{3,Y}^0\\
             \end{pmatrix}.
\end{align*}
The covariant derivative is
\begin{align*}
D_\mu \phi_{3,Y} &= \partial_\mu \phi_{3,Y} - i g [W_\mu, \phi_{3,Y}] - i g' Y B_\mu \phi_{3,Y}.
\end{align*}
The kinetic term reads
\begin{align*}
    \mathcal L_{\phi_{3,Y, \rm{kin}}} =& 2\Tr(D_\mu \phi_{3,Y})^\dagger (D^\mu \phi_{3,Y}) \\
    =&\partial_\mu \phi_{3,Y}^{a*} \partial^\mu \phi^a + i e q^a A^\mu \phi_{3,Y}^{a*} \dlr \phi_{3,Y}^a + i \frac{e}{s_w c_w} (t_3^a - q^a s_w^2) Z^\mu \phi_{3,Y}^{a*} \dlr \phi_{3,Y}^a\\
    &+ i g W^{\mu, -} \left( \phi_{3,Y}^{-*} \dlr \phi_{3,Y}^{0} - \phi_{3,Y}^{0*} \dlr \phi_{3,Y}^{+} \right) + \hc + \ord{g^2}
\end{align*}
where $q^a = t_3^a + Y$ is the electric charge of the field $\phi_{3,Y}^a$ and $t_3^a$ is the corresponding eigenvalue of $T^3$. Here, $a \in \{+, 0, -\}$ indicates the $T^3$ quantum numbers.
Note that for a complex representation, $\phi^{+*} \neq \phi_{3,Y}^-$ and $\phi_{3,Y}^{*0} \neq \phi_{3,Y}^0$. Comparing this with Eq. \eqref{eq:L_SSV} and expressing the fields via the electric charge yields the coefficients
\begin{align} \label{app:coeffstrip}
    K_{W}^{\phi_{3,Y}^{1-Y} \phi_{3,Y}^{0+Y}} &= 1\, ,\\
    K_{W}^{\phi_{3,Y}^{0-Y} \phi_{3,Y}^{1+Y}} &= -1\, ,\\
    K_Z^{\phi_{3,Y}^{(t_3^a + Y)} \phi_{3,Y}^{-(t_3^a + Y)}} &=-(t_3^a - s^2_w q^a).
\end{align}
Note that for a real representation with $Y=0$ we have $\phi^{+*} = \phi^{-}$ and $\phi^{0*} = \phi^{0}$. The kinetic term is then given by $\mathcal L \phi_{3,Y} = \Tr (D_\mu \phi_{3,Y})^\dagger D^\mu \phi_{3,Y}$, which yields the same coefficients as \cref{app:coeffstrip} with $Y=0$.
\bigskip

As a second example, we present the $K^{SS}_V$ coefficients of the pNGBs from $\SU(5)/\SO(5)$ breaking. They are determined analogously, have been presented in \cite{Banerjee:2022xmu} and are listed in the following for completeness:
\begin{equation*}
\def\arraystretch{1.4}
\begin{array}{cccccccccc}
\toprule 
&\multicolumn{2}{c}{K^{S^0_iS^+_j}_W} & K^{S^-_iS^{++}_j}_W\\
\midrule  
& \eta^+_3 & \eta^+_5 & \eta_5^{++}\\ 
\midrule 
h &  0 & 0 & \multirow{5}{*}{$-$}\\
\eta^0_3 & -\frac{i}{2} & \frac{c_\theta}{2} &\\ 
\eta^0_5 & -\frac{c_\theta}{2\sqrt{3}} & \frac{i\sqrt{3}}{2} &\\
\eta^0_1 & \sqrt{\frac{2}{3}}c_\theta & 0 &\\
\eta &  0 & 0 &\\
\eta^-_3 & \multicolumn{2}{c}{\multirow{2}{*}{$-$}} & \frac{c_\theta}{\sqrt{2}} \\ 
\eta^-_5 & & & -\frac{i}{\sqrt{2}}\\
\vphantom{--}\\
\bottomrule
\end{array}
\quad
\begin{array}{ccccccccccccc}
\toprule 
&\multicolumn{5}{c}{K^{S^0_iS^0_j}_Z} & \multicolumn{2}{c}{K^{S^+_iS^{-}_j}_Z} & K^{S^{++}_iS^{--}_j}_Z \\
\midrule 
& h & \eta^0_3 & \eta^0_5 & \eta^0_1 & \eta & \eta^-_3 & \eta^-_5 & \eta_5^{--}\\ 
\midrule  
h &  0 & 0 & 0 & 0 & 0 & \multicolumn{2}{c}{\multirow{5}{*}{$-$}} & \multirow{5}{*}{$-$}\\
\eta^0_3 & & 0 & \frac{i c_\theta}{\sqrt{3}} & i\sqrt{\frac{2}{3}}c_\theta & 0\\ 
\eta^0_5 & & & 0 & 0 & 0 \\
\eta^0_1 & & & & 0 & 0 \\
\eta &  & & & & 0 \\
\eta^+_3 & \multicolumn{5}{c}{\multirow{2}{*}{$-$}} & -\frac{c_{2w}}{2} & -\frac{ic_\theta}{2} \\ 
\eta^+_5 & & & & & & & -\frac{c_{2w}}{2} \\
\eta^{++}_5 & \multicolumn{7}{c}{-} & -c_{2w} \\
\bottomrule 
\end{array}
\end{equation*}
where $c_\theta={\rm cos}(\theta)$ with $\sin (\theta) = v/f_\psi$ and $c_{2w}=\rm{cos}(2\theta_W)$ is the cosine of twice the Weinberg angle.

\newpage

\bibliographystyle{utphys}
\bibliography{literature}

\end{document}